\documentclass[trackchanges,twocolumn]{aastex7}
\usepackage{amsmath} 
\usepackage{bm}
\usepackage{booktabs}
\usepackage{multirow}
\usepackage{makecell} 
\usepackage{tabularx}

\begin{document}

\title{Cosmological Prediction from the joint observation of MeerKAT and CSST at $z$ = 0.4 $\sim$ 1.2}

\correspondingauthor{Yan Gong}
\email{Email: gongyan@bao.ac.cn}

\author[orcid=0009-0001-5169-3519]{Yu-Er Jiang}
\affiliation{National Astronomical Observatories, Chinese Academy of Sciences, 20A Datun Road, Beijing 100101, China}
\affiliation{School of Astronomy and Space Science, University of Chinese Academy of Sciences(UCAS), \\Yuquan Road No.19A Beijing 100049, China}
\affiliation{Department of Physics, Stellenbosch University, Matieland 7602, South Africa} 
\email{yejiang@bao.ac.cn}

\author[orcid=0000-0003-0709-0101]{Yan Gong*}
\affiliation{National Astronomical Observatories, Chinese Academy of Sciences, 20A Datun Road, Beijing 100101, China}
\affiliation{School of Astronomy and Space Science, University of Chinese Academy of Sciences(UCAS), \\Yuquan Road No.19A Beijing 100049, China}
\affiliation{Science Center for Chinese Space Station Survey Telescope, National Astronomical Observatories, \\Chinese Academy of Science, 20A Datun Road, Beijing 100101, China}
\email{gongyan@bao.ac.cn}

\author{Qi Xiong}
\affiliation{National Astronomical Observatories, Chinese Academy of Sciences, 20A Datun Road, Beijing 100101, China}
\affiliation{School of Astronomy and Space Science, University of Chinese Academy of Sciences(UCAS), \\Yuquan Road No.19A Beijing 100049, China}
\email{xiongqi@bao.ac.cn}

\author{Wenxiang Pei}
\affiliation{National Astronomical Observatories, Chinese Academy of Sciences, 20A Datun Road, Beijing 100101, China}
\affiliation{School of Astronomy and Space Science, University of Chinese Academy of Sciences(UCAS), \\Yuquan Road No.19A Beijing 100049, China}
\email{wxpei@nao.cas.cn}

\author{Yun Liu}
\affiliation{National Astronomical Observatories, Chinese Academy of Sciences, 20A Datun Road, Beijing 100101, China}
\affiliation{School of Astronomy and Space Science, University of Chinese Academy of Sciences(UCAS), \\Yuquan Road No.19A Beijing 100049, China}
\email{liuyun@nao.cas.cn}

\author{Furen Deng}
\affiliation{National Astronomical Observatories, Chinese Academy of Sciences, 20A Datun Road, Beijing 100101, China}
\affiliation{School of Astronomy and Space Science, University of Chinese Academy of Sciences(UCAS), \\Yuquan Road No.19A Beijing 100049, China}
\email{frdeng@bao.ac.cn}

\author[orcid=0000-0002-2485-5762]{Zi-yan Yuwen}
\affiliation{Institute of Theoretical Physics, Chinese Academy of Sciences (CAS), Beijing 100190, China}
\affiliation{School of Physics, University of Chinese Academy of Sciences(UCAS), Yuquan Road No.19A Beijing 100049, China}
\affiliation{Department of Physics, Stellenbosch University, Matieland 7602, South Africa} 
\email{yuwenziyan@itp.ac.cn}

\author{Meng Zhang}
\affiliation{National Astronomical Observatories, Chinese Academy of Sciences, 20A Datun Road, Beijing 100101, China}
\affiliation{School of Astronomy and Space Science, University of Chinese Academy of Sciences(UCAS), \\Yuquan Road No.19A Beijing 100049, China}
\email{zhangmeng@nao.cas.cn}

\author{Xingchen Zhou}
\affiliation{National Astronomical Observatories, Chinese Academy of Sciences, 20A Datun Road, Beijing 100101, China}
\affiliation{School of Astronomy and Space Science, University of Chinese Academy of Sciences(UCAS), \\Yuquan Road No.19A Beijing 100049, China}
\email{xczhou@nao.cas.cn}

\author{Xuelei Chen}
\affiliation{National Astronomical Observatories, Chinese Academy of Sciences, 20A Datun Road, Beijing 100101, China}
\affiliation{School of Astronomy and Space Science, University of Chinese Academy of Sciences(UCAS), \\Yuquan Road No.19A Beijing 100049, China}
\affiliation{Department of Physics, College of Sciences, Northeastern University, Shenyang 110819, China}
\affiliation{Centre for High Energy Physics, Peking University, Beijing 100871, China}
\affiliation{State Key Laboratory of Radio Astronomy and Technology, China}
\email{xuelei@bao.ac.cn}

\author{Yin-Zhe Ma}
\affiliation{Department of Physics, Stellenbosch University, Matieland 7602, South Africa}
\email{mayinzhe@sun.ac.za}

\author{Qi Guo}
\affiliation{National Astronomical Observatories, Chinese Academy of Sciences, 20A Datun Road, Beijing 100101, China}
\affiliation{School of Astronomy and Space Science, University of Chinese Academy of Sciences(UCAS), \\Yuquan Road No.19A Beijing 100049, China}
\email{guoqi@nao.cas.cn}

\author{Bin Yue}
\affiliation{National Astronomical Observatories, Chinese Academy of Sciences, 20A Datun Road, Beijing 100101, China}
\affiliation{State Key Laboratory of Radio Astronomy and Technology, China}
\email{yuebin@nao.cas.cn}

\begin{abstract}

Cross-correlating neutral hydrogen (H\textsc{i}) 21cm intensity mapping with galaxy surveys provides an effective probe of astrophysical and cosmological information. This work presents a cross-correlation analysis between MeerKAT single-dish H\textsc{i} intensity mapping and Chinese Space Station Survey Telescope (CSST) spectroscopic galaxy surveys in $z=0.4\sim1.2$, which will share a survey area of several thousand square degrees. Utilizing Jiutian-1G cosmological simulation, we simulate the observational data of MeerKAT and CSST with survey areas from $\sim1600$ to $600$ deg$^2$ at $z=0.5$, 0.7, and 1. The effects of beam pattern, polarization leakage, and different foregrounds in the MeerKAT H\textsc{i} intensity mapping are considered in the simulation. After employing foreground removal with principal component analysis (PCA) method and performing signal compensation, we  derive the cross-power spectra of MeerKAT and CSST. We perform the joint constraint using the CSST galaxy auto-power spectra and MeerKAT-CSST cross-power spectra with the least-squares fitting method. The constraint results show that, in the simulated survey area, the relative accuracy can achieve $6\%\sim 8\%$ for the parameter products $\Omega_{\rm H\textsc{i}}b_{\rm H\textsc{i}}b_{g}r_{\mathrm{H\textsc{i}},g}$ and $\Omega_{\rm H\textsc{i}}b_{\rm H\textsc{i}}r_{\mathrm{H\textsc{i}},g}$ at the three redshifts, which is $3\sim4$ times smaller than the current result. These findings indicate that the full MeerKAT-CSST joint observation with thousands of square degrees overlapping survey area can be a powerful probe of large-scale structure, and has the ability to provide information of cosmic evolution of H\textsc{i} and galaxies in a wide redshift range.

\end{abstract}

\keywords{Cosmology --- large-scale structure of universe --- cosmological parameters}

\section{Introduction}

Probing the large scale structure (LSS) is the main approach to understand the cosmic evolution. In the decades since the cosmic microwave background (CMB) has been detected, cosmologists have made significant strides in mapping and studying  the Universe using different methods and techniques. Line intensity mapping (LIM) has been proposed as a novel and efficient technique to probe our Universe, by leveraging emission lines of atoms and molecules, such as H\textsc{i}, [C\textsc{ii}], [O\textsc{ii}], [O\textsc{iii}], CO, Ly$\alpha$, H$\alpha$, H$\beta$, etc.\citep{2010JCAP...11..016V, 2011ApJ...728L..46G, 2011ApJ...730L..30C, Lidz_2011, Gong_2012,  Silva_2013, Gong_2014, Pullen_2014, Uzgil_2014, Silva_2015, 10.1093/mnras/stw2470, Gong_2017}. In the LIM, huge spatial volumes can be efficiently explored by precisely detecting the cumulative signal within voxels, and the redshift can be accurately measured via the frequency shift of emission line. Furthermore, different emission lines tracing different physical processes and atoms or molecules can provide a multi-probe perspective to study the cosmic evolution.

Among various emission lines, 21cm line from the atomic neutral hydrogen (H\textsc{i}) is of particular interest \citep{pointsource, fitting1, noise, Villaescusa-Navarro_2018}. 
Many radio telescopes have planned or performed the 21cm intensity mapping projects, including Square Kilometer Array (SKA) \citep{SKA1, MeerKAT19cali_Wang, MeerKAT19, SKA2}, Parkes \citep{Parkes_2dF}, 
Green Bank Telescope (GBT) \citep{GBT_deep2, GBT_WiggleZ, GBT_eBOSS}, 
Canadian Hydrogen Intensity Mapping Experiment (CHIME)\citep{CHIME1, CHIME2}, 
Five-hundred-meter Aperture Spherical radio Telescope (FAST) \citep{FAST1, dfr}, 
Tianlai \citep{Tianlai1, Tianlai2}, 
Baryon Acoustic Oscilations from Integrated Neutral Gas Observations (Bingo) \citep{Bingo1, Bingo2}, etc.

In H\textsc{i} intensity mapping observation, the foreground contamination removal or reduction is one of the main challenges. While the Galactic emission can exceed the 21cm signal by four to five orders of magnitude, the continuum emission from extragalactic radio sources can further contaminate the observed intensity maps. In order to extract the 21cm signal, various foreground removal algorithms have been developed, such as the blind separation techniques like principal component analysis (PCA) \citep{PCA1, PCA2, PCA3} 
and independent component analysis (ICA) \citep{ICA1, ICA2}, 
which exploit different frequency smoothness of foreground and signal, the polynomial/parametric-fitting method, which model the foreground emission based on its physical properties \citep{fitting1}, and machine learning (ML) methods \citep[e.g.][]{ML1,DL}.

Furthermore, the instrumental systematics, like the non-gaussian beam patterns and polarization leakage, will increase the complexity of contamination mixed with the 21cm signal, leading to inevitable signal loss and foreground residuals. To extract the 21cm signal and improve the signal-to-noise ratio (SNR), many experiments conduct cross-correlation of 21cm intensity mapping with the optical galaxy survey in the same sky area, and positive results have been obtained. 
For instance, the GBT has successfully detected cross-correlations between 21cm intensity maps and optical galaxy surveys including DEEP2 \citep{GBT_deep2}, WiggleZ \citep{GBT_WiggleZ}, and eBOSS \citep{GBT_eBOSS}. Similarly, the Parkes telescope has reported 21cm-galaxy correlations using the 2dF galaxy survey \citep{Parkes_2dF}. Most recently, MeerKAT achieved a new milestone by correlating 21cm intensity maps with the WiggleZ survey \citep{MeerKAT_WiggleZ} and GAMA survey \citep{MeerKAT_GAMA}.
All these experiments provide their constraint on H\textsc{i}-galaxy cross-correlation parameter product $\Omega_{\mathrm{H\textsc{i}}} b_{\mathrm{H\textsc{i}}} r_{\mathrm{H\textsc{i}},g}$ at different redshifts, where $\Omega_{\mathrm{H\textsc{i}}}$, $b_{\mathrm{H\textsc{i}}}$ and $r_{\mathrm{H\textsc{i}},g}$ are the H\textsc{i} energy density parameter, H\textsc{i} bias, and correlation coefficient of H\textsc{i} and galaxy, respectively.

In this work, we utilize N-body simulation and semi-analytical model to study the cross-correlation between MeerKAT H\textsc{i} intensity mapping and Chinese Space Station Survey Telescope (CSST) spectroscopic galaxy survey, and forecast the cosmological constraint power. MeerKAT is a pathfinder project of the SKA \citep{MeerKAT1, MeerKAT2}. 
It consists 64 dishes equipped with state-of-the-art receivers, which are capable of observing in both single-dish and interferometric modes. 
And one of the flagship missions of MeerKAT, which is called MeerKLASS (MeerKAT Large Area Synoptic Survey), aims to conduct H\textsc{i} intensity mapping in single-dish mode over survey area of 4000 deg$^2$ \citep{MeerKAT1}.

The galaxy survey we consider is the Chinese Space Station Optical Survey (CSS-OS) \citep{CSST1, SNR1, CSST3, CSST4, CSST6}. CSS-OS is the major observation project of CSST, targeting to obtain a high-quality galaxy photometric and spectroscopic catalogs over 17,500 $\mathrm{deg^2}$ survey area during its 10-year mission. Here, we focus on the CSST slitless spectroscopic survey, which can determine the redshift of galaxies accurately. The overlapping survey area of MeerKAT H\textsc{i} intensity mapping and CSST spectroscopic galaxy survey can reach several thousand square degrees, and their target redshift are basically in the same range. Thus we anticipate they would make promising cross-correlation detection in the near future.

The paper is structured as follows: in Section.\ref{sec:simulation}, we introduce our methods of generating mock data of MeerKAT H\textsc{i} intensity mapping and CSST spectroscopic galaxy survey; in Section.\ref{sec:power spectrum}, we describe the estimators of the galaxy auto- and 21cm-galaxy cross-power spectra; in Section.\ref{sec:PCA&TF}, we discuss the details of the methodology of signal extraction, including both foreground removal of H\text{i} intensity maps and signal compensation for the cross-power spectra; in Section.\ref{sec:Obbr}, we present  the predicted constraint results on the relevant cosmological parameters; we summarize our work in Section.\ref{sec:conclusion}

\section{Mock Map Making}\label{sec:simulation}

\subsection{Simulation}

We employ Jiutian-1G simulation to generate the mock observational maps for both MeerKAT H\textsc{i} intensity mapping and CSST spectroscopic galaxy survey \citep{Jiutian}. Jiutian-1G is the 1 $h^{-1}\rm Gpc$ box of Jiutian simulation suite,  
a state-of-the-art hybrid cosmological simulation prepared for data analysis of CSST extragalactic surveys. 
The simulation is carried out using the \textsc{LGadget-3} code with $6144^3$ particles. Dark matter halos and subhalos are identified through the halo finding code Friends-of-friends (\textsc{FoF}) \citep{fof} and \textsc{SubFind} \citep{subfind}. 
Additionally, merger trees of halos are constructed using the B-Tree code. Jiutian-1G adopts the $\Lambda$ Cold Dark Matter ($\Lambda$CDM) cosmological model with parameters from Planck-2018 results.
The values of the relevant simulation and cosmological parameters are $\Omega_{m}= 0.3111$, $\Omega_{\Lambda}= 0.6889$, $\Omega_{b}= 0.049$, $n_{s}= 0.9665$, $\sigma_{8}= 0.8102$ and $h=0.6766$.

To simulate the properties of galaxies, semi-analytical model (SAM) is applied on the dark matter only simulation. 
Jiutian-1G simulation employs the \textsc{LGalaxies} code \citep{LGalaxies}, which includes various baryonic processes. Subsequently, hydrogen model \citep{HI_model} is applied to obtain the H\textsc{i} mass of each galaxy from cold gas distribution. 
With the mass resolution of $3.723\times10^8 M_{\odot}$, Jiutian-1G simulation is sufficient enough to resolve the low mass halos that contain H\textsc{i}.
Furthermore, \cite{emission_lines} utilizes the public code \textsc{CLOUDY}~\citep{2025arXiv250801102G} to develop an emission lines model for simulation of galaxies. This model enables us to obtain the luminosities of 13 emission lines. At this stage, Jiutian-1G simulation contains all the necessary cosmological information to simulate the cross-correlation between galaxy survey and H\textsc{i} intensity mapping. 

To match the observational capabilities of both MeerKAT and CSST, we set the simulated radio observational frequency bands to be $900-1015$ MHz, $770-900$ MHz and $650-770$ MHz. The first band is in the L-band of MeerKAT receiver and the last two are in the UHF-band. The central redshifts of these three bands are $z=0.5$, 0.7, and 1, respectively, which are also the main target redshifts for the CSST spectroscopic galaxy survey.

\subsection{MeerKAT H\textsc{i} Intensity Mapping} \label{sec:MeerKAT}
Firstly, to simulate MeerKAT H\textsc{i} intensity maps, we start with generating time-ordered data (TOD). Currently, two single-dish H\textsc{i} intensity mapping surveys run by MeerKAT, i.e. the MeerKAT L-band pilot survey (hereafter MeerKAT19) \citep{MeerKAT19cali_Wang, MeerKAT_WiggleZ} and MeerKLASS L-band deep field survey (hereafter MeerKAT21) \citep{MeerKAT_GAMA}. Both of them employ the on-the-fly mode in the observation \citep{on-the-fly}.
This mode requires antennas to maintain a fixed elevation angle and move in azimuth, in order to minimize the effect of leakage from ground and airmass in signal calibration.

\begin{figure}[ht]
  \centering
  \includegraphics[width=\columnwidth]{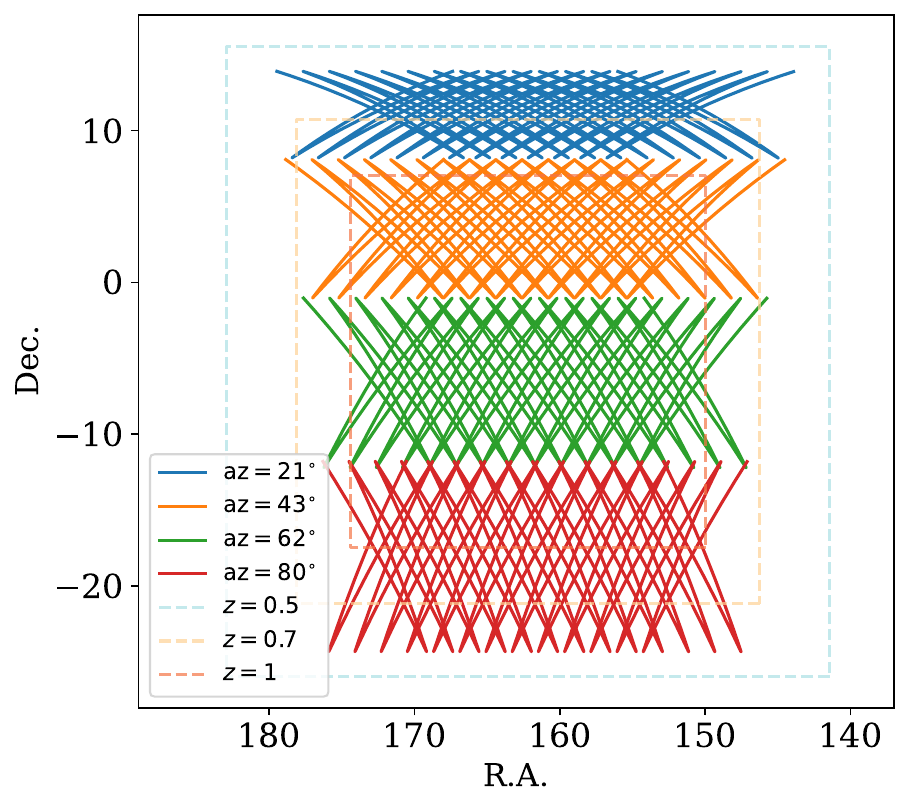} 
  \caption{The survey trails in the middle day of the MeerKAT observational plan.  The blue, orange, green, and red lines are the azimuth angles at $21^{\circ}$, $43^{\circ}$, $62^{\circ}$, and $80^{\circ}$, respectively. The blue, orange, and red dashed lines denote the sky area which the box size of Jiutian-1G corresponds to at $z=0.5$, $z=0.7$ and $z=1$, respectively.}
  \label{fig:survey_trails}
\end{figure}

Here we basically followed the observation strategy of the MeerKAT19, where each scan lasts 1.56 hours with two symmetrical scans performed each night. All 64 dishes of MeerKAT work in the same way and TOD are taken every 2 seconds. Moreover, to adequately cover the survey area corresponding to 1 $h^{-1}\rm Gpc$ at $z=0.5$, we slightly modified the azimuth and elevation angle to obtain four distinct scan trails and extended the observational time to 25 consecutive nights. 
In Figure~\ref{fig:survey_trails}, we show the four scan trails at one night and their corresponding antenna settings. 
The brightness temperature of every TOD is generated by
\begin{equation}
T_{\mathrm{D}}(\bm{x},\nu) = \int T_{\mathrm{b}}(\bm{r},\nu) \mathcal{B}(\bm{x} - \bm{r},\nu) {\rm d}\bm{x} + n_T,
\label{eq:convolution}
\end{equation}
where $\mathcal{B}(\bm{x},\nu)$ is the beam pattern at frequency $\nu$ pointing to the sky position $\bm{x}$, $T_{\mathrm{b}}(\bm{r})$ is the brightness temperature from direction $\bm{r}$. 
In our work, $T_{\mathrm{b}}$ includes three components: the H\textsc{i} signal $T_{\mathrm{H\textsc{i}}}$, the foreground emission from Milky Way $T_{\mathrm{MW}}$, and bright radio point sources $T_{\mathrm{ps}}$. 
Terrestrial and instrumental effects will also contribute to the observed signal, including radio frequency interference (RFI), ground spill, as well as the polarization leakage, thermal noise and $1/f$ noise of the antenna. Here we only consider the polarization leakage and thermal noise $n_{T}$ in the simulation, under the assumption that RFI and ground spill could be calibrated. Pervious works \citep[e.g.][]{1/f} indicates that, $1/f$ noise caused by the time dependency of the gain fluctuation, could be problematic in H\textsc{i} intensity mapping. However, recent studies of MeerKAT \citep{MeerKAT19cali_Wang, 1/f_liyichao} shows that, $1/f$ noise can be under good control by applying fast scan speed strategy, noise diode injection or SVD algorithm in TOD. Hence, we assume that $1/f$ noise would not have significant impact on the MeerKAT H\textsc{i} intensity mapping survey.

\begin{figure*}[ht]
  \centering
  \newlength{\figwidth}
  \setlength{\figwidth}{0.32\textwidth} 
  \newlength{\figheight}
  \setlength{\figheight}{0.2\textheight} 

  \begin{minipage}{\figwidth}
    \includegraphics[width=\linewidth, height=\figheight, keepaspectratio]{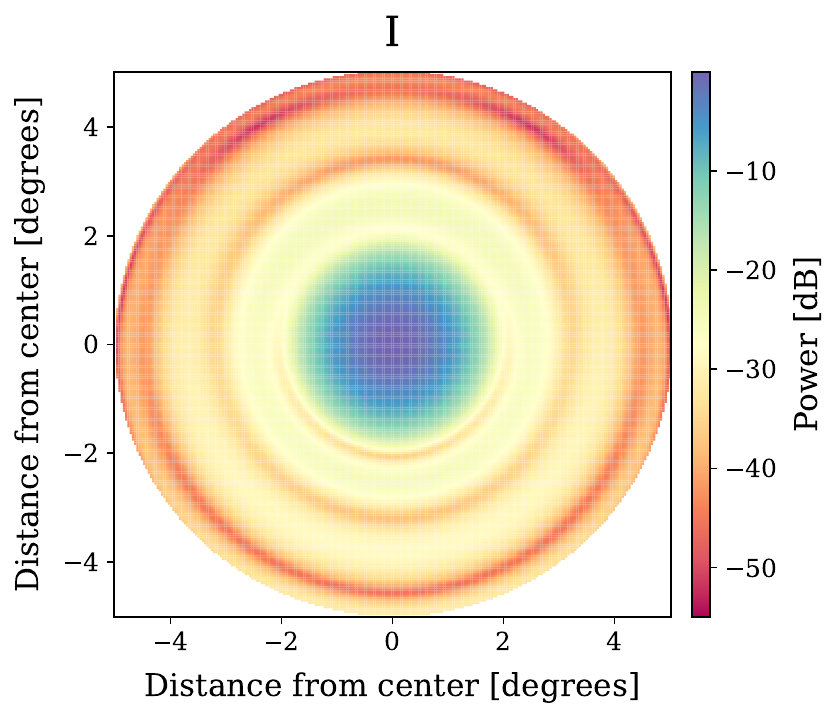}
    \label{fig:a}
  \end{minipage}
  \hfill
  \begin{minipage}{\figwidth}
    \includegraphics[width=\linewidth, height=\figheight, keepaspectratio]{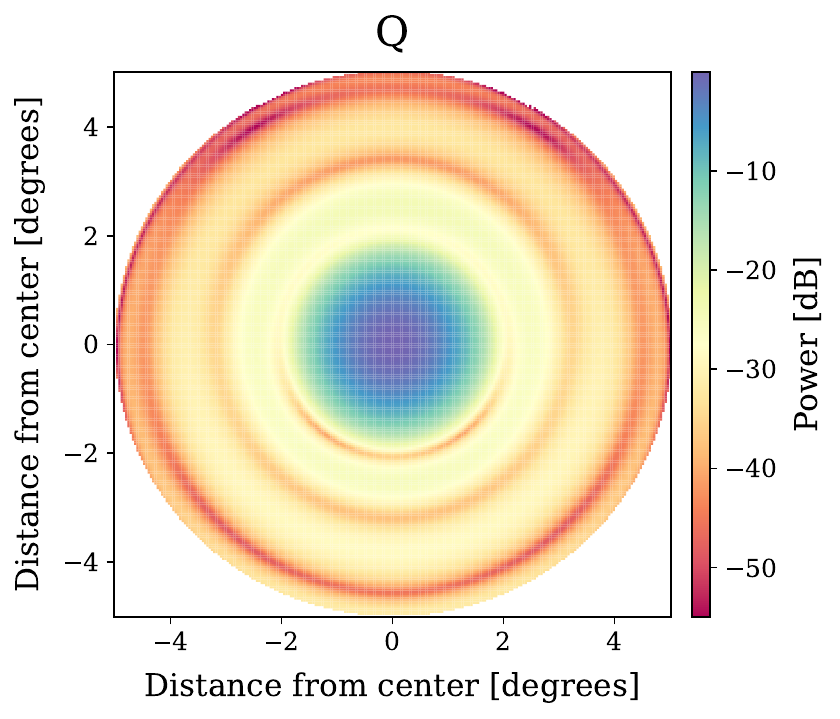}
    \label{fig:b}
  \end{minipage}
  \hfill
  \begin{minipage}{\figwidth}
    \includegraphics[width=\linewidth, height=\figheight, keepaspectratio]{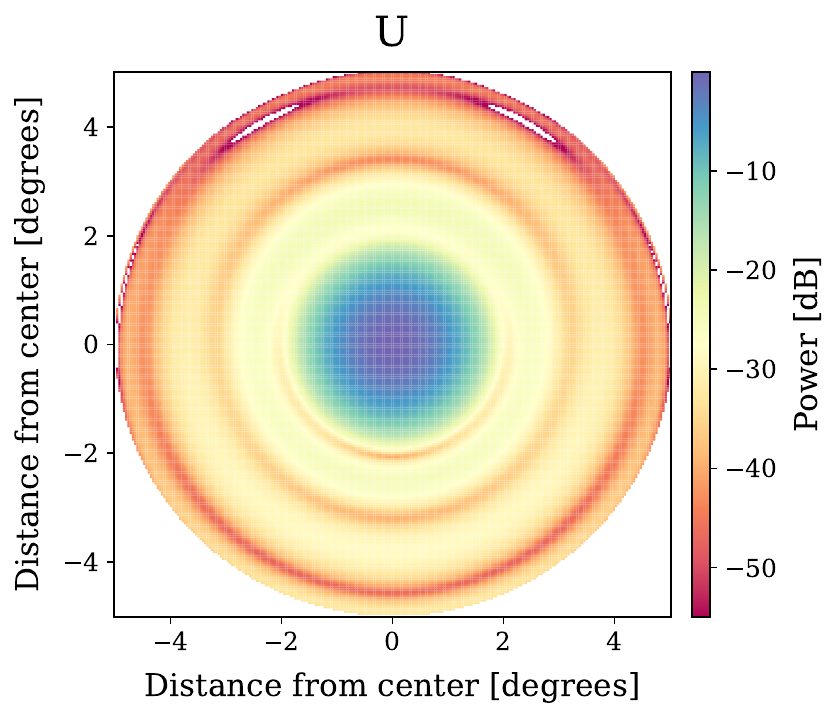}
    \label{fig:c}
  \end{minipage}

  \vspace{1em} 

    \begin{minipage}{0.48\textwidth}
    \centering
    \includegraphics[width=\linewidth, height=\figheight, keepaspectratio]{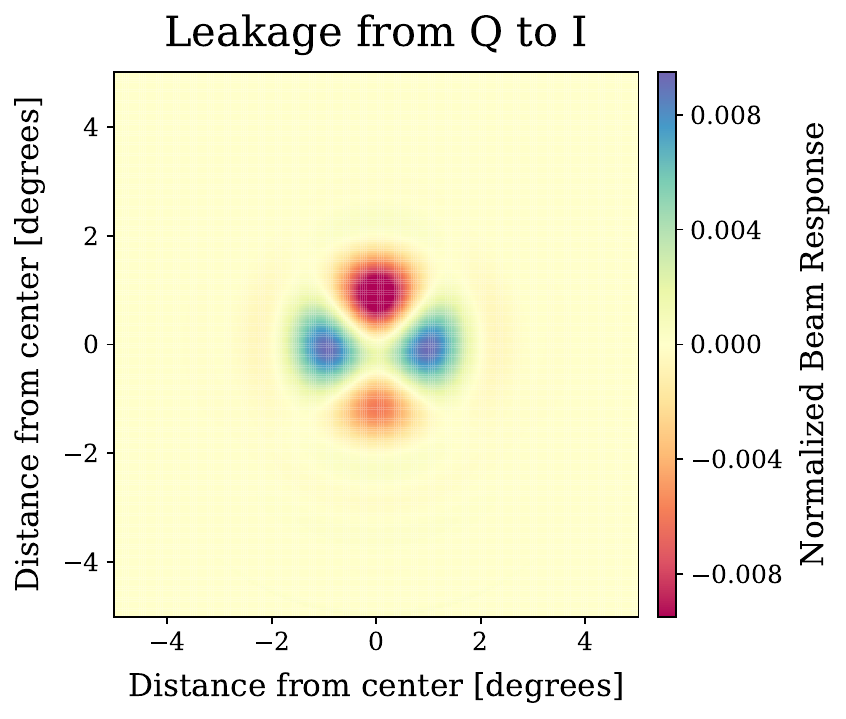}
    \label{fig:d}
  \end{minipage}
  \hspace{0.1pt}
  \begin{minipage}{0.48\textwidth}
    \centering
    \includegraphics[width=\linewidth, height=\figheight, keepaspectratio]{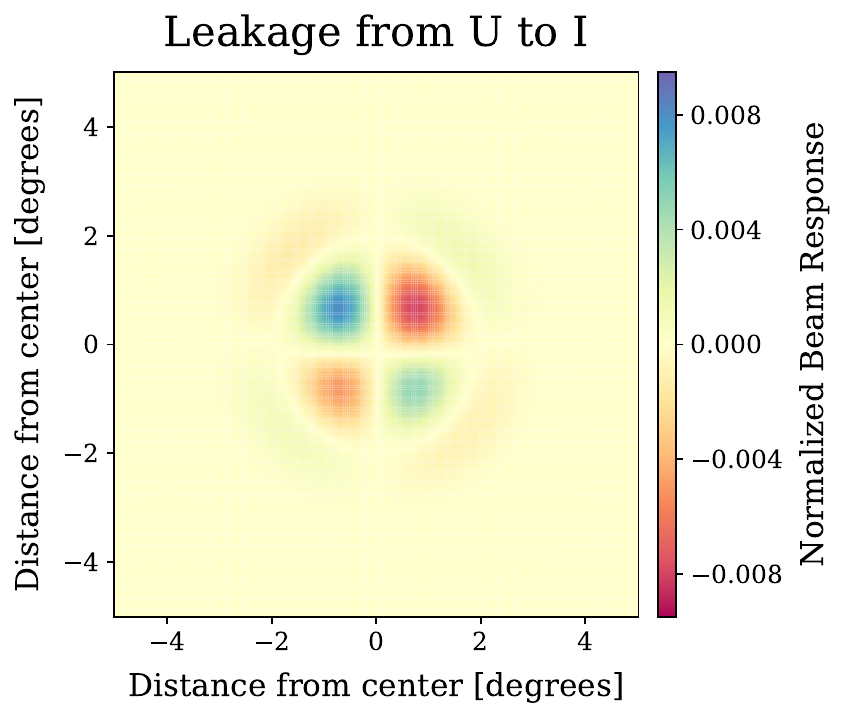}
    \label{fig:e}
  \end{minipage}

  \caption{The patterns of the MeerKAT primary beam model at 930 MHz obtained using the \textsc{EIDOS} package. The upper row are the beam patterns of I and polarization Q and U. The lower row are the beam patterns of polarization leakage from Q and U to I. }
  \label{fig:beam_patterns}
\end{figure*}

The beam pattern at each frequency is generated via the Python package \textsc{EIDOS}\footnote{\url{https://github.com/ratt-ru/eidos}} \citep{eidos}. 
It employs Zernike polynomials to model the primary beam of MeerKAT L-band and UHF-band within a maximum diameter of 10 degrees, based on the holographic measurement of the antenna. The model accurately reproduce the beam reponse of both mainlobe and sidelobes.
Furthermore, it can simulate the beam pattern of polarization leakage from Q, U, V to I, which enables us to quantify this kind of signal contamination from the polarization of galactic emission and radio point sources (since the \rm H\textsc{i} signal is non-polarized). 
The patterns of primary beam and polarization leakage beam are shown in Figure~\ref{fig:beam_patterns}. 
It may seem non-trivial that the polarization leakage beams are signed response function. This arises from coherent field-level interference encoded in the Mueller formalism, representing the coherence cancellation of polarization signal, rather than negative intensity.
We also note that the raw MeerKAT beam patterns could be quite asymmetric \citep[e.g.][]{2022AJ....163..135D}, and the patterns shown here can be treated as the ones after eliminating most of the asymmetrical effects by instrumental calibration. 
Besides, there will be some limitations caused by the maximum diameter range of EIDOS. As \cite{2022AJ....163..135D} has mentioned that, there is a far out sidelobe which has a 0.01\% power level, and its effect are not considered in our simulation.

After the brightness temperature of all the components above is simulated, they are combined to get the observational temperature of each TOD. 
The total survey area at $z=0.5$, $0.7$ and $1$ are obtained to be $1597 ~\mathrm{deg^2}$, $1024 ~\mathrm{deg^2}$ and $595 ~\mathrm{deg^2}$ using the flat-sky approximation.
Then we set the resolution of intensity maps to be $0.4$ degree \citep{MeerKAT_WiggleZ, MeerKAT_GAMA}, 
and generate the intensity maps from TOD via map-making process:
\begin{align}
    T_{\mathrm{obs}}=(A^{\top}N^{-1}A)^{-1}A^{\top}N^{-1}T_{\mathrm{D}},
\end{align}
where $A$ is the pointing matrix and $N$ is the diagonal covariance matrix of noise. We discuss the details of simulating the  H\textsc{i} signal, foregrounds, and instumental thermal noise in the following subsections.

\subsubsection{$\mathrm{H\textup{\textsc{i}}}$ signal} \label{sec:HI}
As mentioned above, Jiutian-1G simulates H\textsc{i} mass in each galaxy with the semi-analytical model described in \cite{HI_model}. 
Assuming that the cold gas mostly exists in axially symmetric flat disks of galaxies and hydrogen in the cold gas exists in two forms, i.e. H\textsc{i} and $\rm H_{2}$, the H\textsc{i} mass can be expressed as
\begin{equation}
M_{\rm H\textsc{i}} = (M_{\rm cg}-M_{\rm Z}) \times \beta \times (1 + R_{\rm{H_{2}}})^{-1},
\label{eq:HI fraction}
\end{equation}
where $M_{\rm cg}$ is the cold gas mass in galaxies and $M_{\rm Z}$ is the metal mass in cold gas. $\beta=0.75$ is the hydrogen fraction and $R_{\rm H_{2}} \equiv M_{\rm H_{2}}/M_{\rm H\textsc{i}}$ is the mass ratio of $\rm H_{2}$ and $\rm H\textsc{i}$ . From the observational results \citep{Blitz_2006, Leroy_2008}, $R_{\rm{H_{2}}}$ is obtained by integrating the assumed exponential profile of gas in galaxy disks
\begin{equation}
R_{\rm{H_{2}}}\approx({3.44 R_{\rm{H_{2}}}^{c}}^{ -0.506} + 4.82 {R_{\rm{H_{2}}}^{c}}^{-0.1054})^{-1},
\label{eq:H2/HI}
\end{equation}
with central value of the radial profile of the $\rm H_{2}/H\textsc{i}$ ratio
\begin{equation}
R^{c}_{\rm{H_{2}}} = \left[ \frac{G}{8\pi P_{\rm mid}} r_{\rm disk}^{-4} M_{\rm cg}(M_{\rm cg} + \langle f_{\sigma}\rangle M_{*,\rm disk} ) \right]^{0.8},
\label{eq:central value}
\end{equation}
where
$G$ is the gravitational constant, 
$P_{\rm mid} = 2.35\times10^{-13}$ is a empirical parameter of the kinematic midplane pressure \citep{Leroy_2008} and
$\langle f_{\sigma} \rangle$ is the constant parameter of the ratio between the vertical velocity dispersions of gas and stars from both theoretical and observational studies \citep{1990ApJ...352..522D, 1993A&A...275...16B, Leroy_2008}. 
$r_{\rm disk}$ is the scale length of the gas disk 
and $M_{*,\rm disk}$ is the stellar mass in the disk, which can be obtained from Jiutian simulation.

\begin{figure*}[ht]
  \centering
  \setlength{\figwidth}{0.32\textwidth} 
  \setlength{\figheight}{0.2\textheight} 

  \begin{minipage}{\figwidth}
    \includegraphics[width=\linewidth, height=\figheight, keepaspectratio]{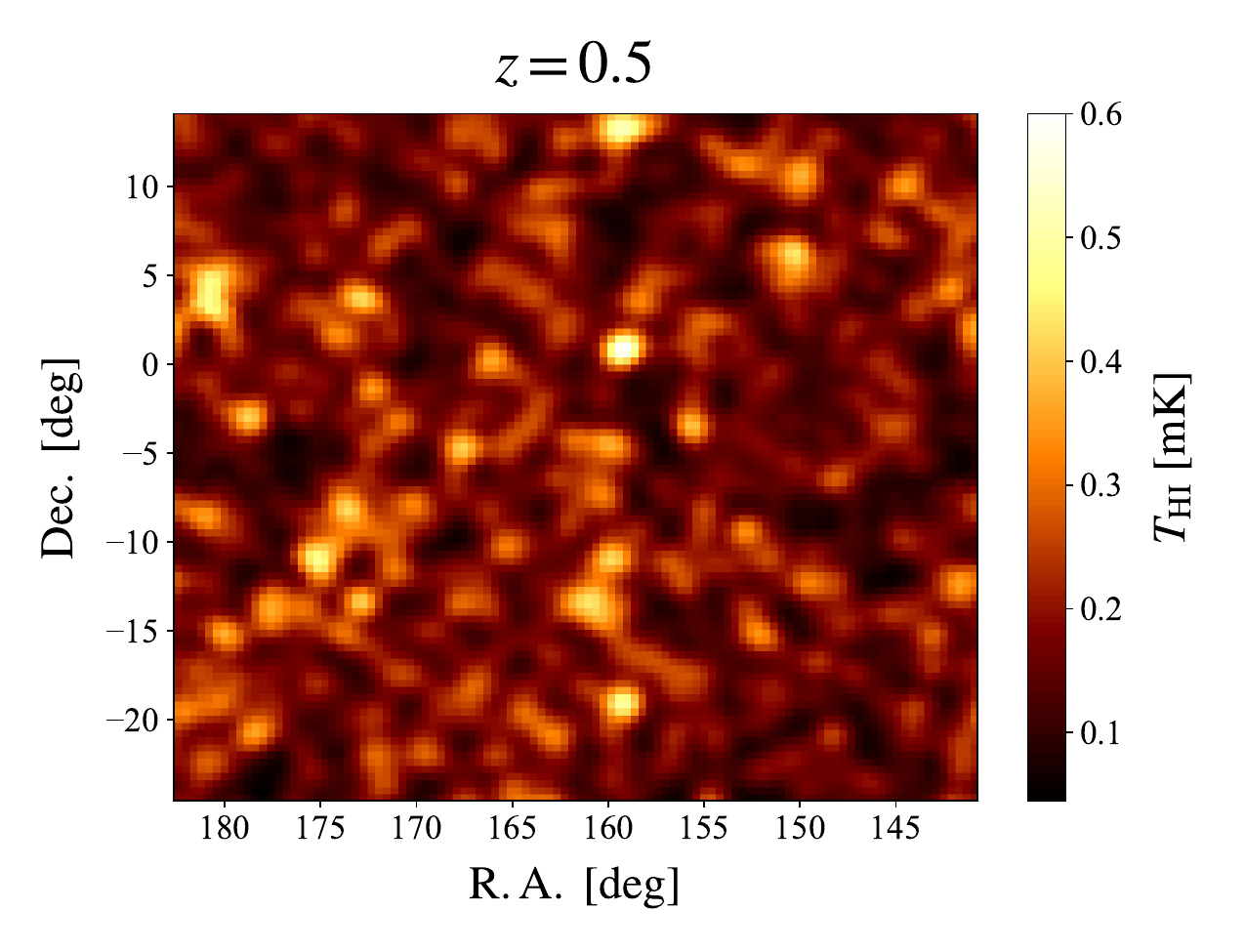}
    \label{fig:a}
  \end{minipage}
  \hfill
  \begin{minipage}{\figwidth}
    \includegraphics[width=\linewidth, height=\figheight, keepaspectratio]{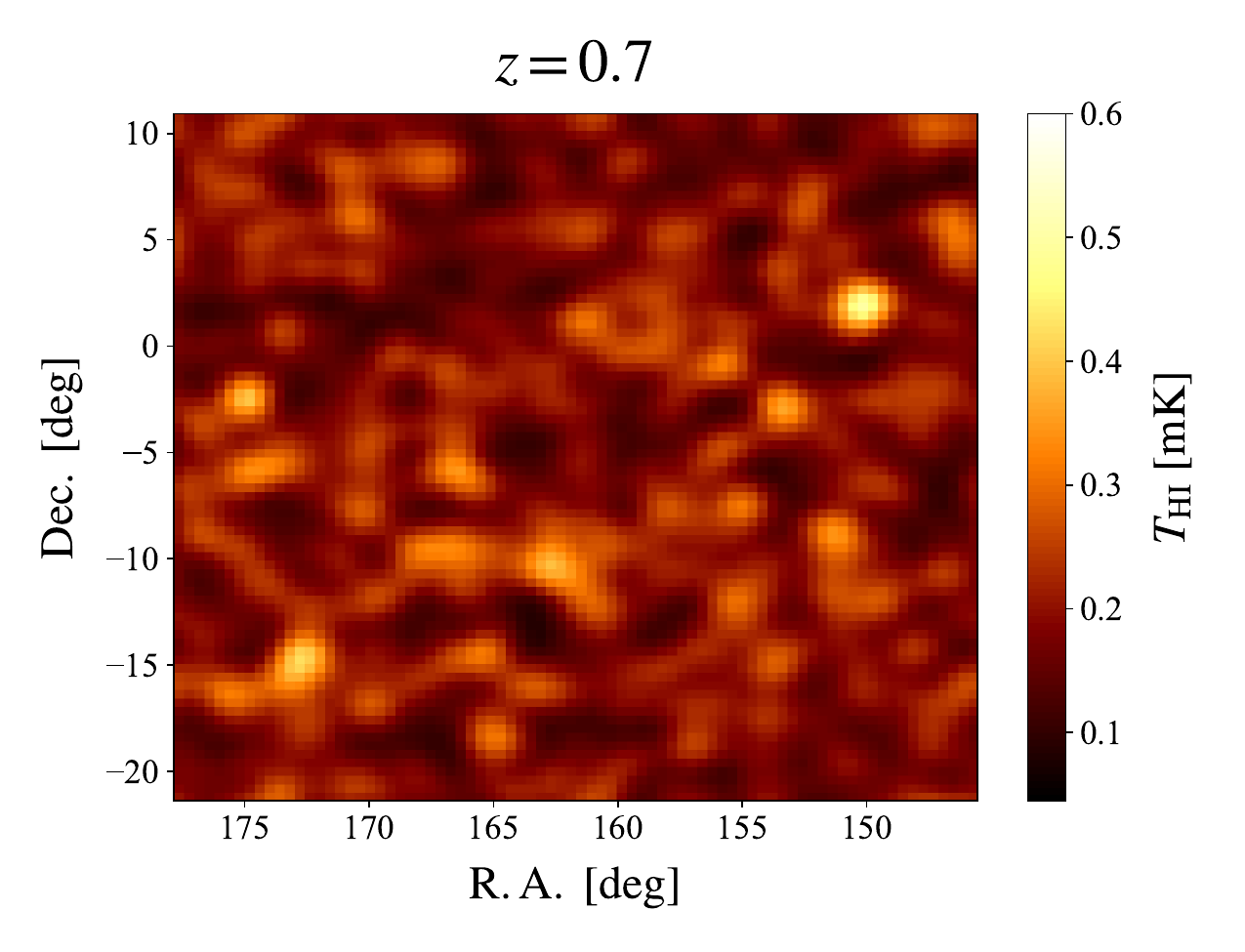}
    \label{fig:b}
  \end{minipage}
  \hfill
  \begin{minipage}{\figwidth}
    \includegraphics[width=\linewidth, height=\figheight, keepaspectratio]{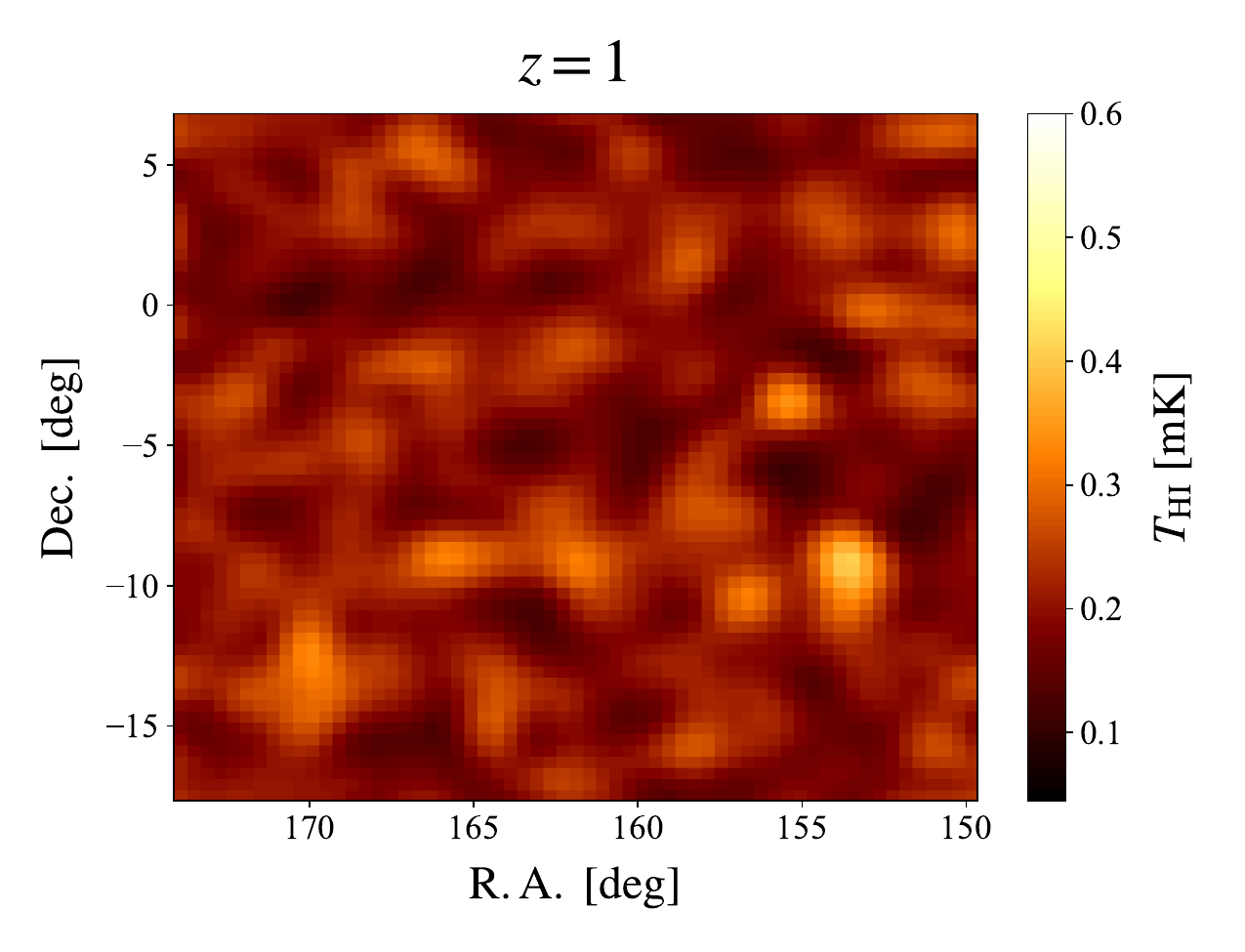}
    \label{fig:c}
  \end{minipage}
  \caption{The simulated HI signal map in MeerKAT intensity mapping for the central frequency slice of each simulation box at z=0.5, 0.7, and 1.}
  \label{fig:HI map}
\end{figure*}

After pre-processing the simulation boxes on the line-of-sight direction, The H\textsc{i} brightness temperature from position $\bm{r}$ and redshift $z$ can be obtained by \citep{Villaescusa-Navarro_2018}
\begin{align}
    T_{\rm b}(\bm{r},z) & = 189\frac{h}{E(z)}\Omega_{\rm H\textsc{i}}(\bm{r},z)(1+z)^{2} \  [\mathrm{mK}] \nonumber \\
    & =T_0(z) \Omega_{\rm H \textsc{i}}(\bm{r},z).
\end{align}
Here $E(z)=H(z)/H_{0}$ represents the evolution of the Hubble parameter, and the H\textsc{i} energy density parameter is defined as $\Omega_{\rm H\textsc{i}}(z) = \rho_{\mathrm{H\textsc{i}}}(z)/ \rho^{0}_{c}$ , where $\rho_{\rm H\textsc{i}}(z) $ is the H\textsc{i} energy density at $z$ and $\rho^{0}_{c}$ is the critical energy density of the present Universe. $T_0(z)$ is a redshift-dependent term, which is defined as $T_0=189\frac{h}{E(z)}(1+z)^{2} \mathrm{mK}$. As an example, the middle frequency bin of the simulated H\textsc{i} data cube at each redshift are shown in Figure~\ref{fig:HI map}. 
By averaging the $\Omega_{\rm H\textsc{i}}$ from different positions in the simulation boxes, we can estimate the mean H\textsc{i} brightness temperature at our target redshifts $z=0.5$, $0.7$ and $1$ to be $0.181\, \rm mK$, $0.194\,  \rm mK$ and $0.198\,  \rm mK$, respectively.

\subsubsection{Foreground {\rm\uppercase\expandafter{\romannumeral 1}}: Galactic emission} \label{sec:galaxy}

Foreground contamination has always been the major challenge for intensity mapping observations. In this work, we include two  foreground components, one is the emission from the Milky Way, and the other one is the continuum emission of the extra-galactic radio point sources. Here the foreground of Galactic emission is absolutely dominant, since its brightness temperature could be over 4 orders of magnitude higher than the H\textsc{i} signal. We simulate the Galactic emission using the \textsc{GSM2016} model \citep{GSM2016}, which is the advanced version of the original \textsc{GSM} (Global Sky Model)\citep{GSM}. 
This comprehensive framework accounts for six components of the Galactic emission, i.e., synchrotron emission, free-free emission, cold and warm dust thermal emission, and the CMB. 
For the $T_{\mathrm{MW}}$ of TOD, we generate the high resolution full-sky map with \textsc{GSM2016} at each frequency bin, then make interpolation with TOD's R.A. and Dec. to obtain the brightness temperature of Galactic emission.

In order to better approximate real observations, we further include the polarization leakage effect of the foreground into our simulation \citep{polarization_leakage1, polarization_leakage2, 2021MNRAS.504..208C, 2021MNRAS.508.5556M, 2022MNRAS.510.5872S}. Polarization leakage is an instrumental effect caused by the imperfect calibration of the beam response, allowing polarized foreground signals to contaminate the total intensity measurement. Although the H\textsc{i} signal itself remains unaffected due to its unpolarized nature, and the absolute level of polarization leakage is relatively small compared to the total signal, its intensity can become comparable to that of the H\textsc{i} signal. This effect introduces two significant complications: first, it creates additional mixing between different signal components, and second, it substantially increases the complexity of the foreground emission. Consequently, polarization leakage presents non-negligible challenges for accurate H\textsc{i} signal extraction during data analysis.

Accurate simulation of polarization leakage in MeerKAT observation requires two key components: the beam pattern characterizing leakage from all polarizations to the total intensity, and the polarization maps of the foreground emission. As previously noted, EIDOS has done accurate measurements of MeerKAT beam and is able to generate both polarized beam pattern and polarization leakage beam pattern from Q, U to I. The simulation of polarization maps requires us to have a good understanding of the Galactic magnetic field structure and electron distribution. 

\begin{figure*}[ht]
  \centering
  \setlength{\figwidth}{0.32\textwidth} 
  \setlength{\figheight}{0.2\textheight} 
  \begin{minipage}{\figwidth}
    \includegraphics[width=\linewidth, height=\figheight, keepaspectratio]{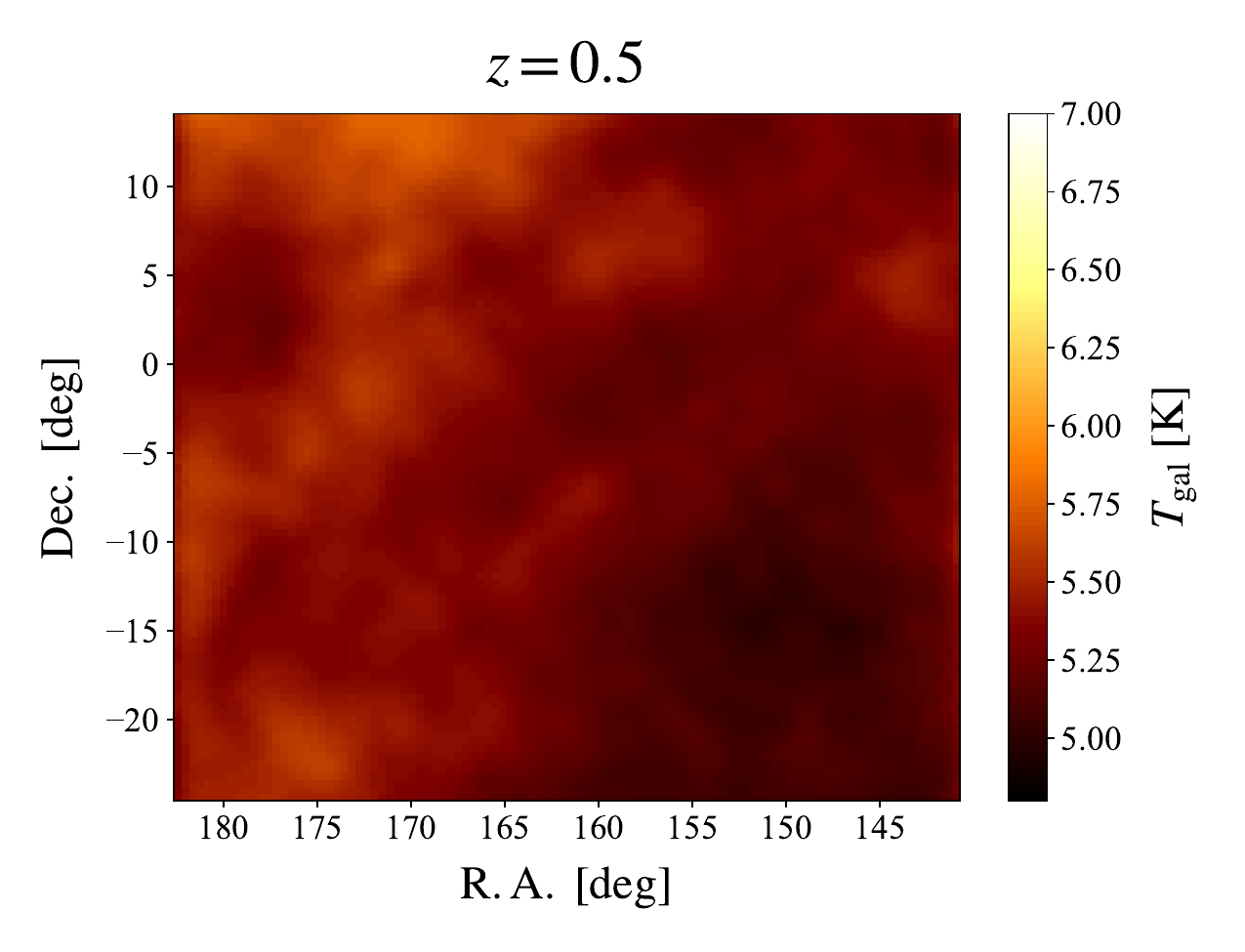}
  \end{minipage}
  \hfill
  \begin{minipage}{\figwidth}
    \includegraphics[width=\linewidth, height=\figheight, keepaspectratio]{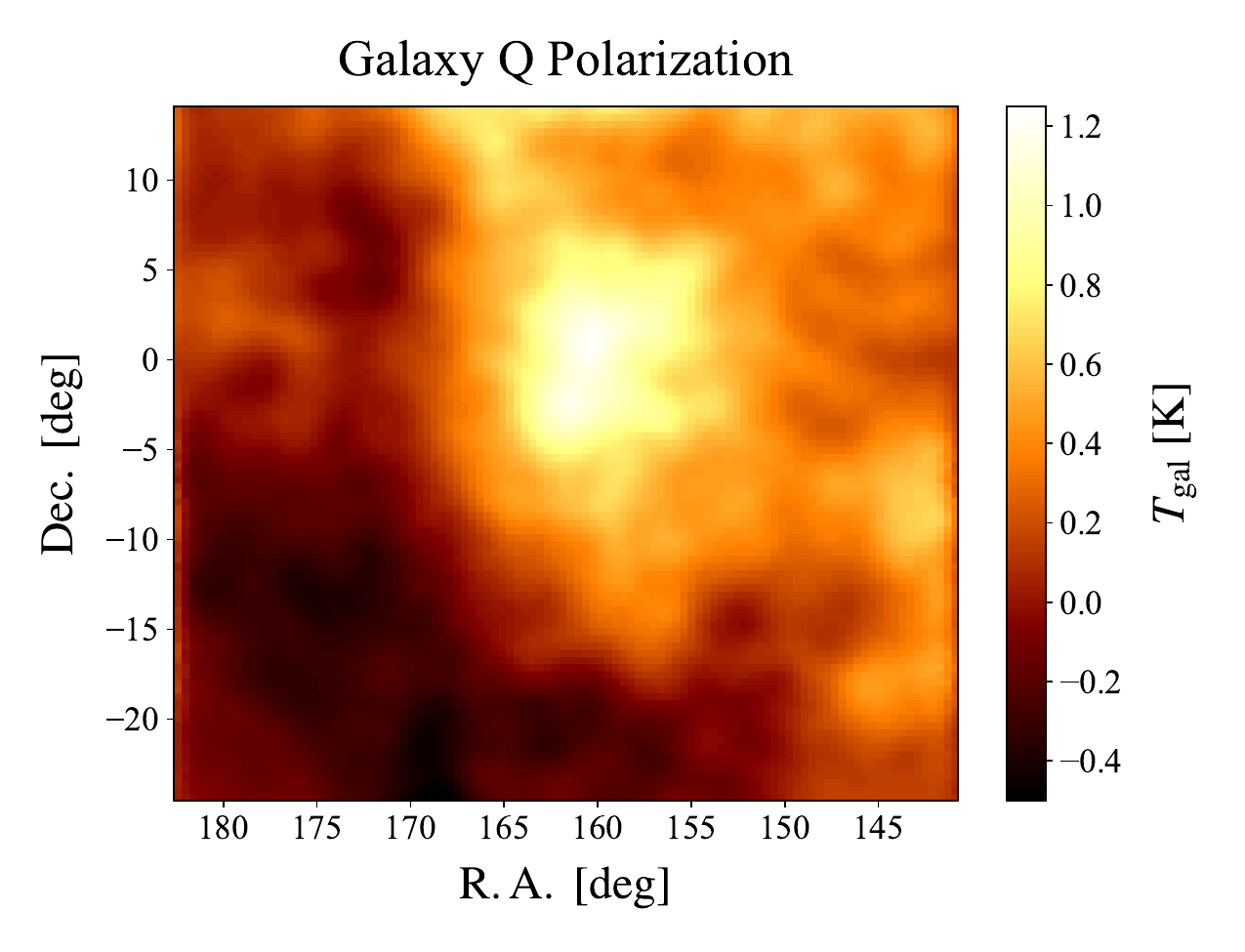}
  \end{minipage}
  \hfill
  \begin{minipage}{\figwidth}
    \includegraphics[width=\linewidth, height=\figheight, keepaspectratio]{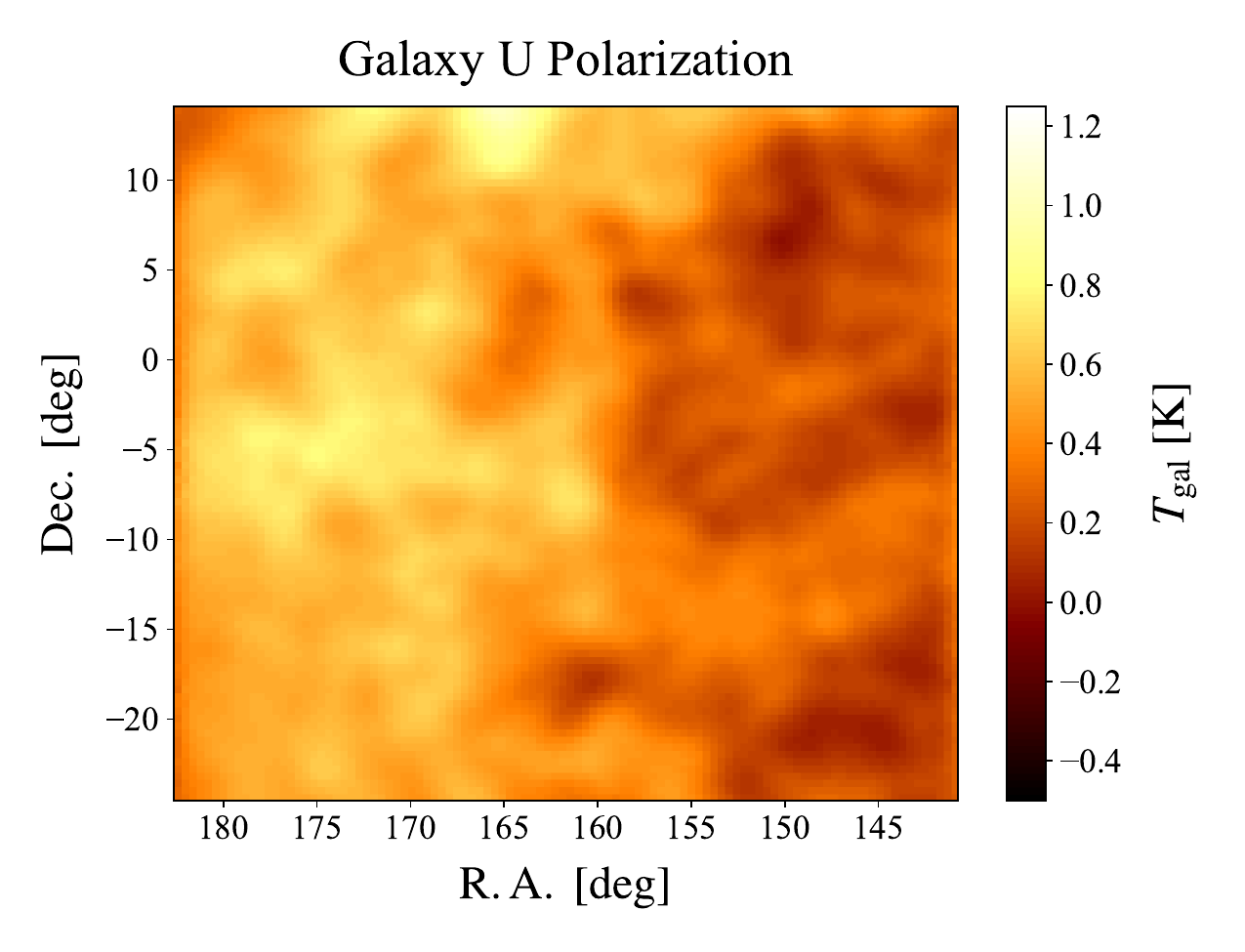}
  \end{minipage}

  \vspace{1em} 

  \begin{minipage}{0.48\textwidth}
    \centering
    \includegraphics[width=\linewidth, height=\figheight, keepaspectratio]{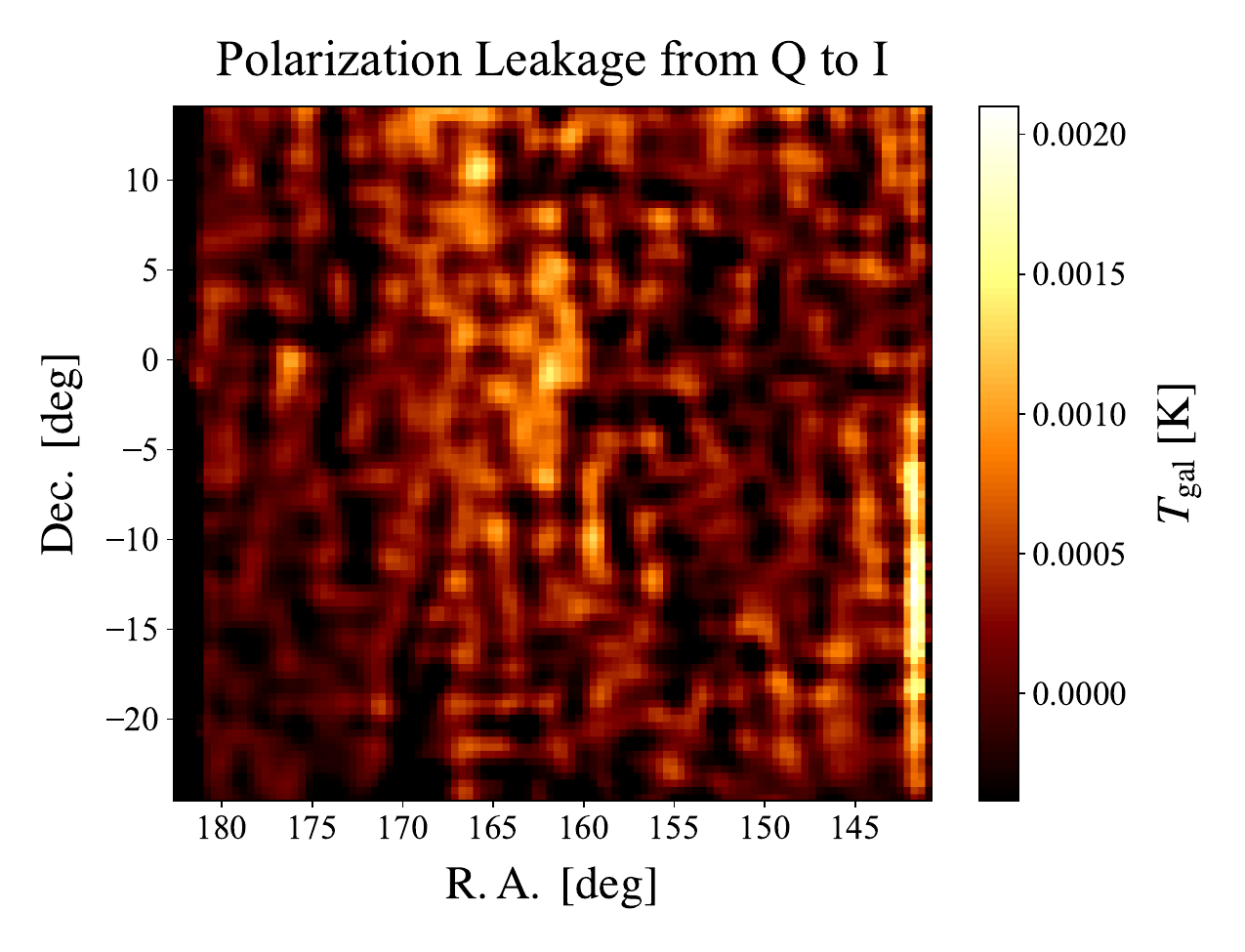}
  \end{minipage}
  \hspace{0.1pt}
  \begin{minipage}{0.48\textwidth}
    \centering
    \includegraphics[width=\linewidth, height=\figheight, keepaspectratio]{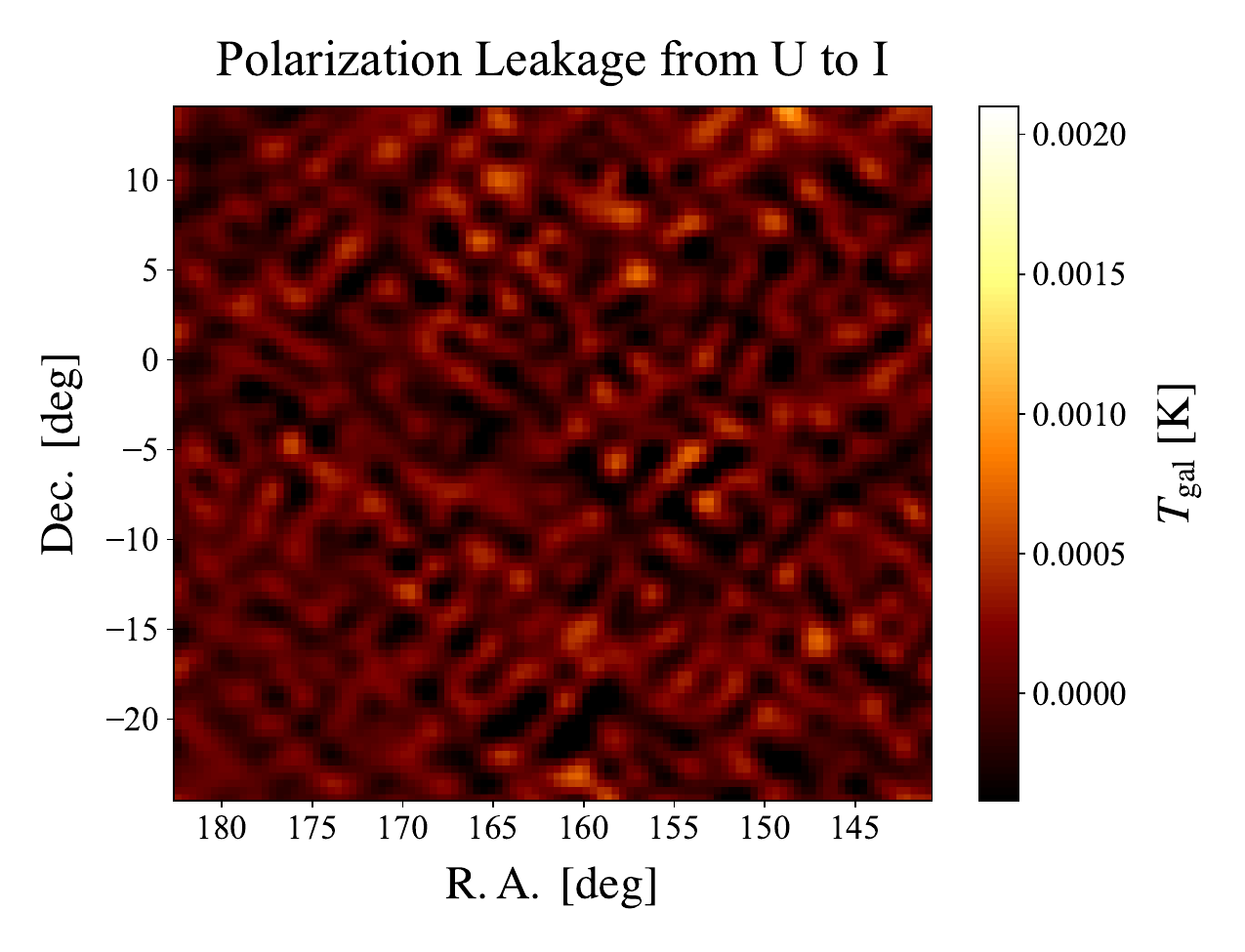}
  \end{minipage}
  \caption{The brightness temperature maps of the Galactic emission at 930 MHz. The upper row are the maps of I and polarization Q and U. The lower row are the maps of polarization leakage from Q and U to I. }
  \label{fig:galactic emission}
\end{figure*}

To improve the efficiency, we use the algorithm provided in a Python package \textsc{cora}\footnote{\url{https://github.com/radiocosmology/cora}} \citep{shaw_2024_13181020}. \textsc{cora} is designed for simulating skies of all components of H\textsc{i} intensity mapping surveys. And it applied the Faraday Rotation Measure Synthesis \citep{FRMS}
to model polarization structure. \textsc{cora} allows users to generate galactic emission with 3 different models including \textsc{GSM}, but \textsc{GSM2016} has not been included yet. 
To overcome this limitation, our polarization maps have to be done manually. At the code level, we simply load the results generated by \textsc{PyGDSM \footnote{\url{https://github.com/telegraphic/pygdsm}} \citep{2016ascl.soft03013P}} into the \textsc{cora} source code \citep{GSM, GSM2016}. Here we briefly illustrate the method of Faraday Rotation Measure Synthesis \citep{FRMS} and our map-generating procedure using \textsc{cora}:
\begin{itemize}
    \item When polarized radiation which emits from sky position $\bm{r}$ and distance $s$ passes through the magnetized plasma in the interstellar medium, its observed polarization angle $\theta_{\rm obs}$ will be different from its initial polarization angle $\theta_0$ by amount relates to the wavelength of the radiation $\lambda$ and Faraday depth $\phi$:
    \begin{align}
    \theta_{\rm obs} = \theta_0 + \Delta\theta = \theta_0 + \phi\lambda^2,
    \end{align}
    and $\phi$ at sky position $\bm{r}$ and distance $s$ is defined as 
    \begin{align}
    \phi(\bm{r}, s) = \int_{0}^{s} n_{e}(\bm{r}, s')B_{\parallel}(\bm{r}, s') \, ds' ,
    \end{align}
    where $n_{e}$ is the electron density and $B_{\parallel}$ is the line-of-sight component of the magnetic intensity.
    So the intensity of polarized radiation $I_p$ can be inferred as 
    \begin{align}
    I_p &= p I \nonumber \\
    & = p I_0 e^{2i\theta} \nonumber\\
    & =p I_0 e^{2i(\theta_0+\phi\lambda^2)}  ,
    \end{align}
    where I is the total intensity and $p=p_{\mathrm{Q}}+ip_{\mathrm{U}}$ is the polarized fraction coefficient. For the polarized galactic emission, the total intensity $I_p$ we observed from position $(\bm{r},s)$ at wavelength $\lambda$, is contributed by the radiation from all the point along the line-of-sight. So $I_p$ can be further inferred as 
    \begin{align}
    I_p(\bm{r},s) = &\int_{0}^{s}p(\bm{r},s') j_{I}(\bm{r}, s', \lambda) \times \nonumber \\
    &\hspace{6em} e^{2i\left(\theta_0(\bm{r},s')+\phi(\bm{r},s')\lambda^2\right)}ds' \nonumber \\
    = &\int_{0}^{s} F(\bm{r},s',\lambda)e^{2i\phi\lambda^2}ds' \label{Ip_ns}
    \end{align}
    where $j_I$ is the emission coefficient and $F(\bm{r},s,\lambda)=p(\bm{r},s) j_{I}(\bm{r}, s, \lambda)e^{2i\theta_0(\bm{r},s)}$ is the polarization emission.
    
    \item The definition of Faraday depth $\phi$ indicates that it's equivalent to the distance along line-of-sight $s$. So Equation~\eqref{Ip_ns} can be transferred from $s$ space into $\phi$ space, and we can obtain:
    \begin{align}
    I_p(\bm{r},s) = \int F(\bm{r},\phi,\lambda)e^{2i\phi\lambda^2}d\phi \label{Ip_nphi}.
    \end{align}
    The key idea of Faraday Rotation Measure Synthesis is to model the galactic emission in $\phi$ space in stead of $s$ space, which does not need accurate measurement of the electron profile and magnetic field of the Milky Way. $F(\bm{r},\phi,\lambda)$ is presumed to be separable in spectral dependence, which suggests $F(\bm{r},\phi,\lambda)=A(\bm{r},\phi)I_{\mathrm{MW}}(\bm{r},\lambda)$. $A(\bm{r},\phi)$ is the distribution of complex amplitude of the polarized radiation from the emission region $(\bm{r}, \phi)$, which represents the polarization structure in Faraday space. $I_{\mathrm{MW}}(\bm{r},\lambda)$ is the intensity of galactic emission as a function of wavelength and could be generated by \textsc{GSM2016} as mentioned above.
    \item Faraday Rotation Measure Synthesis models $A(\bm{r},\phi)$ further in two parts:
    \begin{align}
    A(\bm{r},\phi) = w(\bm{r},\phi)\delta_{p}(\bm{r},\phi) ,
    \end{align}
    where $w(\bm{r},\phi)$ is a positive envelope function which defines the position of emission region in Faraday depth
    \begin{align}
    w(\bm{r},\phi)\propto \frac{\alpha}{\sqrt{4\pi\sigma_{\phi}}}e^{-\frac{1}{4}(\frac{\phi}{\sigma_{\phi}})^2} .
    \end{align}
    where $\sigma_{\phi}$ is the standard deviation of the Faraday depth and it is determined from the Faraday rotation map  \citep{faraday_map}. 
    $\delta_{p}(\bm{r},\phi)$ is a random field that gives fluctuations in the complex polarization as a function of Faraday depth 
    \begin{align}
    C_\ell(\phi,\phi') \propto p^2 \left(\frac{\ell}{100}\right)^{-\beta} \rm exp \left(-\frac{(\phi-\phi')^2}{2\phi_{c}^2}\right)  ,
    \end{align}
    where $\beta=2.8$ is the spectra index of angular distribution of emission regions and $\phi_{c}$ is the correlation length in Faraday space. The value of the parameters
    $\alpha$, $p$ and $\phi_{c}$ in \textsc{cora}'s model is obtained from real sky observations \citep{cora}. 
\end{itemize}

Following the method above, full-sky maps of polarized Galactic emissions at each frequency can be generated by \textsc{GSM2016} and \textsc{cora}. After that, the intensity of polarizations Q and U and their polarization leakage are obtained by convolving the polarization maps with corresponding beam patterns. In Figure~\ref{fig:galactic emission}, the maps at $930$ MHz are shown as examples for polarization maps and polarization leakage of Galactic emission. 
The negative values in the polarization leakage maps are the consequence of convolution with signed polarization beams, reflecting the coherent cancellation of polarization signal rather than negative intensity.

\begin{figure*}[ht]
  \centering
  \setlength{\figwidth}{0.48\textwidth}
  \setlength{\figheight}{0.22\textheight} 

  \begin{minipage}[t]{\figwidth}
    \centering
    \includegraphics[width=0.95\linewidth, keepaspectratio]{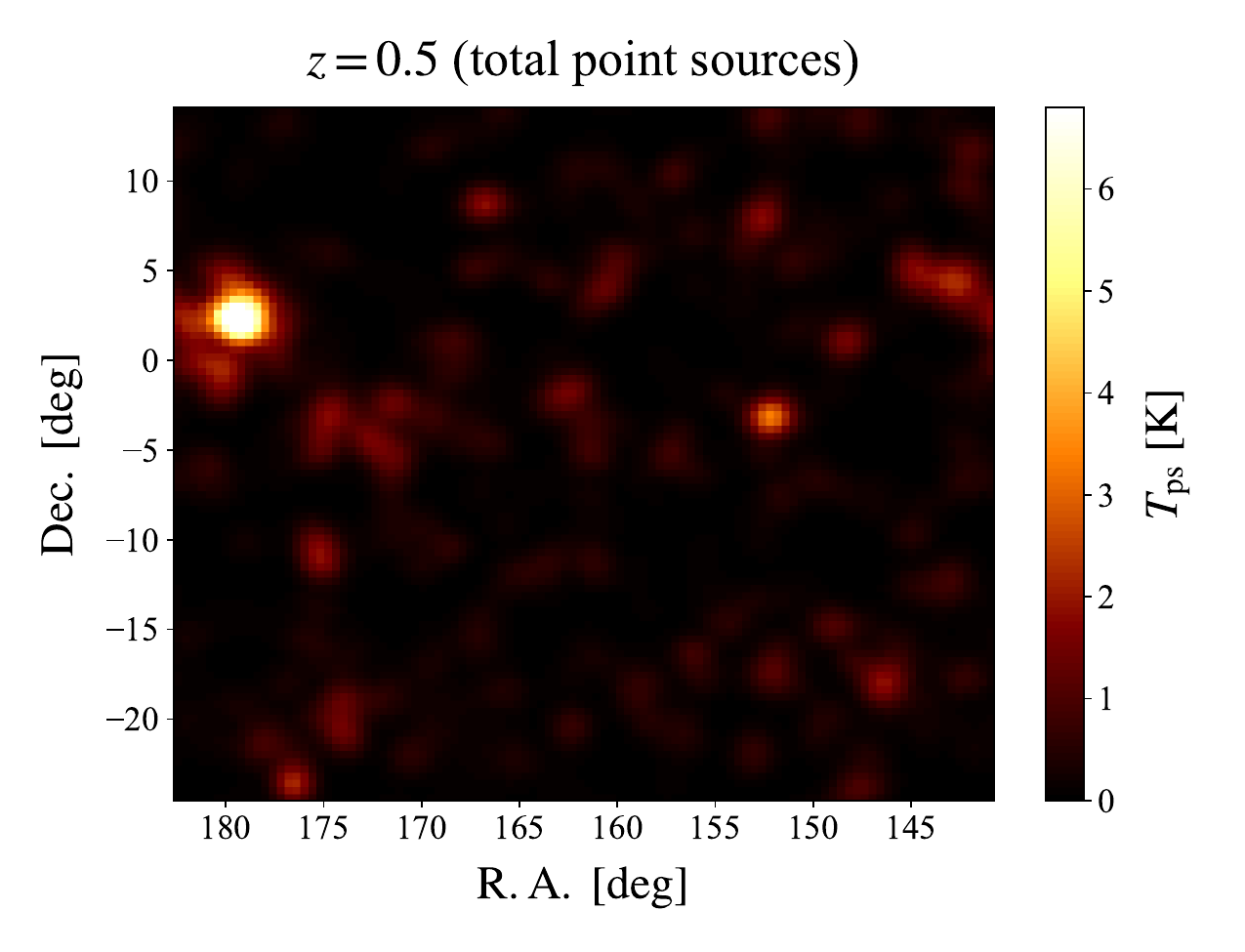}
    \label{fig:subfig1}
  \end{minipage}
  \hfill
  \begin{minipage}[t]{\figwidth}
    \centering
    \includegraphics[width=0.95\linewidth, keepaspectratio]{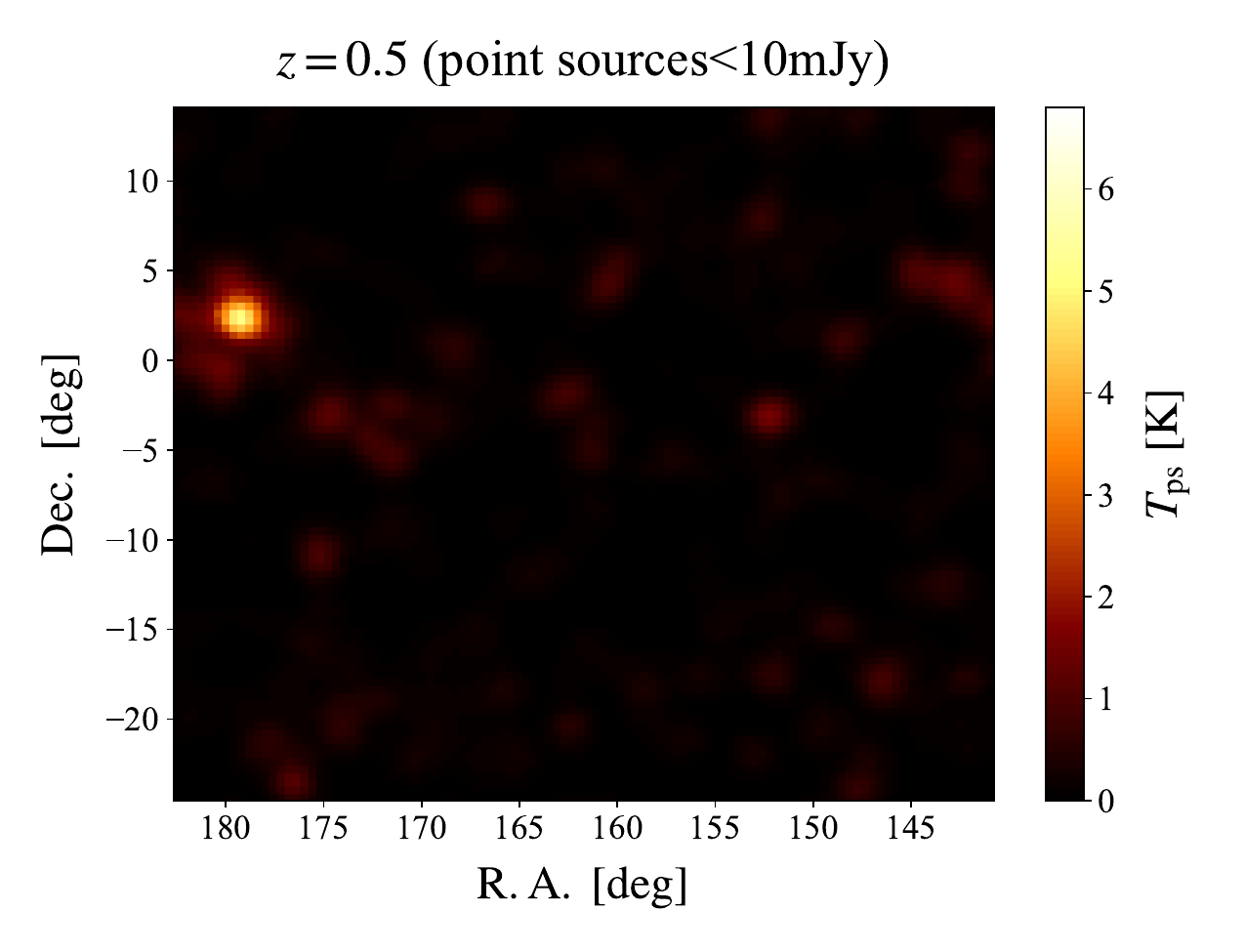}
    \label{fig:subfig2}
  \end{minipage}

  \vspace{0.5em} 

  \begin{minipage}[t]{\figwidth}
    \centering
    \includegraphics[width=0.95\linewidth, keepaspectratio]{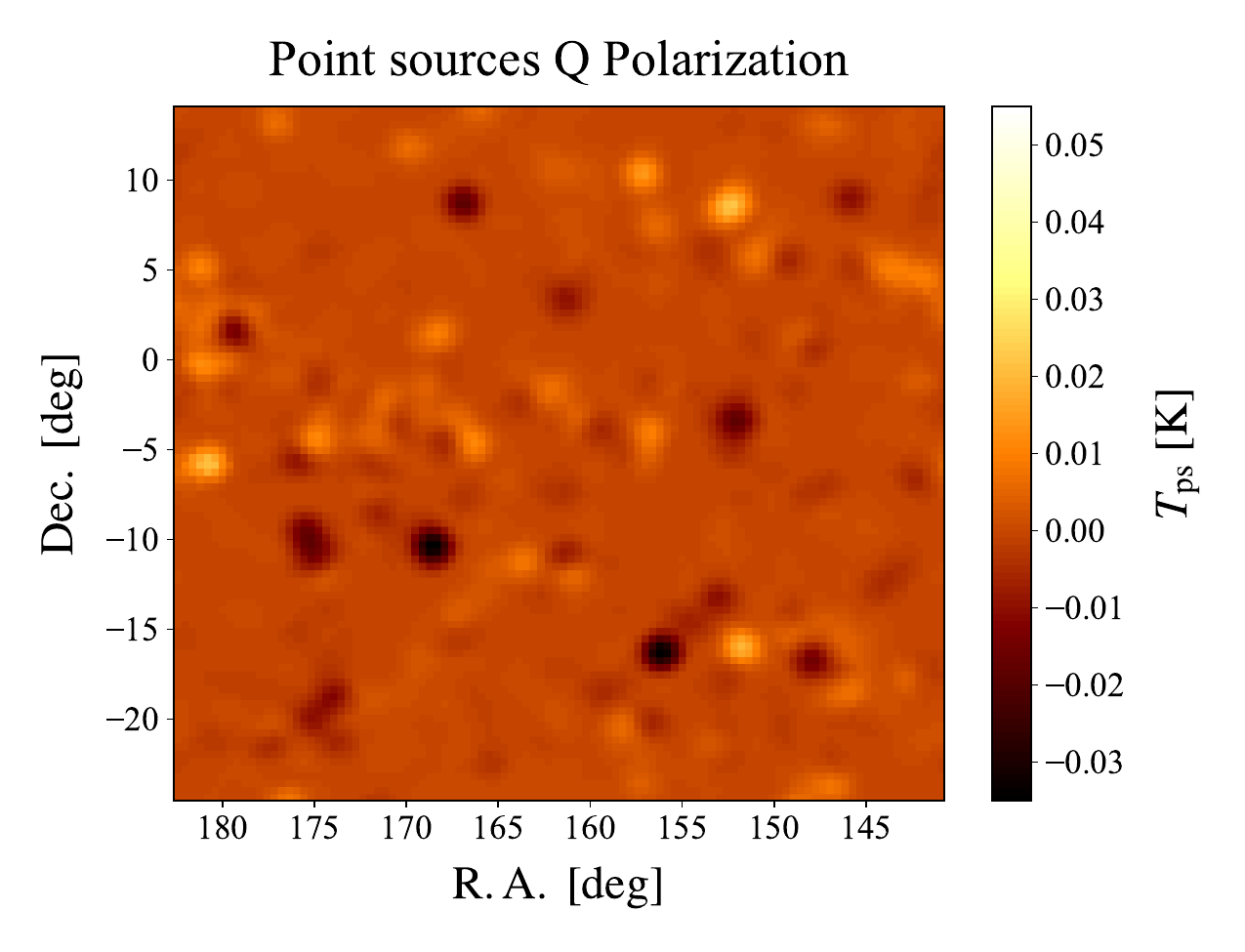}
    \label{fig:subfig3}
  \end{minipage}
  \hfill
  \begin{minipage}[t]{\figwidth}
    \centering
    \includegraphics[width=0.95\linewidth, keepaspectratio]{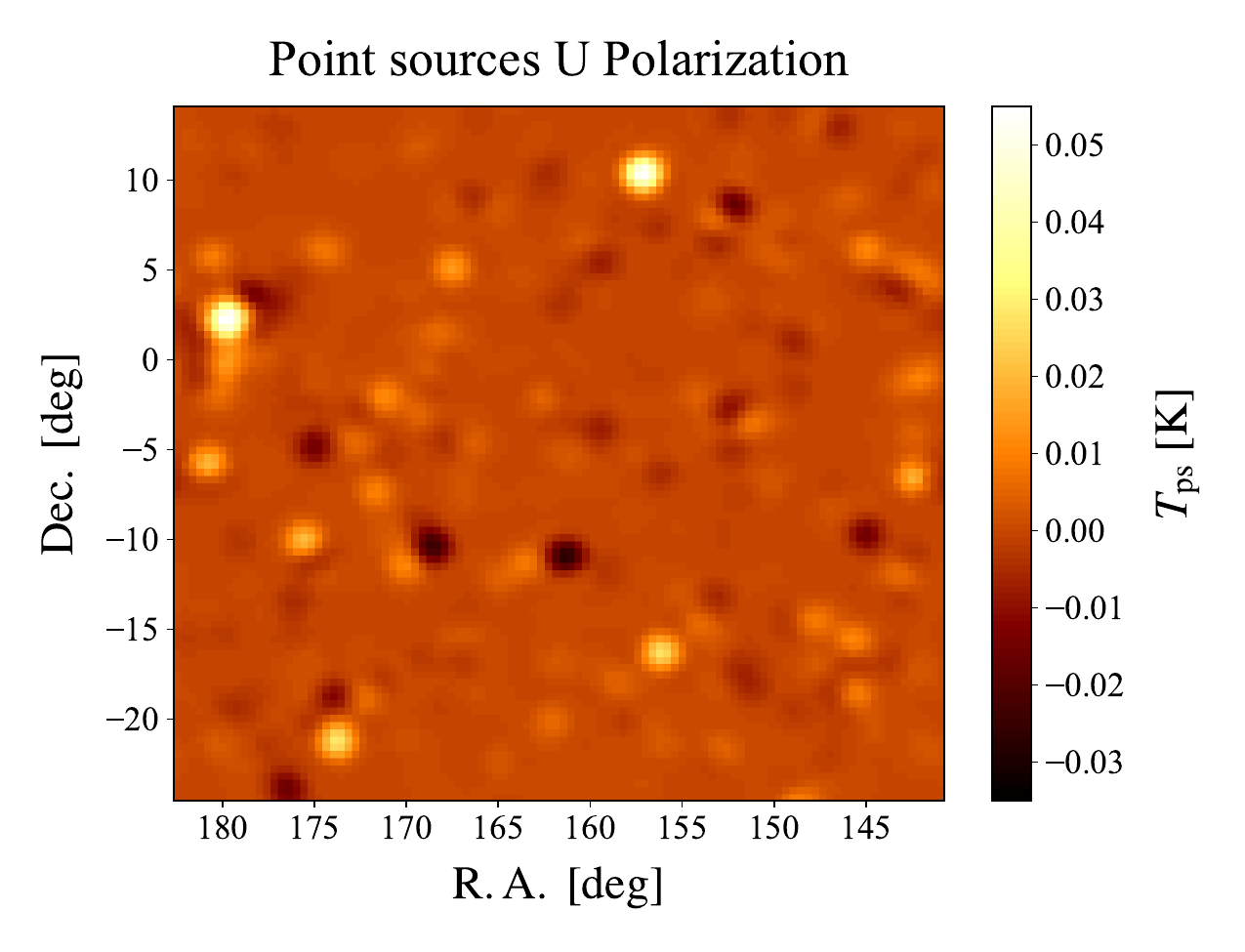}
    \label{fig:subfig4}
  \end{minipage}

  \vspace{0.5em} 

  \begin{minipage}[t]{\figwidth}
    \centering
    \includegraphics[width=0.95\linewidth, keepaspectratio]{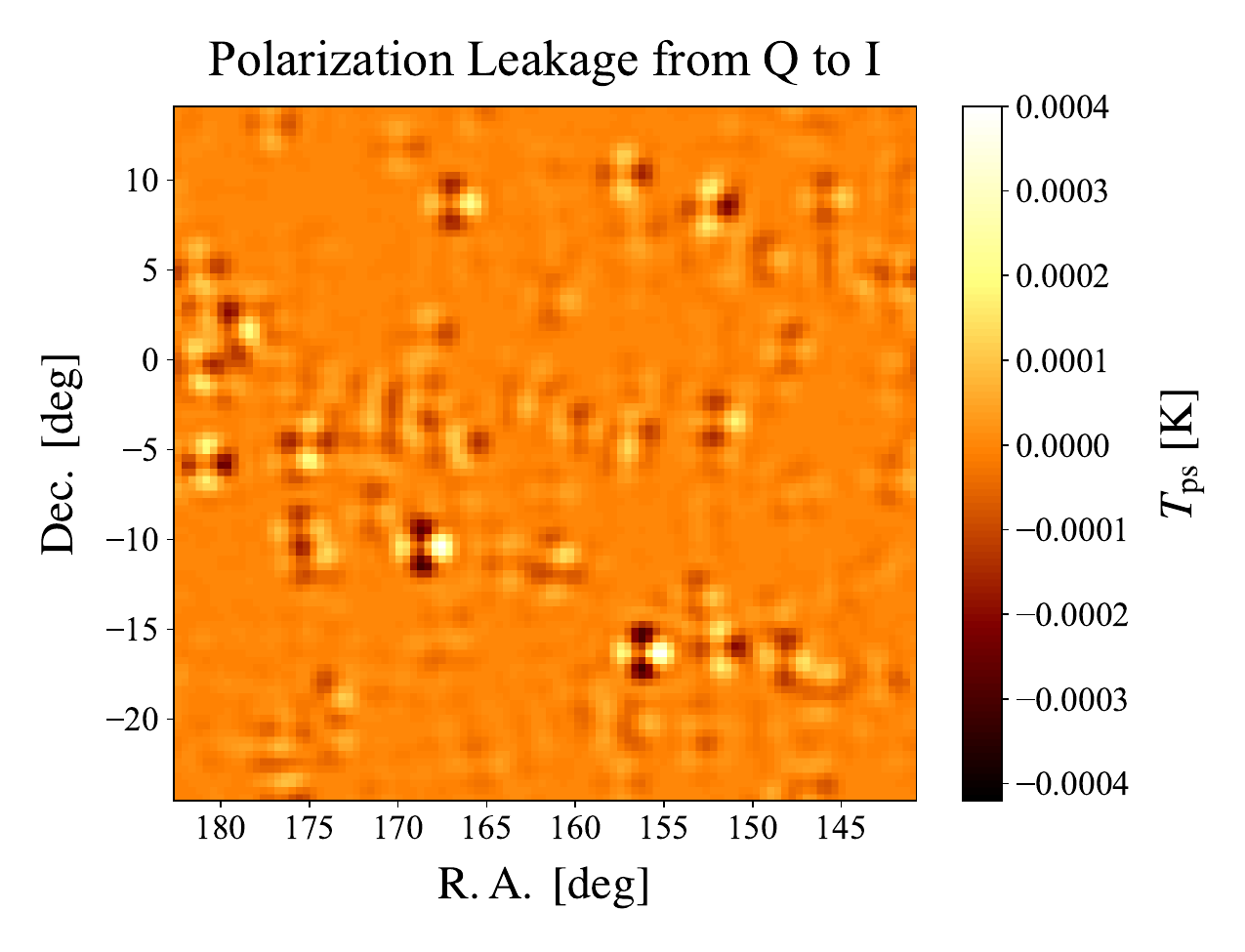}
    \label{fig:subfig5}
  \end{minipage}
  \hfill
  \begin{minipage}[t]{\figwidth}
    \centering
    \includegraphics[width=0.95\linewidth, keepaspectratio]{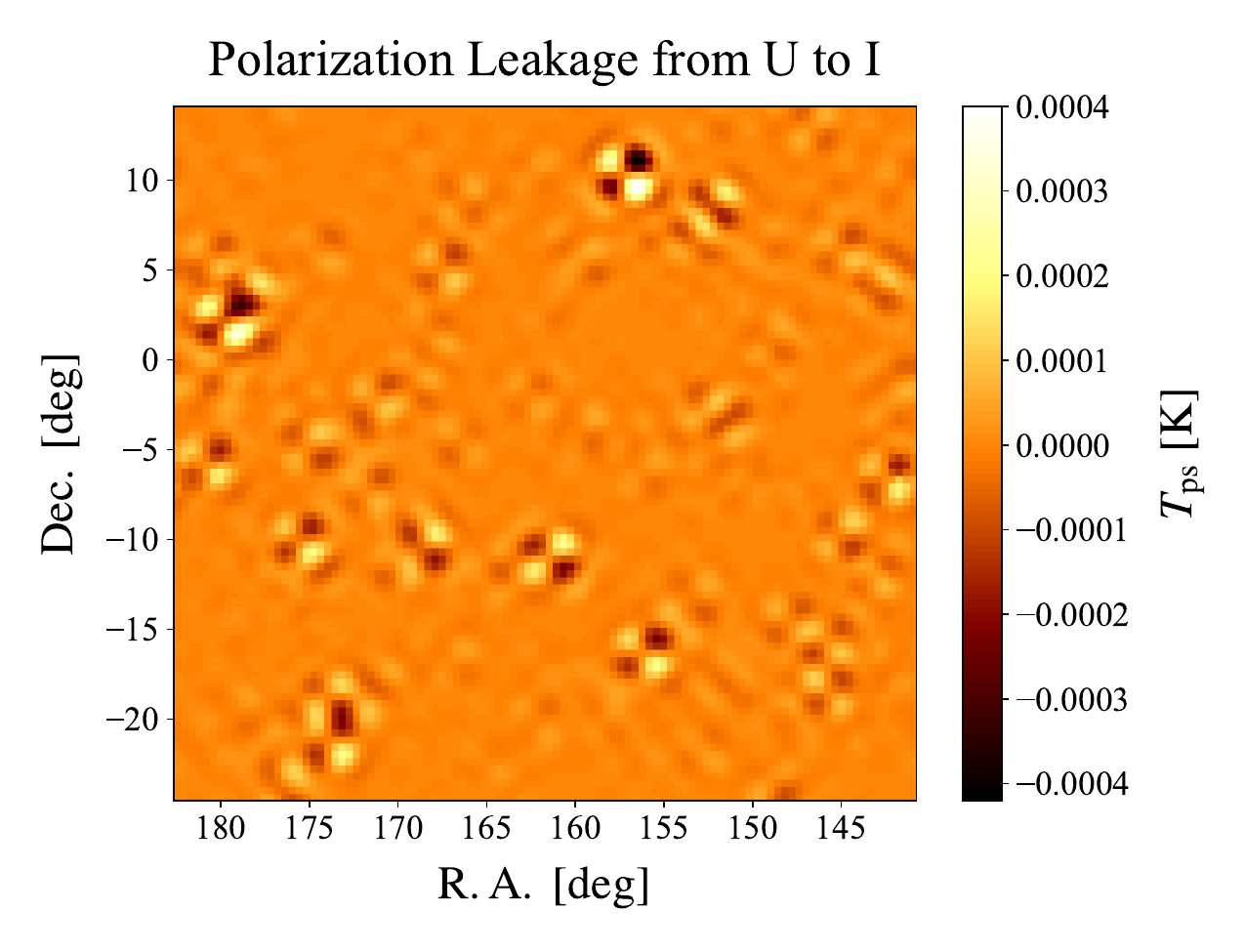}
    \label{fig:subfig6}
  \end{minipage}

  \vspace{0.5em}
  \caption{The brightness temperature maps of the extragalactic point sources at 930 MHz.
           The first row are total intensity maps of point sources, the left panel is the non-masking map and the right panel is the map masked point sources $>10$ mJy.
           The second row are the polarization Q and U of non-masking point sources map and the third row are the corresponding polarization leakage maps from Q, U to I. 
           }
  \label{fig:pointsources}
\end{figure*}

\subsubsection{Foreground {\rm\uppercase\expandafter{\romannumeral 2}}: Radio Point Sources} \label{sec:ps}
Extragalactic point sources are a combination of quasars, radio galaxies, starburst galaxies and other objects, which will leak their signal into radio observational band by continuum emission. While these sources may lie at redshifts either in front of or behind our target H\textsc{i} emission, they are conventionally classified as foreground contamination. It has been studied, that the point sources which dominate in H\textsc{i} intensity mapping survey have a steep spectrum, whose flux density spectral index is $\alpha\approx -0.75$ (where $S\propto \nu^{\alpha}$) \citep{1968ApL.....2..105K}. To simulate the distribution and intensity of point sources in our H\textsc{i} intensity mapping survey, we use the flux function described in \cite{pointsource}. It is a fifth-order polynomial fitted from the observational data of continuum surveys at 1.4 GHz, which gives
\begin{align}
    \mathrm{log} \left( \frac{S^{2.5}\mathrm{d}N/\mathrm{d}S}{N_0} \right)= \sum_{i=0}^5 \left[ \mathrm{log}\left(\ \frac{S}{S_0}\right) \right] ^{i} ,
    \label{flux function}
\end{align}
where $N$ is the number of sources per steradian and $S$ is the flux. The value of fitting coefficients are $a_0=2.593$, $a_1=0.093$, $a_2=-0.0004$, $a_3=0.249$, $a_4=0.090$ and $a_5=0.009$, and the normalization constants are $N_0=1\, \mathrm{Jy~sr^{-1}} $ and $S_0=1\, \mathrm{Jy}$.

\begin{figure*}[ht]
  \centering
  \setlength{\figwidth}{0.32\textwidth}
  \setlength{\figheight}{0.2\textheight}

  \begin{minipage}{\figwidth}
    \includegraphics[width=\linewidth, height=\figheight, keepaspectratio]{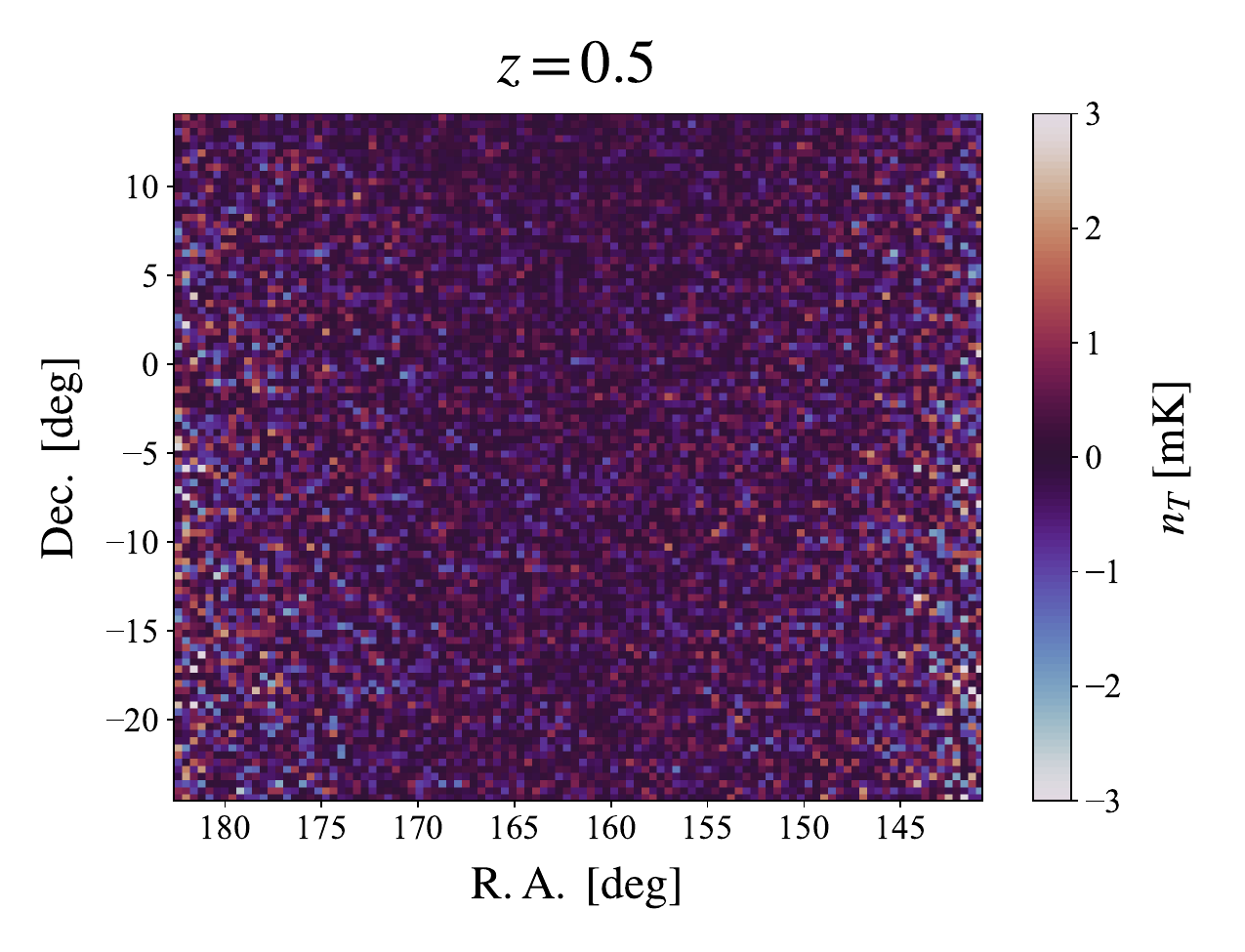}
    \label{fig:a}
  \end{minipage}
  \hfill
  \begin{minipage}{\figwidth}
    \includegraphics[width=\linewidth, height=\figheight, keepaspectratio]{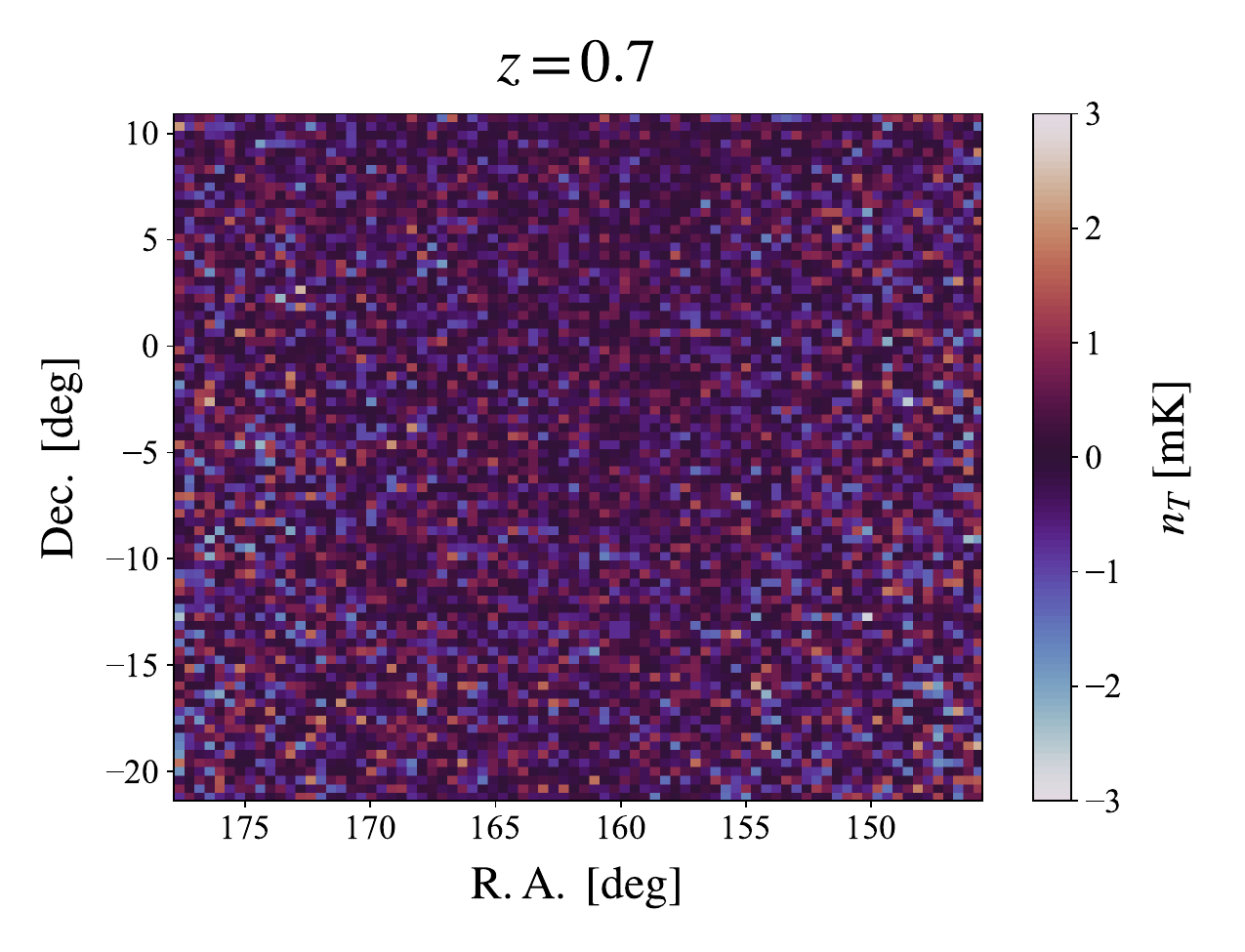}
    \label{fig:b}
  \end{minipage}
  \hfill
  \begin{minipage}{\figwidth}
    \includegraphics[width=\linewidth, height=\figheight, keepaspectratio]{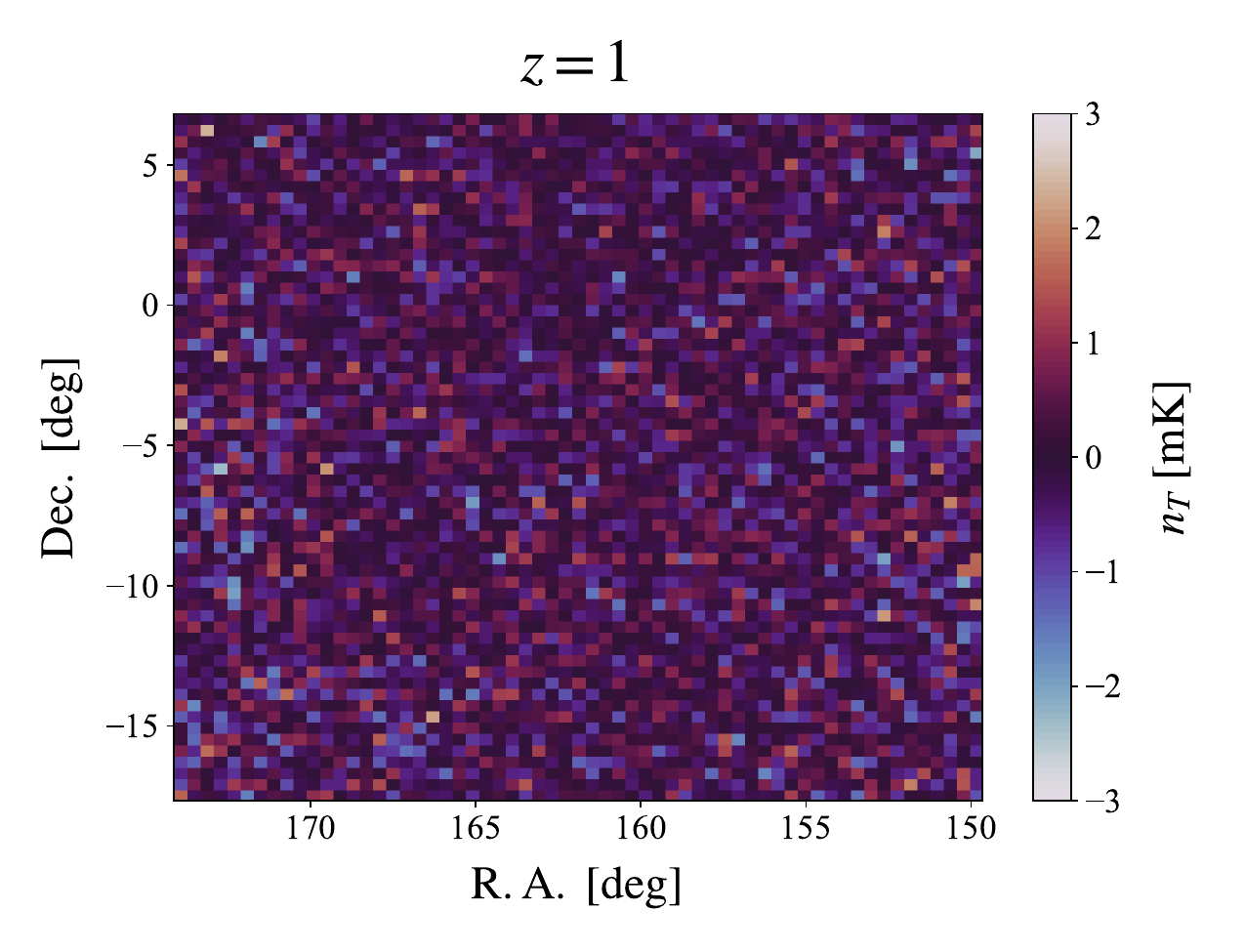}
    \label{fig:c}
  \end{minipage}
  \caption{The instrumental thermal noise maps in MeerKAT H\textsc{i} intensity mapping simulation for the central slice of each simulation box at $z=0.5$, 0.7, and 1.}
  \label{fig:noise}
\end{figure*}

We assume that the point sources should have clustering property. Since we know little about the two-point correlation function of the point sources distribution in H\textsc{i} intensity mapping surveys, we roughly set the power spectrum to be $P_{\mathrm{ps}}(k) = b_{\mathrm{ps}}^2P_{\rm m}(k)$, where we assume $b_{\mathrm{ps}}=1$ and $P_{\rm m}(k)$ is the matter power spectrum of the present Universe. Then we use $P_{\mathrm{ps}}(k)$ to generate a random surface distribution of point sources in the survey area corresponds to $1 (\mathrm{Gpc}/h)^2$. The number density of sources can be obtained by $N=\int \frac{\mathrm{d}N}{\mathrm{d}S}\mathrm{d}S$. After that, flux at $1400\, \mathrm{MHz}$ is assigned to each point source according to the Equation~\eqref{flux function}. The observed brightness temperature at each frequency $\nu$ of point sources can be calculated as
\begin{align}
    T_{\mathrm{ps}}(\nu)=2k_{\rm B}\nu^2\frac{S_{\nu}}{\Omega}=2k\nu^2 \frac{S_0}{\Omega} \left(\frac{\nu}{\nu_{1400}}\right)^{\alpha},
\end{align}
where $k_{\rm B}$ is the Boltzmann constant, $\Omega$ is the solid angle of the point source and $\alpha=-0.75$ is the spectral index. The brightness temperature map of point sources at $930$ MHz is shown in Figure~\ref{fig:pointsources}. 
Furthermore, it is expected that point sources above $S_{\rm max}=10$ mJy are able to be identified by other obervations, such as H\textsc{i} galaxy survey, and subtracted from the map, and masking bright point sources may reduce the complexity of the foreground to some extent. So we also simulate the point source distribution without sources above $S_{\rm max}=10$ mJy as shown in Figure~\ref{fig:pointsources} (top right panel). We find that the mean brightness temperature of point sources at $z=0.5$, $z=0.7$ and $z=1$ are $0.281$, $0.386$, and $0.553$~K, respectively, and will decrease to $0.155$, $0.204$, and $0.292$~K after masking the bright sources.

The emission from radio point sources also has polarized components, which suggests that they can contribute to the polarization leakage. We make use of the model provided in \textsc{cora} to simulate the polarization map of point sources. For each point source, a Gaussian random initial polarization fraction $\alpha_p=\alpha_{\mathrm{Q}}+i\alpha_{\mathrm{U}}$ is assigned with $\sigma=0.03$. Then the observed flux of polarization emission at frequency $\nu$ is obtained by
\begin{align}
    S_p(\nu)=p S_{\nu} e^{2i\phi\frac{c^2}{\nu^2}},
\end{align}
where $S_{p}=S_{\mathrm{Q}}+iS_{\mathrm{U}}$ and $\phi$ is the Faraday depth which is obtained by the same Faraday rotation map we used in simulating the Galactic emission \citep{faraday_map}. Finally, we obtain the point source components in the observed brightness temperature by convolving the emission with corresponding beam patterns.
The point sources maps of Q/U polarization and their polarization leakages at $930$ MHz are shown in Figure~\ref{fig:pointsources} as examples. 


\subsubsection{Instrumental Thermal Noise} \label{sec:noise}

We model the instrumental thermal noise of single-dish survey as Gaussian noise for each TOD. The root mean square (rms) noise temperature can be calculated as \citep{noise}
\begin{align}
    \sigma_{\rm T}=\frac{T_{\rm sys}}{\sqrt{\delta\nu t_{\rm scan}}}\frac{\lambda^2}{\theta_{\rm b}^2 A_{e}}\sqrt{\frac{A_{\rm S}}{\theta_{\rm b}^2}},
\end{align}
where $T_{\mathrm{sys}}=20$ mK is the system temperature, $\delta \nu$ is the frequency resolution of MeerKAT ($208.9$ kHz for L-band and $132.8$ kHz for UHF-band), $t_{\mathrm{scan}}=2$ s is the observational time of each TOD, $A_e$ is the effective collecting area of a dish, $A_{\mathrm{S}}$ is the survey area, and $\theta_{\rm b}$ is the full width at half maximum (FWHM) of the beam. The frequency dependence of $\theta_{\rm b}$ is essential in intensity mapping simulation. Technically, the FWHM of the beam at each frequency can be obtained by integrating the beam pattern. In this work, we use the ripple model described in \cite{L-band_FWHM}, which is an 8th degree polynomial to accurately model the FWHM in MeerKAT L-band, and it is given by
\begin{align}
    \theta_{b}=\frac{c}{\nu D} \left[ \sum_{d=0}^8 a_d \nu^d + A\,{\rm sin}\left( \frac{2\pi\nu}{T}\right)\right],
\end{align}
where $A=0.1$ arcmin and $T=10$ MHz is the period and amplitude of the ``ripples'', respectively. The values of fitting parameter $a_d$ are $a_{0}=6.7\times 10^3$, $a_{1}=-50.3$, $a_{2}=0.16$, $a_{3}=-3.0\times10^{-4}$, $a_{4}=3.5\times10^{7}$, $a_{5}=-2.6\time10^{-10}$, $a_{6}=1.2\times10^{-13}$, $a_{7}=-3.0\times10^{-17}$, $a_{8}=3.4\times10^{-21}$. 
For UHF-band, we make use of the measurement data from \textsc{katbeam}\footnote{\url{https://pypi.org/project/katbeam/}}, which is also a Python package for MeerKAT beam simulation. It contains the FWHM measurement data of MeerKAT dish at sampled frequencies, and interpolation is made to obtain the FWHM at each observation frequency.

In Figure~\ref{fig:noise}, we show the central slice of each simulation box at $z=0.5$, 0.7 and 1. Since the noise level of each is straightly related to $t_{\mathrm{tot}}=N_{\mathrm{scan}}t_{\mathrm{scan}}$, and the scan trail of on-the-fly observation will cause inhomogeneous $N_{\mathrm{scan}}$ on different pixels. So it can be inferred that, as shown in Figure~\ref{fig:noise}, the central part of the survey area clearly has lower noise than the regions on the left or right side at all frequencies or redshifts. 

\begin{figure*}[ht]
  \centering
  \setlength{\figwidth}{0.32\textwidth} 
  \setlength{\figheight}{0.188\textheight} 

  \begin{minipage}{\figwidth}
    \includegraphics[width=\linewidth, height=\figheight, keepaspectratio]{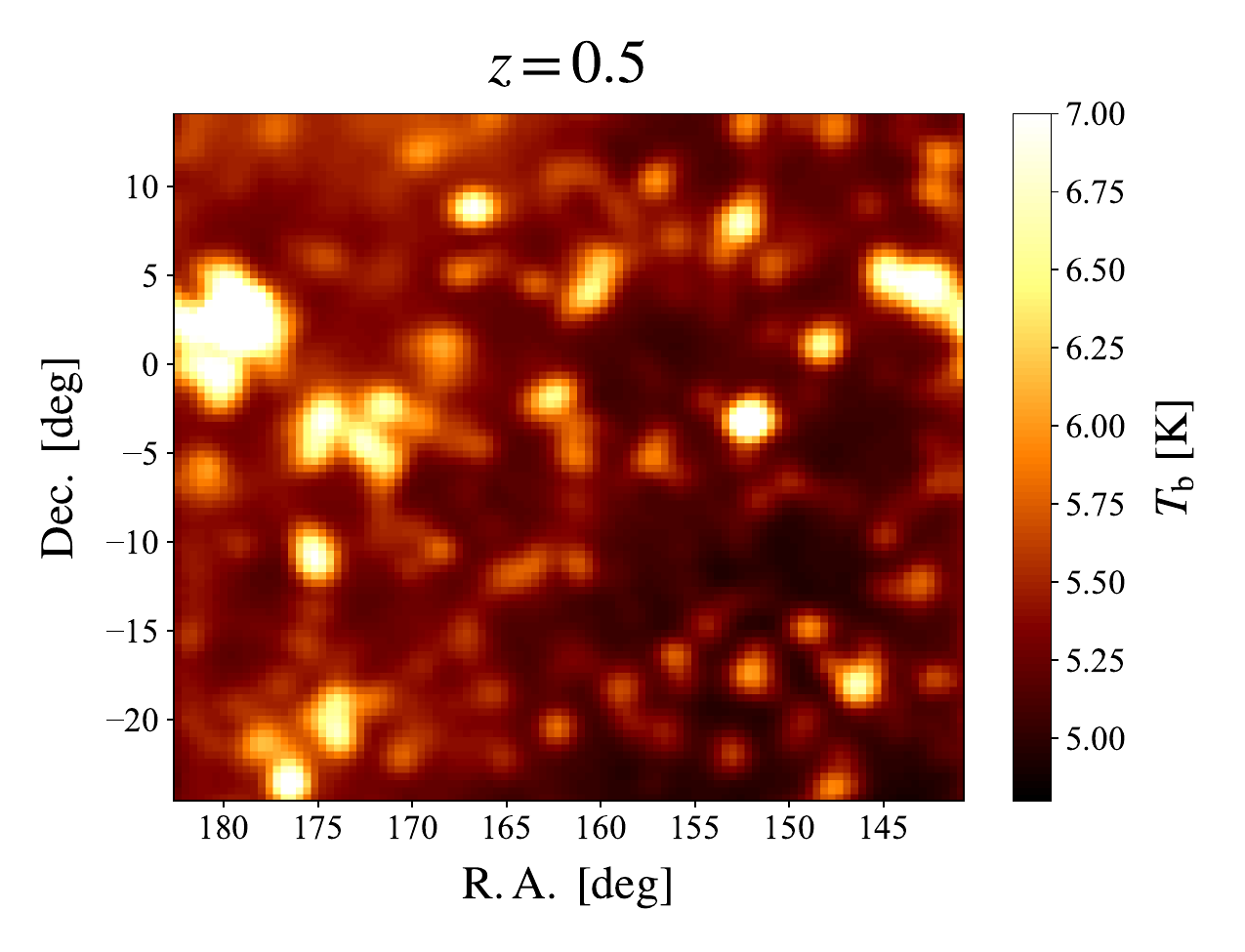}
  \end{minipage}
  \hfill
  \begin{minipage}{\figwidth}
    \includegraphics[width=\linewidth, height=\figheight, keepaspectratio]{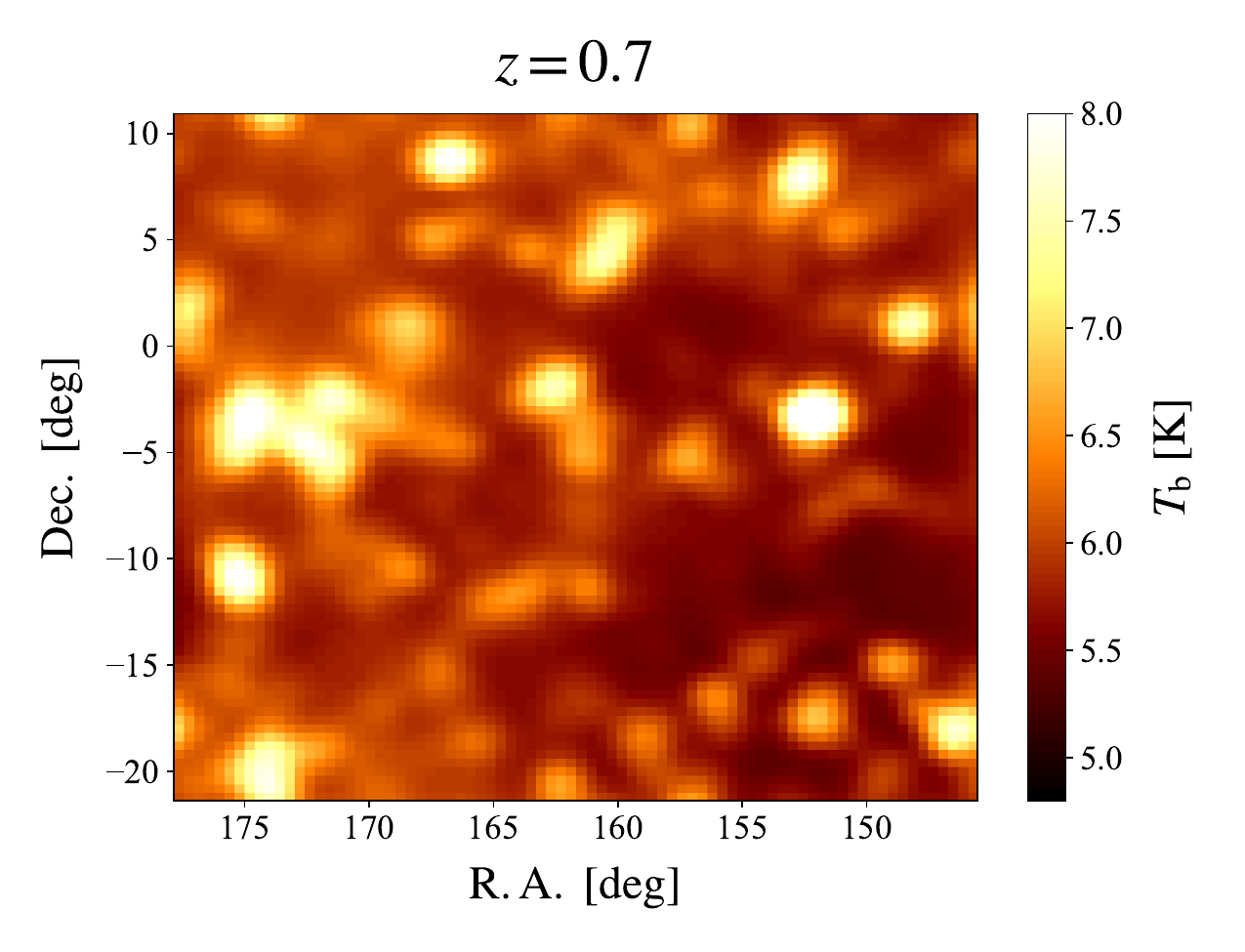}
  \end{minipage}
  \hfill
  \begin{minipage}{\figwidth}
    \includegraphics[width=\linewidth, height=\figheight, keepaspectratio]{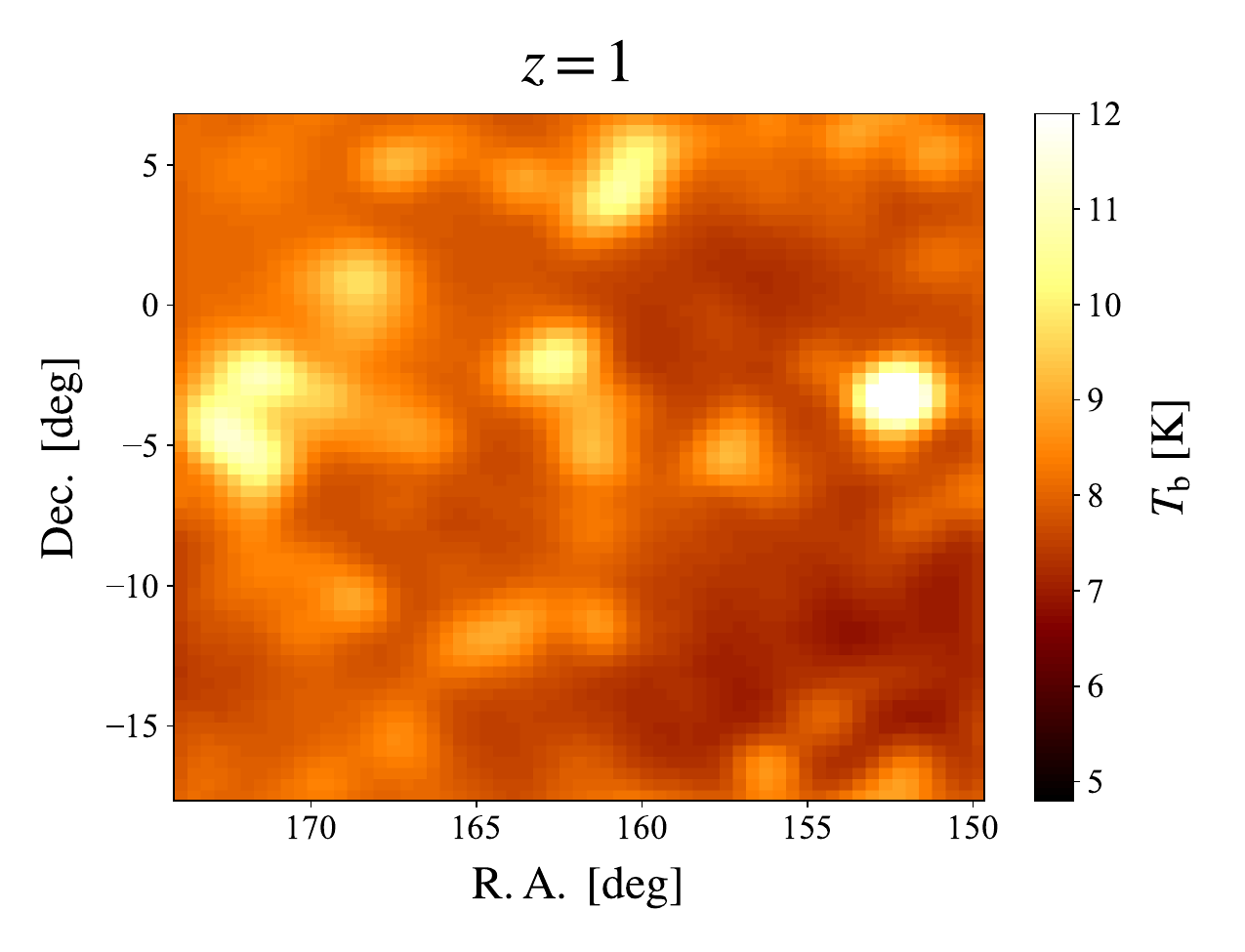}
  \end{minipage}


  \begin{minipage}{\figwidth}
    \includegraphics[width=\linewidth, height=\figheight, keepaspectratio]{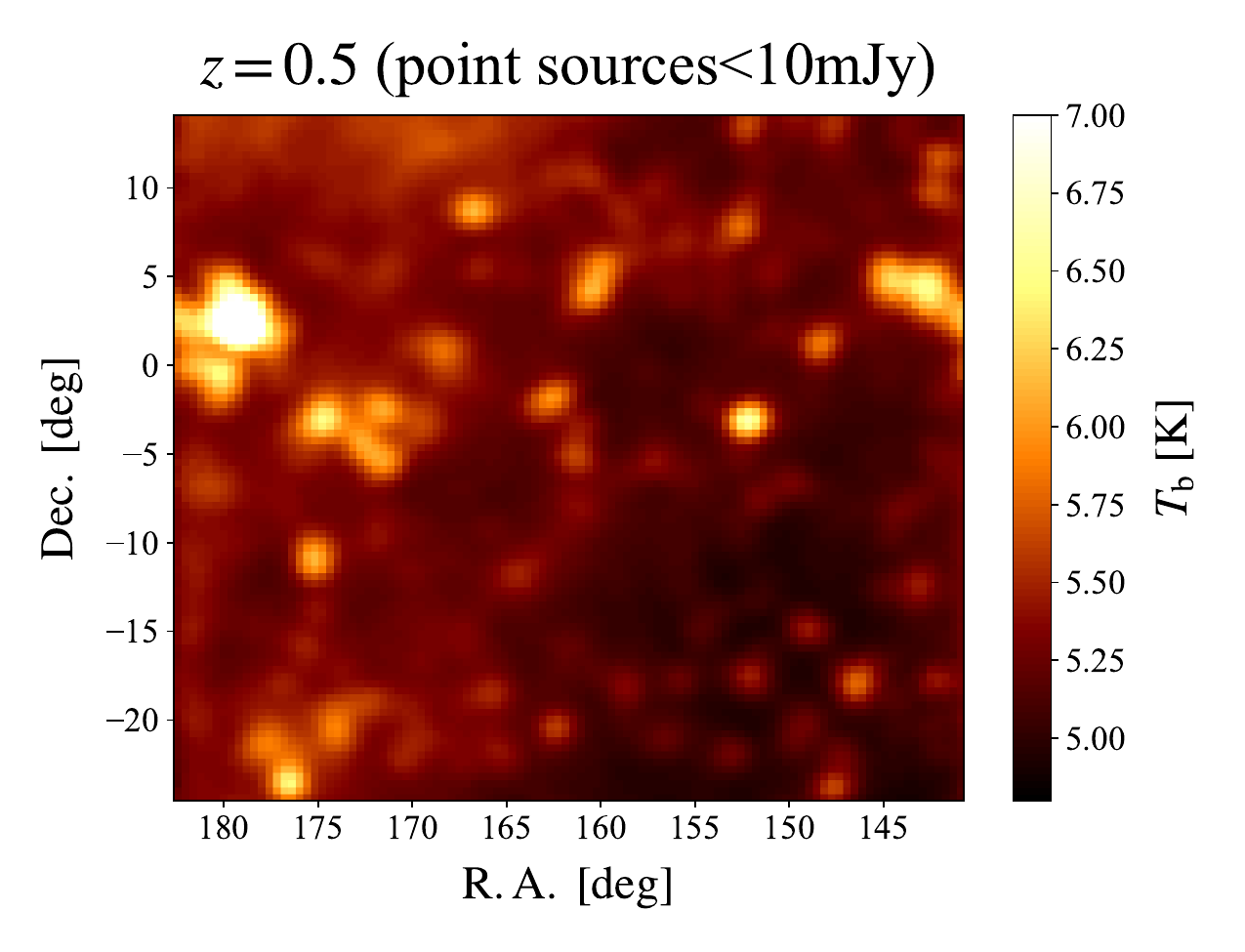}
  \end{minipage}
  \hfill
  \begin{minipage}{\figwidth}
    \includegraphics[width=\linewidth, height=\figheight, keepaspectratio]{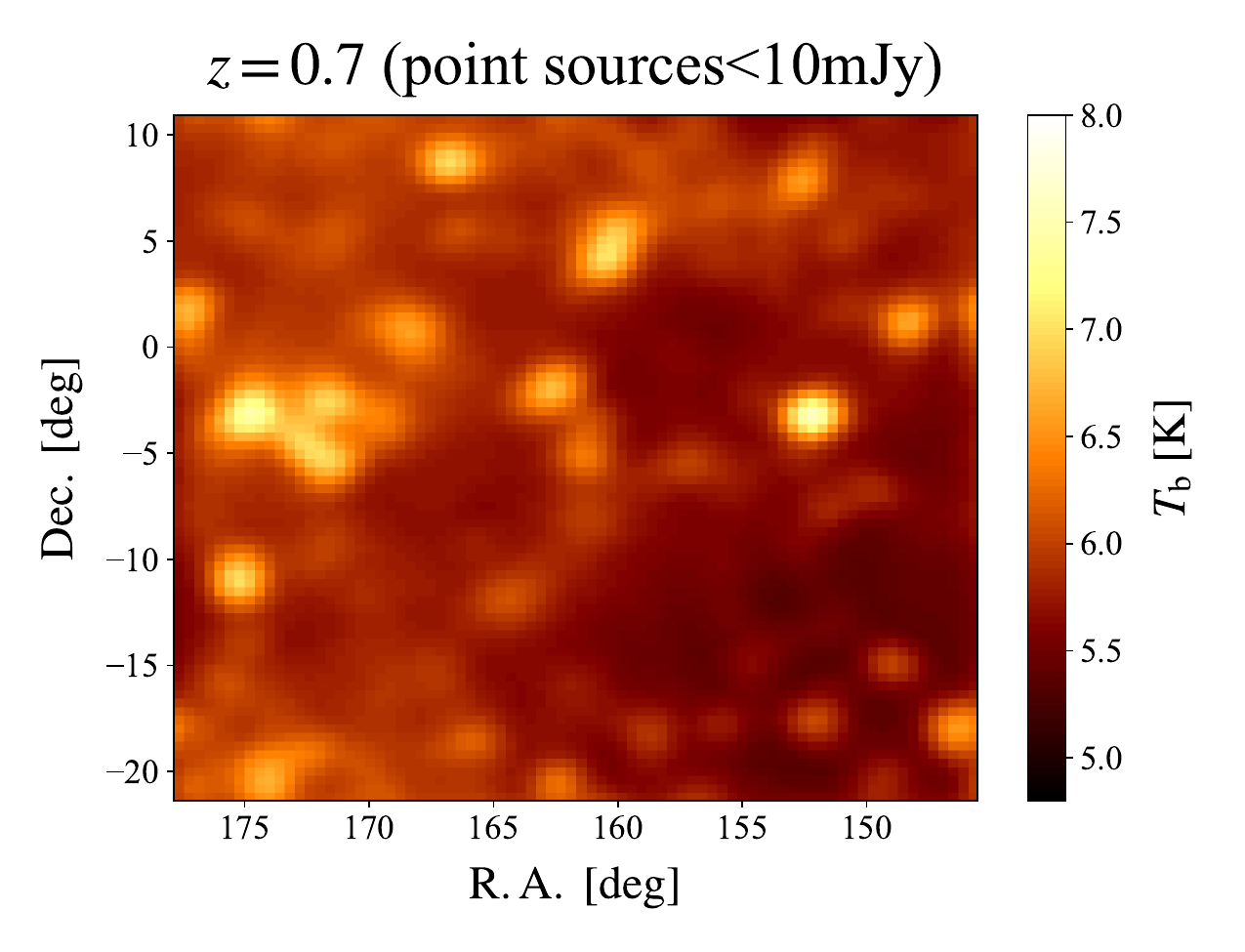}
  \end{minipage}
  \hfill
  \begin{minipage}{\figwidth}
    \includegraphics[width=\linewidth, height=\figheight, keepaspectratio]{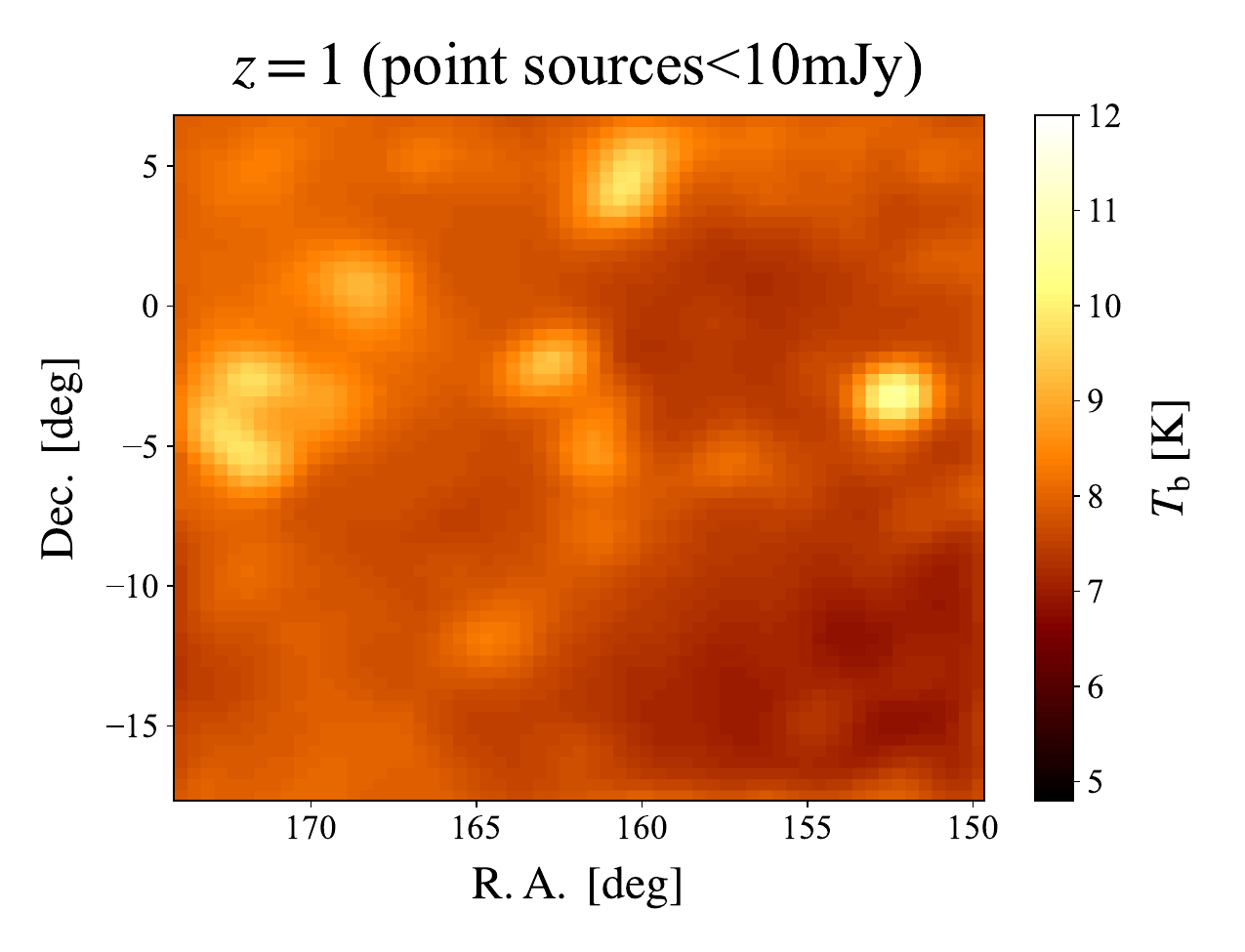}
  \end{minipage}
  
  \caption{The simulation results of MeerKAT H\textsc{i} intensity mapping survey at $z=0.5$, 0.7, and 1. 
  The upper row are the non-masking maps and the lower row are the maps masking bright point sources with flux $>10$ mJy.}
  \label{fig:full map}
\end{figure*}

Upon completion of simulating all component maps for the H\textsc{i} observation, we combine them to generate the final  observational results of MeerKAT H\textsc{i} intensity mapping.
Figure~\ref{fig:full map} presents the mid-frequency maps for all three frequency bands. For each redshift, we display both the original map and the map with bright point sources masked (flux $>10$ mJy), facilitating a direct comparison of the masking effect.

\subsection{CSST Spectroscopic Galaxy Survey}
For the CSST slitless spectroscopic survey, it has three bands (i.e. $GU$, $GI$, and $GV$) with spectral resolution $R \gtrsim 200$ and can reach a $5 \sigma$ point-source detection limit of $\sim 23$ AB magnitude.
In Jiutian-1G simulation, the distribution and intrinsic properties of galaxies has already been given by SAM, along with the luminosities of emission lines \citep{emission_lines}. 
We construct the galaxy catalog based on the signal-to-noise ratio (SNR) in the CSST spectroscopic survey \citep{SNR1, dfr}
\begin{align}
    \mathrm{SNR}=\frac{C_{\rm S}t_{\mathrm{exp}}\sqrt{N_{\mathrm{exp}}}}{\sqrt{C_{\rm S}t_{\mathrm{exp}}+N_{\rm pix}[(B_{\rm sky}+B_{\rm det})+R^2_{n}]}},
    \label{eq:SNR}
\end{align}
where $t_{\mathrm{exp}}=150\, \mathrm{s}$ is the exposure time, $N_{\mathrm{exp}}=4$ is the number of exposures, and $N_{\rm pix}=\Delta A/l^{2}_{p}$ is the number of detector pixels covered by an object. $R_n=5 e^{-1}\mathrm{s}^{-1}\mathrm{pixel}^{-1}$ represents the read noise. $B_{\rm det}$ is the dark current of the detector, $B_{\rm sky}$ is the sky background and $C_{\rm S}$ is the counting rate of emission lines from galaxy \citep[e.g.][]{SNR2}.

\begin{figure*}[ht]
  \centering
  \setlength{\figwidth}{0.32\textwidth} 
  \setlength{\figheight}{0.2\textheight} 
  
  \begin{minipage}{\figwidth}
    \includegraphics[width=\linewidth, height=\figheight, keepaspectratio]{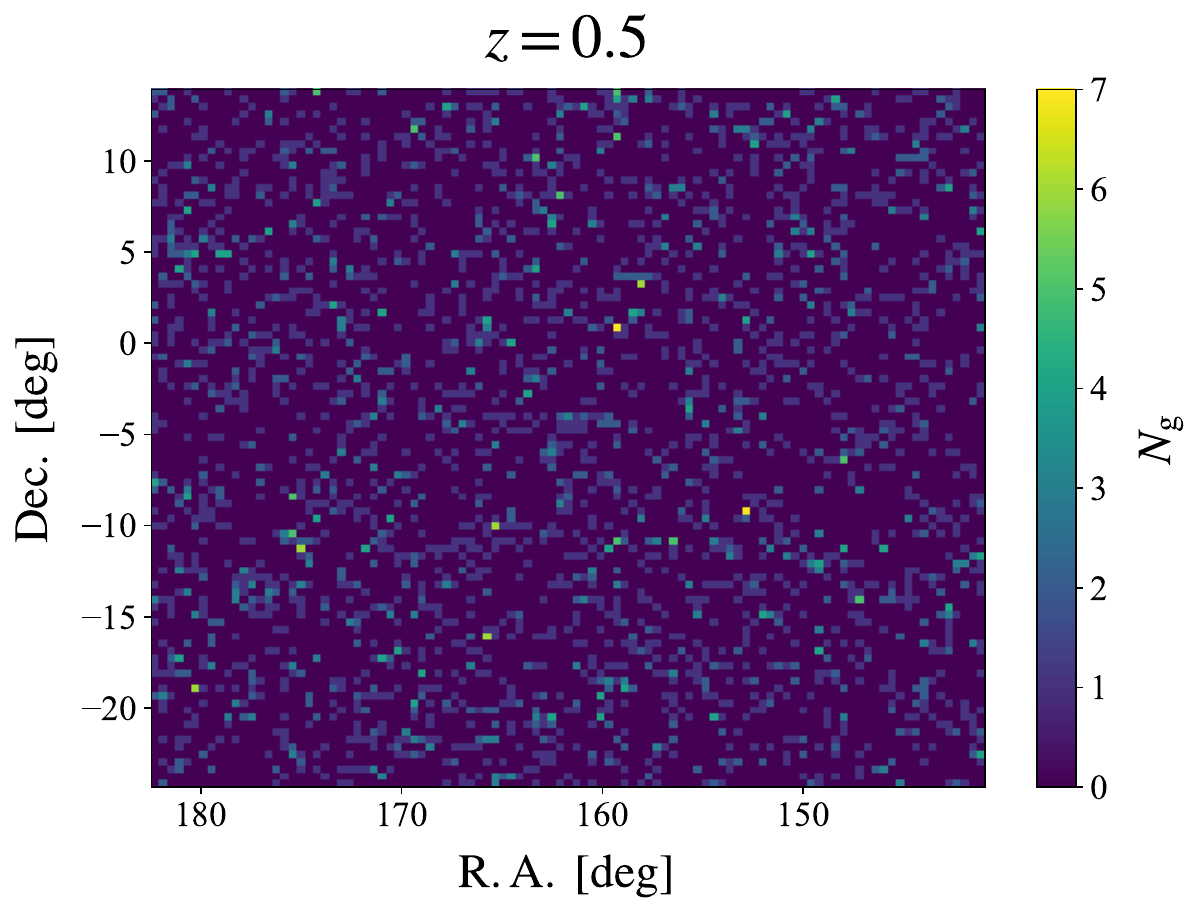}
    \label{fig:a}
  \end{minipage}
  \hfill
  \begin{minipage}{\figwidth}
    \includegraphics[width=\linewidth, height=\figheight, keepaspectratio]{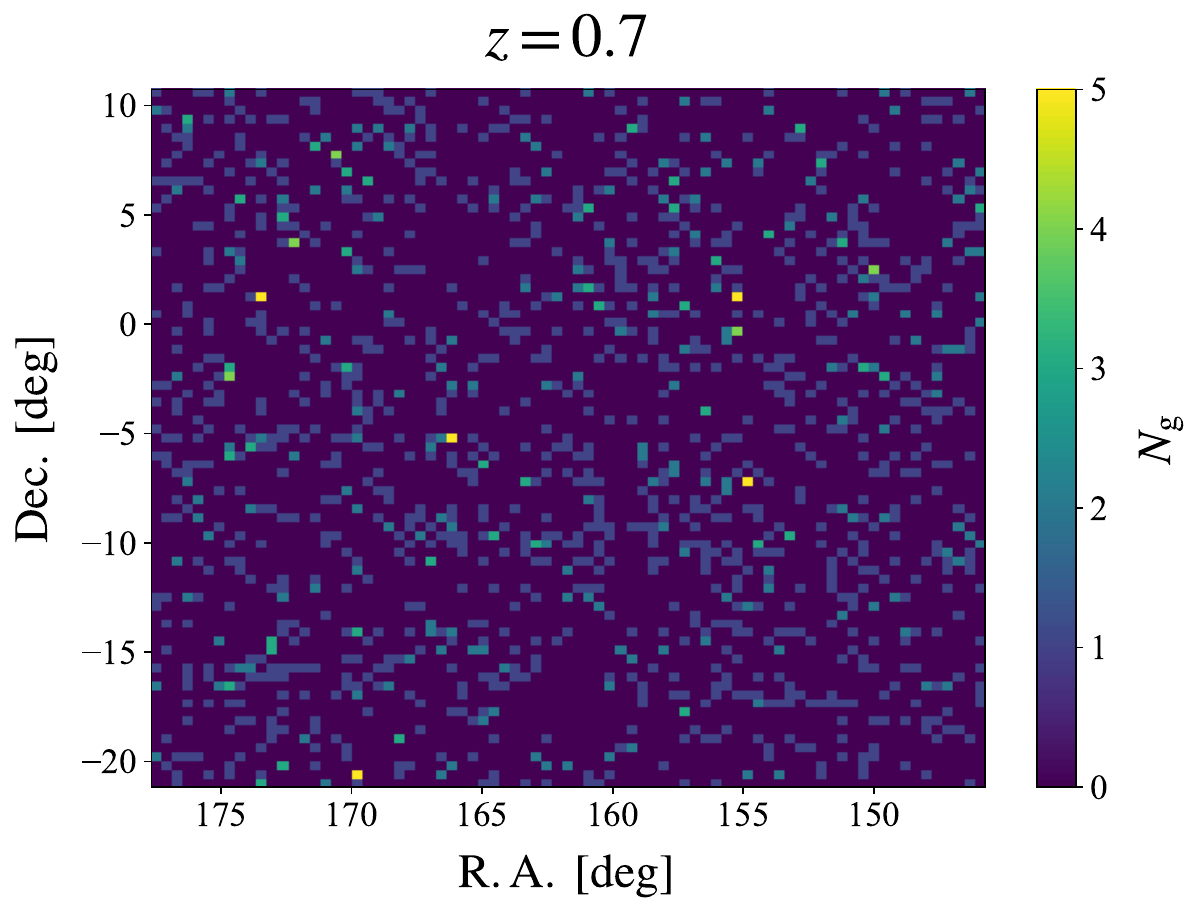}
    \label{fig:b}
  \end{minipage}
  \hfill
  \begin{minipage}{\figwidth}
    \includegraphics[width=\linewidth, height=\figheight, keepaspectratio]{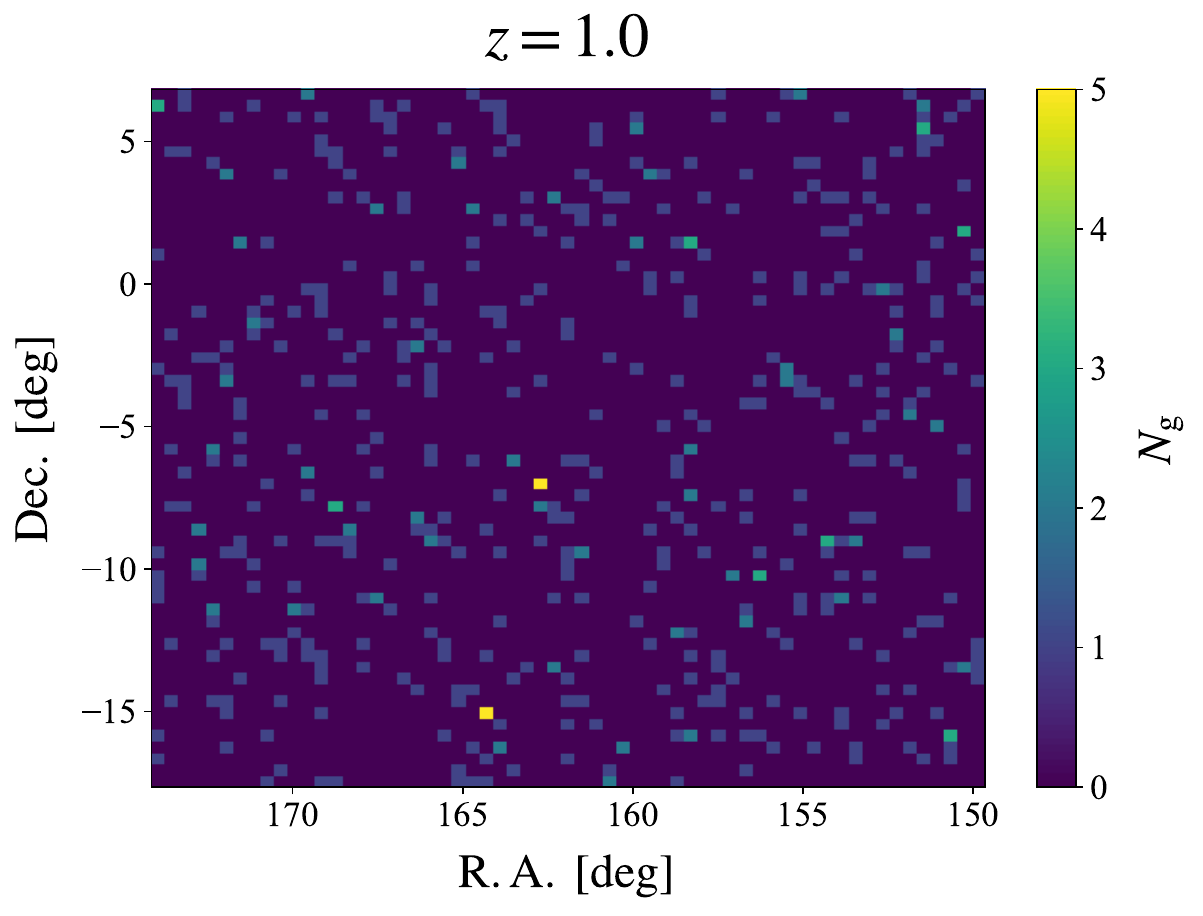}
    \label{fig:c}
  \end{minipage}
  \caption{The simulated galaxy maps in the CSST spectroscopic galaxy survey for the central slice of each simulation box at $z=0.5$, 0.7, and 1, whose resolution is rescaled to match the corresponding MeerKAT H\textsc{i} intensity mapping survey.}
  \label{fig:galaxy}
\end{figure*}

Among all the emission lines that Jiutian-1G can provide, $\mathrm{H}\alpha$, $\mathrm{H}\beta$, $[\mathrm{O\textsc{ii}}]$, and $[\mathrm{O\textsc{iii}}]$ are chosen for the CSST galaxy selection. 
Applying an ${\rm SNR}>5$ threshold in any spectroscopic band, we construct the mock galaxy catalog by including all galaxies where at least one emission line meets this criterion. 
We find that the number density of galaxies are $5.3\times 10^{-3}$, $2.9\times 10^{-3}$, $0.9\times 10^{-3}\, h^{3}\mathrm{Mpc}^{-3}$
for the three snapshots at redshift $z=[0.5, 0.7, 1]$, respectively, which are in agreement with the results in the previous works \citep[e.g.][]{CSST3}. 
Then we regrid the selected galaxies into the same voxels as MeerKAT H\textsc{i} intensity mapping mock data for the power spectrum estimation. In Figure~\ref{fig:galaxy}, we show the corresponding galaxy map of the CSST spectroscopic survey at each redshift.

\section{Power spectrum estimation} \label{sec:power spectrum}

We follow the methods for estimating the cross-correlation between galaxy survey and H\textsc{i} intensity mapping provided by \cite{power_spectrum}. 
The data from the galaxy survey and H\textsc{i} intensity mapping are converted into galaxy over-density field and brightness over-temperature field, respectively, which are given by
\begin{align}
    \delta_{g}(\bm{x})=\frac{N_{g}(\bm{x})-\langle N_{g} \rangle}{\langle N_{g} \rangle},
\end{align}
\begin{align}
    \delta_{\mathrm{H\textsc{i}}}(\bm{x})=T_{\mathrm{obs}}(\bm{x})-\langle T_{\rm obs} \rangle,
\end{align}
where $N_{g}$ is the number of galaxies in each voxel and the angled brackets denote mean values of the observational data. The Fourier transformed fields of $\delta_{g}(\bm x)$ and $\delta_{\mathrm{H\textsc{i}}}(\bm x)$ are given by
\begin{align}
    \tilde{F}_g(\bm{k}) = \sum_{\bm{x}}\delta_g(\bm{x})e^{i\bm{k} \cdot \bm{x}},
\end{align}
\begin{align}
    \tilde{F}_{\mathrm{H\textsc{i}}}(\bm{k}) = \sum_{\bm{x}}\delta_{\mathrm{H\textsc{i}}}(\bm{x})e^{i\bm{k} \cdot \bm{x}}.
\end{align}
Then the estimators for the auto-correlation power spectra of galaxy $P_g$ and H\textsc{i} intensity mapping $P_{\mathrm{H\textsc{i}}}$ and their cross-correlation power spectrum $P_{\mathrm{H\textsc{i}},g}$ are given by
\begin{align}
    P_{g}(k) = \frac{V}{N^2_{\mathrm{voxel}}} \langle \tilde{F}_g(\bm{k}) \tilde{F}^*_g(\bm{k})\rangle,
\end{align}
\begin{align}
    P_{\mathrm{H\textsc{i}}}(k) = \frac{V}{N^2_{\mathrm{voxel}}} \langle \tilde{F}_{\mathrm{H\textsc{i}}}(\bm{k}) \tilde{F}^*_{\mathrm{H\textsc{i}}}(\bm{k})\rangle,
\end{align}
\begin{align}
    P_{\mathrm{H\textsc{i}},g}(k) = \frac{V}{N^2_{\mathrm{voxel}}} \langle \tilde{F}_g(\bm{k}) \tilde{F}^*_{\mathrm{H\textsc{i}}}(\bm{k})\rangle,
\end{align}
where $V$ is the survey volume and $N_{\mathrm{voxel}}$ is the number of voxels in the survey. Note that $P_g(k)$ is composed of the clustering and shot noise terms, i.e. $P_g(k)=P_g^{\rm clus}(k)+P_g^{\rm SN}$, and $P_{g}^{\mathrm{SN}}=1/ \langle N\rangle$ is the shot noise term which can be removed in the data analysis. The errors of the power spectra can be estimated by 
\begin{align}
    \sigma_{g} = \frac{1}{\sqrt{2N_{\rm m}(k)}}\sqrt{P_g(k)},
\end{align}
\begin{align}
  \sigma_{\mathrm{H\textsc{i}}} = \frac{1}{\sqrt{2N_{\rm m}(k)}}\sqrt{P_{\mathrm{H\textsc{i}}}(k)},
\end{align}
\begin{align}
    \sigma_{\mathrm{H\textsc{i}},g} = \frac{1}{\sqrt{2N_{\rm m}(k)}}\sqrt{P^2_{\mathrm{H\textsc{i}},g}(k)+P_{\mathrm{H\textsc{i}}}(k)P_g(k)},
\end{align}
where $N_{\rm m}(k)$ is the number of modes in each $k$ bin.

\section{signal extraction} \label{sec:PCA&TF}
The 21cm signal always suffers from severe foreground contamination. Theoretically, the foreground contamination can be effectively mitigated by cross-correlating with other tracers (e.g. galaxies and other emission lines), since the foregrounds and instrumental effects from different surveys in different frequency bands are not correlated. However, our tests suggest that a direct cross-correlation is not effective enough to eliminate the effects of the foregrounds, and additional foreground removal techniques are needed before the cross-correlation analysis \citep{GBT_eBOSS, 2023ApJ...945...38Z, 2025A&A...703A.222C}.



\subsection{foreground removal} \label{sec:PCA}
In our analysis, we adopt the PCA/SVD (Principle Components Analysis and Singular Value Decomposition) approach for foreground removal. These methods are particularly effective because they leverage the distinct frequency-domain correlations of different components, enabling separation of the spectrally smooth foregrounds from the H\textsc{i} signal. PCA/SVD is especially advantageous as it requires minimal prior knowledge about the data components. While PCA identifies orthogonal components through eigenvalue decomposition, SVD offers a computationally efficient alternative by directly factorizing the data matrix, yielding comparable results with reduced computational steps \citep{PCA2}.

To apply the foreground removal procedure, the intensity mapping are represented by a data matrix $X_{\mathrm{obs}}$ with dimensions $N_{\nu}\times N_{p}$, where $N_{\nu}$ is the number of frequency channels and $N_{p}$ is the number of pixels at each frequency. Then the data matrix $X_{\mathrm{obs}}$ can be decomposed by SVD as
\begin{align}
    X_{\mathrm{obs}}=W^{\top}\Sigma R
\end{align}
where $W^{\top}$ and $R$ are called left and right singular vectors, respectively, and they both are unitary matrices. And $\Sigma$ is a rectangular diagonal matrix of singular values. Singular vectors $W^{\top}$ and singular values $\Sigma$ are equivalent to the eigenvectors and eigenvalues in PCA, respectively. 

\begin{figure*}[ht]
    \centering
    
    \includegraphics[width=0.31\linewidth]{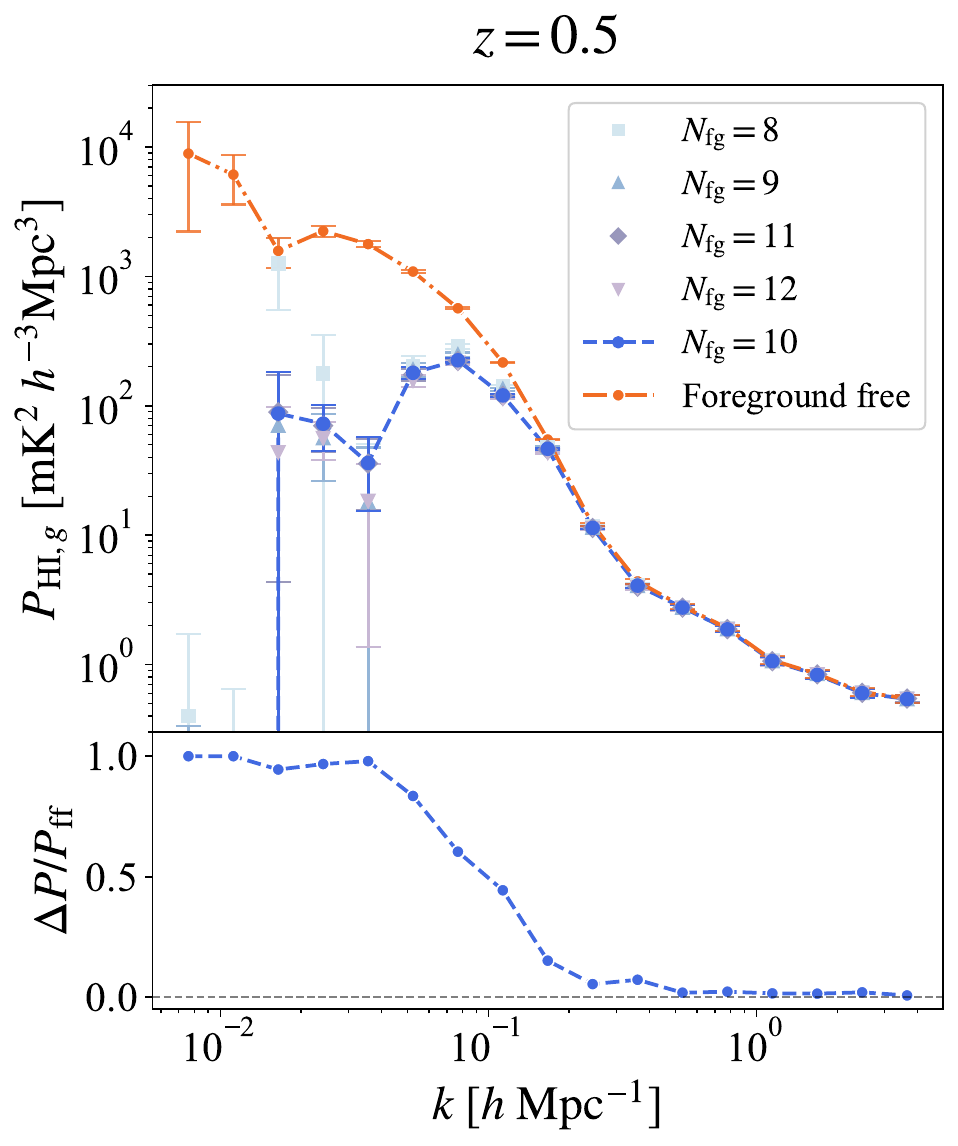}
    ~
    \includegraphics[width=0.31\linewidth]{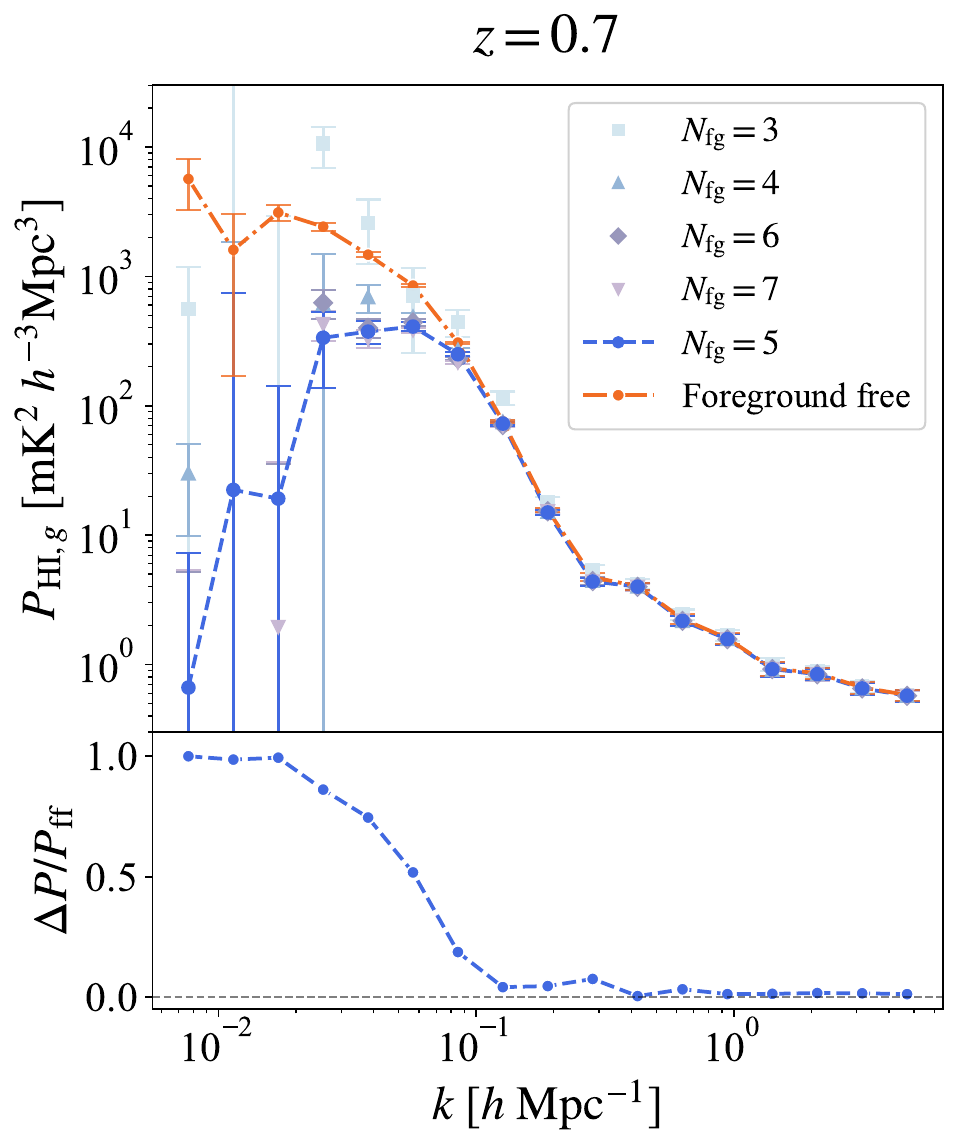}
    ~
    \includegraphics[width=0.31\linewidth]{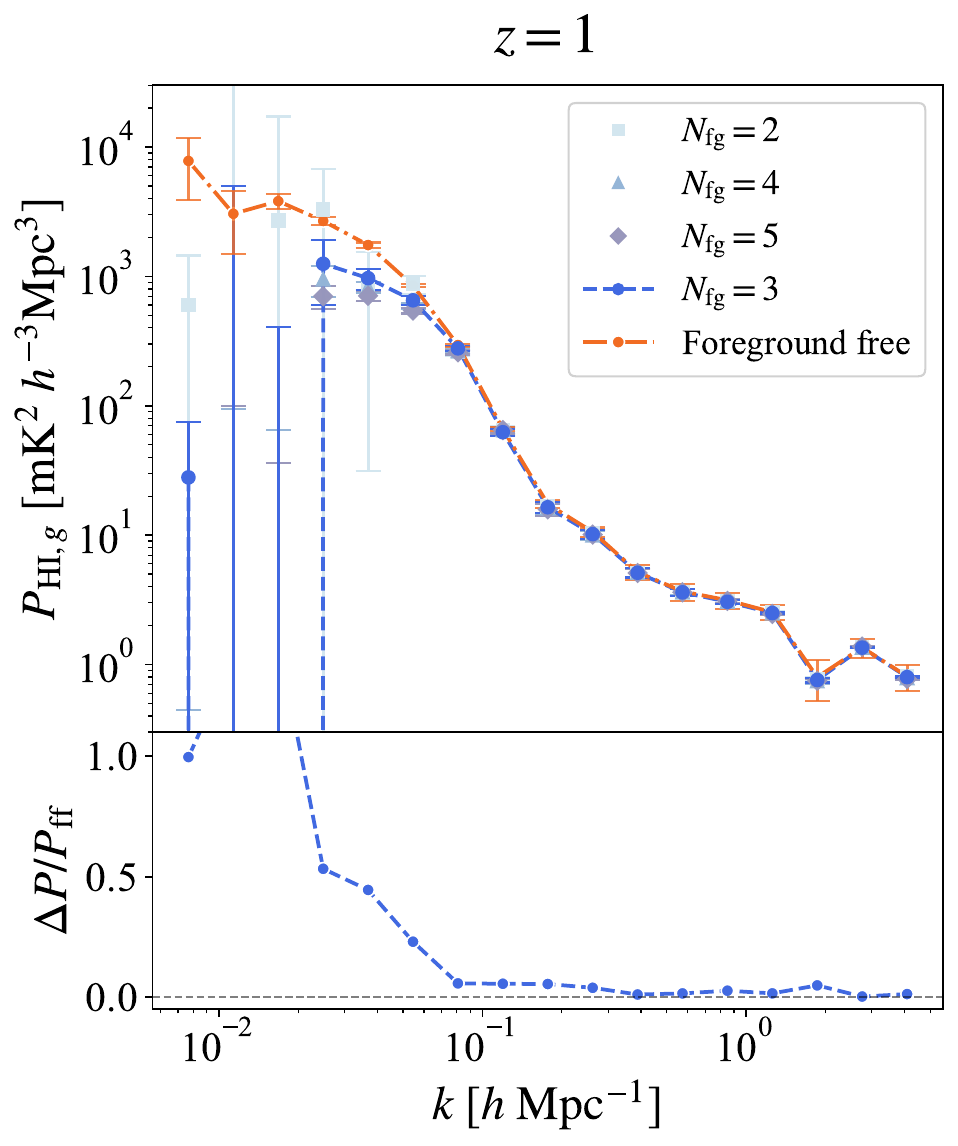} 
    
    \caption{ The cross power spectra comparison among different $N_{\mathrm{fg}}$ values and the results of our final $N_{\mathrm{fg}}$ decision of each redshift (blue dashed curves in the upper panels). The lower panels show the the level of signal loss, where $P_{\mathrm{ff}}$ is the foreground free power spectrum (orange dash-dotted curves in the upper panels) and $\Delta P$ is the difference between the power spectrum of foreground removed map and $P_{\mathrm{ff}}$.}
    \label{fig:PCA}
\end{figure*}

\begin{figure*}[ht]
  \centering
  \setlength{\figwidth}{0.32\textwidth} 
  \setlength{\figheight}{0.2\textheight} 

  \begin{minipage}{\figwidth}
    \includegraphics[width=\linewidth, height=\figheight, keepaspectratio]{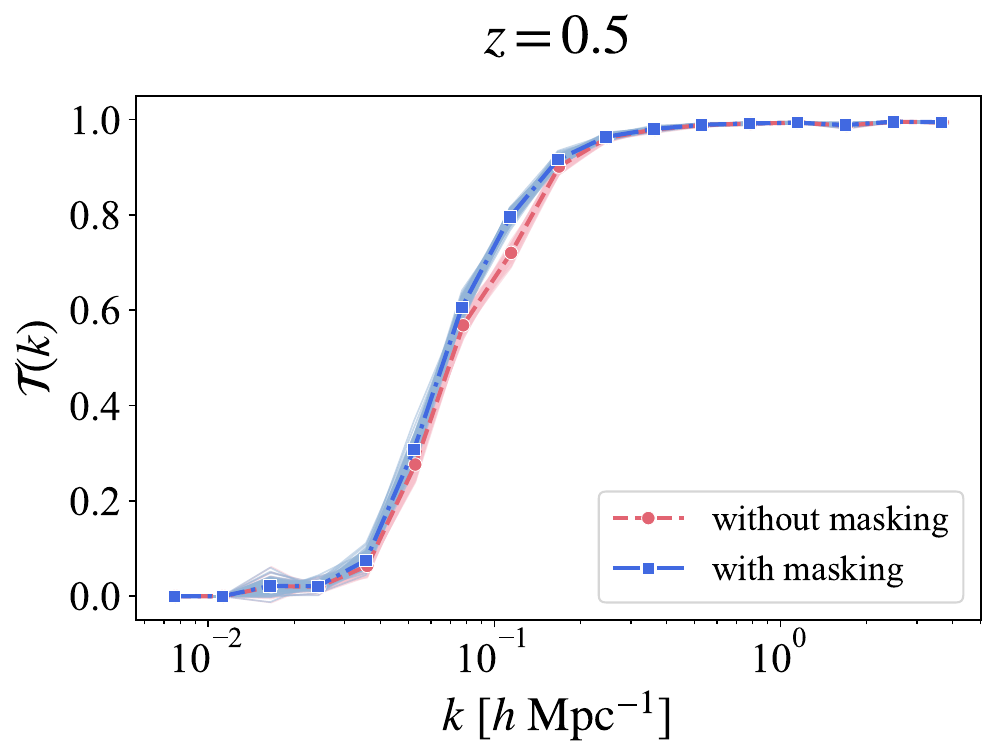}
    \label{fig:a}
  \end{minipage}
  \hfill
  \begin{minipage}{\figwidth}
    \includegraphics[width=\linewidth, height=\figheight, keepaspectratio]{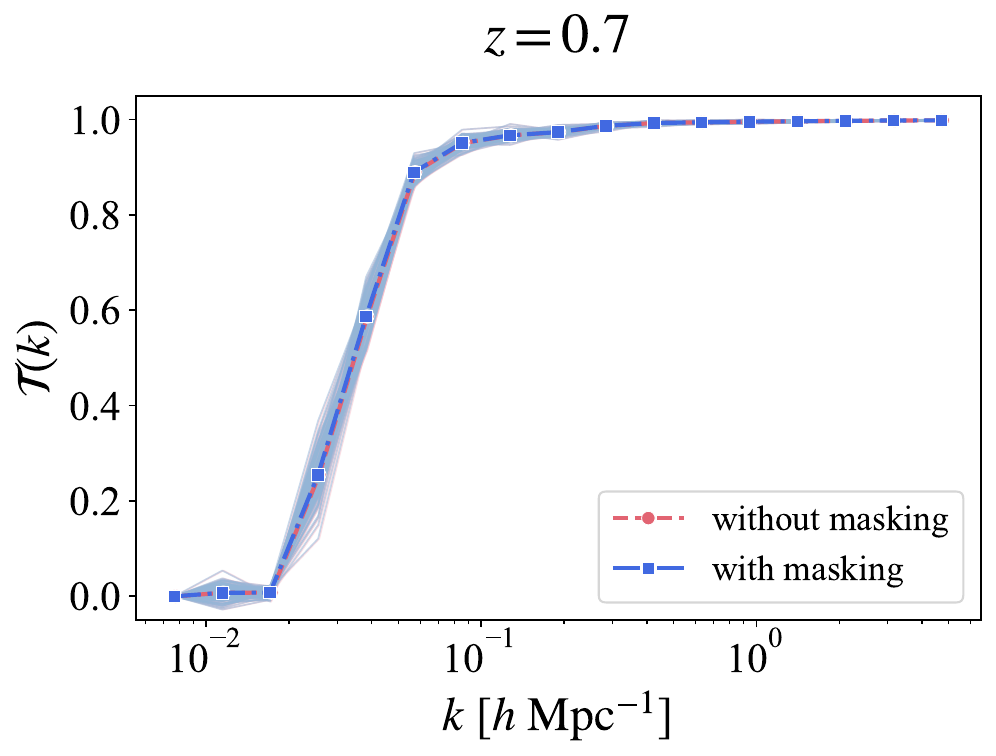}
    \label{fig:b}
  \end{minipage}
  \hfill
  \begin{minipage}{\figwidth}
    \includegraphics[width=\linewidth, height=\figheight, keepaspectratio]{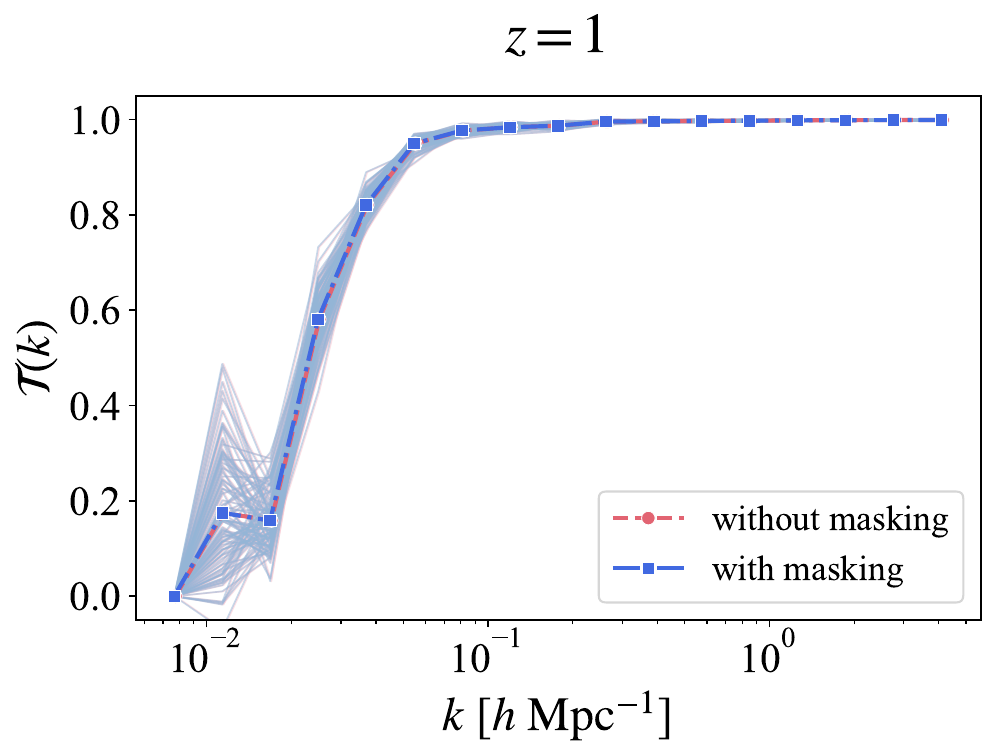}
    \label{fig:c}
  \end{minipage}
  \caption{The transfer functions constructed for signal compensation of MeerKAT-CSST cross-power spectra at $z=0.5$, 0.7, and 1. The red dashed curve and blue dash-dotted curve are for the non-masking and  masking H\textsc{i} intensity maps, respectively.
  The solid curves denote all the transfer functions of 100 mocks, and the dashed curves are the average of these transfer functions.}
  \label{fig:transfer function}
\end{figure*}

Then, for each intensity map, an $N_{\nu} \times N_{\mathrm{fg}}$ projection matrix $W'$ will be composed with the first $N_{\mathrm{fg}}$ columns of $W$, where $N_{\mathrm{fg}}$ is the number of the principal components identified as foreground contamination which will be removed from the map. The determination of $N_{\mathrm{fg}}$ represents a crucial step in the data processing pipeline, as it directly controls the balance between foreground removal and signal preservation. 
 
The dominant principal components will be obtained when the data matrix $X_{\mathrm{obs}}$ is projected onto the projection matrix $W'$ by
\begin{align}
    U=W'^{\top}X_{\mathrm{obs}},
\end{align}
and
\begin{align}
    V=W' U,
\end{align}
where $U$ is the foreground information constructed from the data matrix. And the H\textsc{i} signal will be filtered by
\begin{align}
    S_{\mathrm{H\textsc{i}}} = X_{\mathrm{obs}}-V.
\end{align}
Finally, the filtered signal will be transformed back to the original shape of intensity maps, and the foreground-subtracted maps are obtained. 

The determination of the values of $N_{\mathrm{fg}}$ is an essential part in H\textsc{i} intensity mapping data processing. 
We compared the cross-correlation power spectra of H\textsc{i} maps with different $N_{\mathrm{fg}}$ values and referred to the eigenvalue spectra and eigenvectors obtained from PCA/SVD method, the number of removal modes $N_{\mathrm{fg}}$ is determined to be 10, 5, 3 at $z=0.5$, 0.7, and 1 respectively. We also perform the same tests with the maps masked bright point sources with ${\rm flux} >10\,\mathrm{mJy}$ to check whether masking will decrease the difficulty of foreground removal. We did find that, at $z=0.5$, we can remove one less mode to obtain the similar result as non-masking map. However, at $z=0.7$ and 1, the values of $N_{\mathrm{fg}}$ remain the same.

We also notice that the value of $N_{\mathrm{fg}}$ decreases with increasing redshift. We check the spectra of all the components in our simulation at each redshift, and find that this is mainly due to that the beam patterns that \textsc{EIDOS} generated will be the same at frequency lower than $\sim 870$ MHz. Since the frequency dependency of beam pattern causes oscillations of polarization leakage along the frequency, when it stops, it will lessen the complexity of foreground. Therefore, it becomes more effective for the foreground component separation using PCA/SVD.

\begin{figure*}[ht]
  \centering
  \setlength{\figwidth}{0.32\textwidth} 
  \setlength{\figheight}{0.188\textheight} 

  \begin{minipage}{\figwidth}
    \includegraphics[width=\linewidth, keepaspectratio]{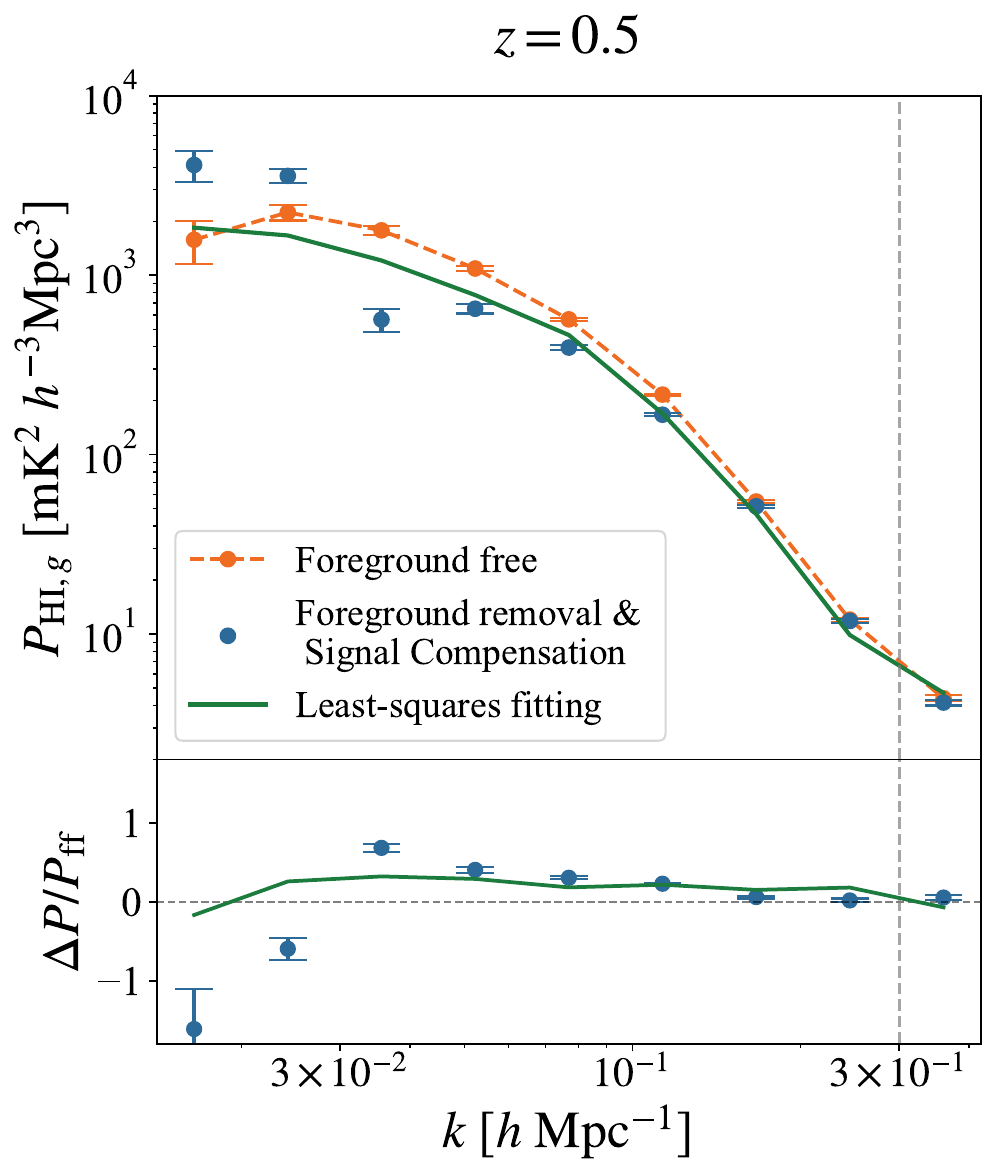}
    \label{fig:a}
  \end{minipage}
  \hfill
  \begin{minipage}{\figwidth}
    \includegraphics[width=\linewidth, keepaspectratio]{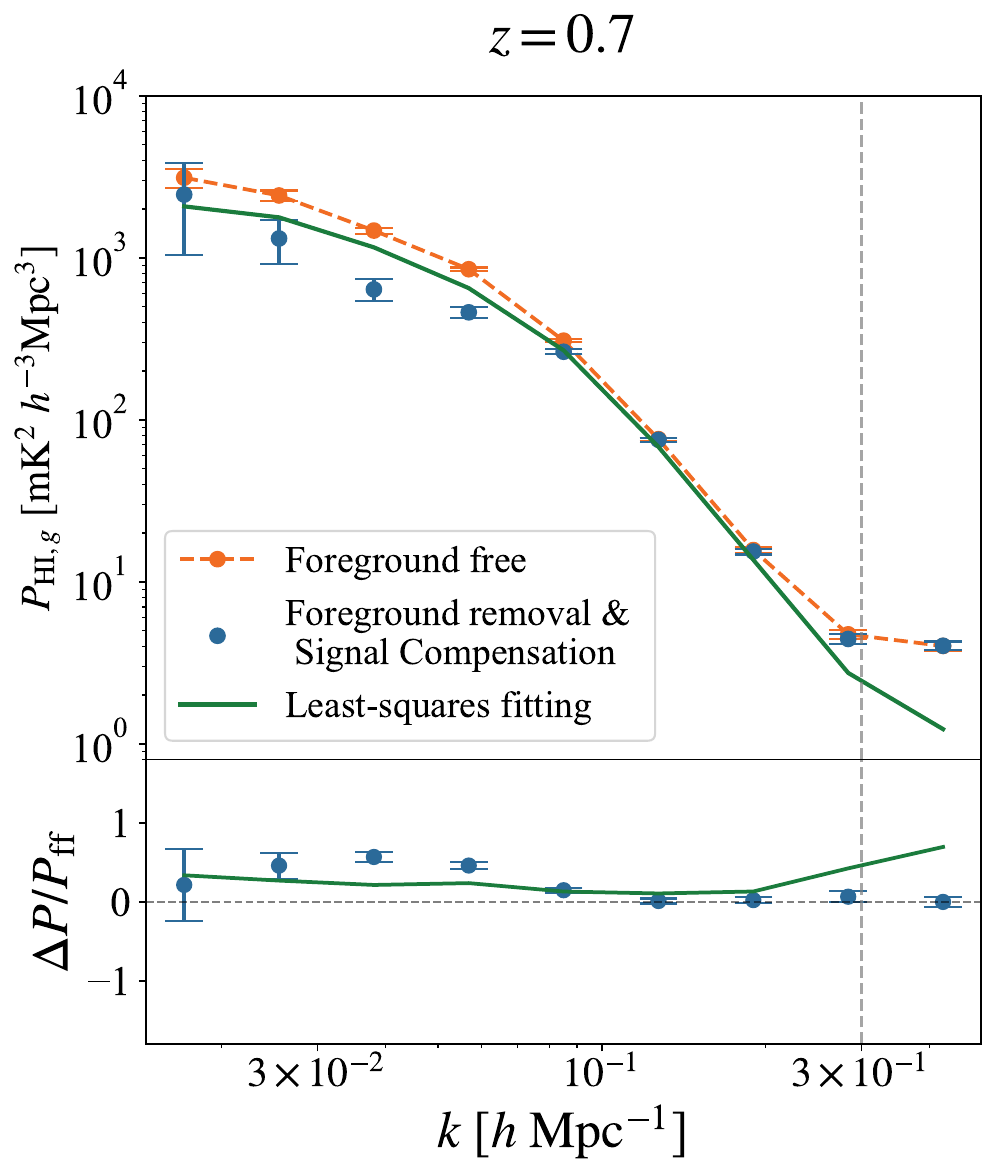}
    \label{fig:b}
  \end{minipage}
  \hfill
  \begin{minipage}{\figwidth}
    \includegraphics[width=\linewidth, keepaspectratio]{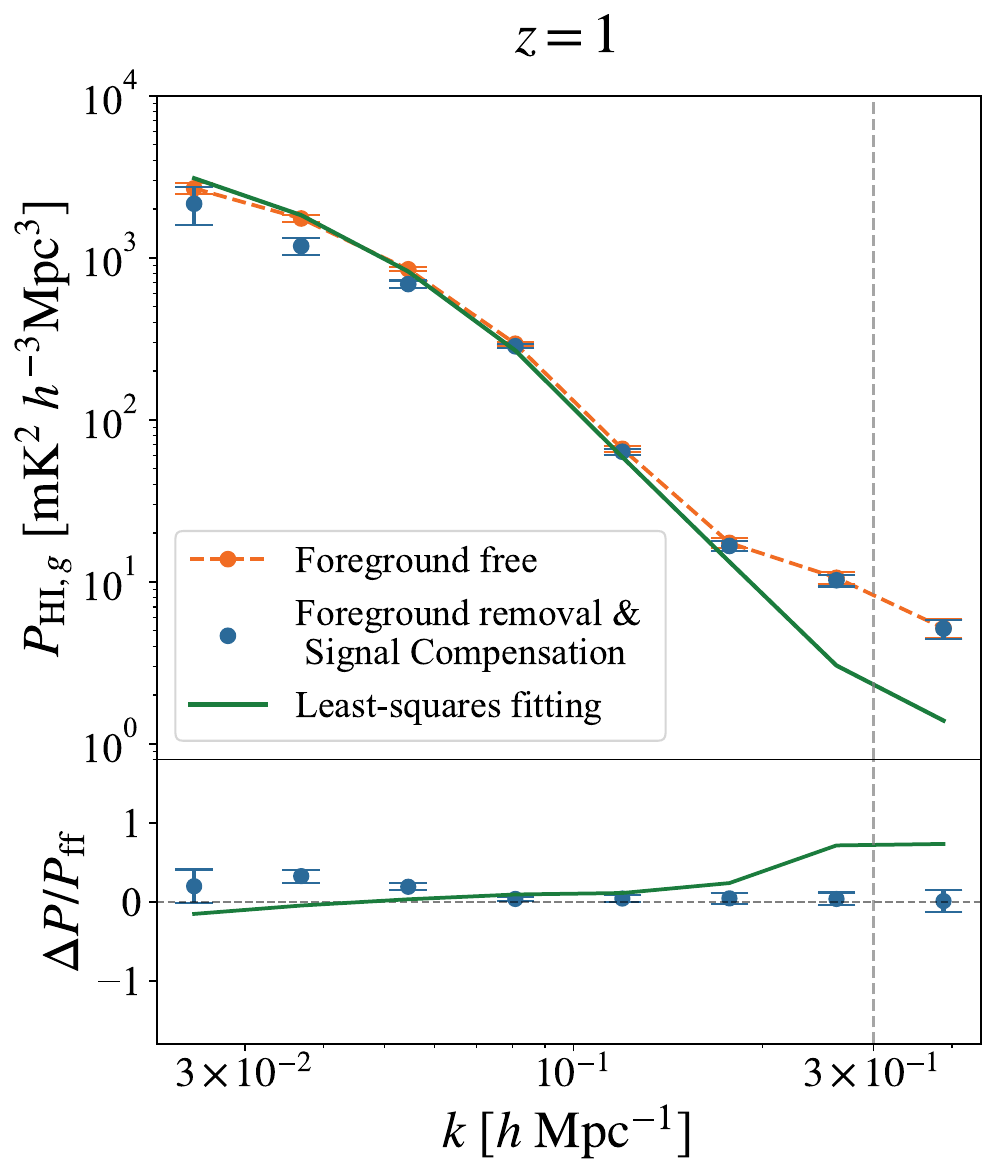}
    \label{fig:c}
  \end{minipage}


  \begin{minipage}{\figwidth}
    \includegraphics[width=\linewidth, keepaspectratio]{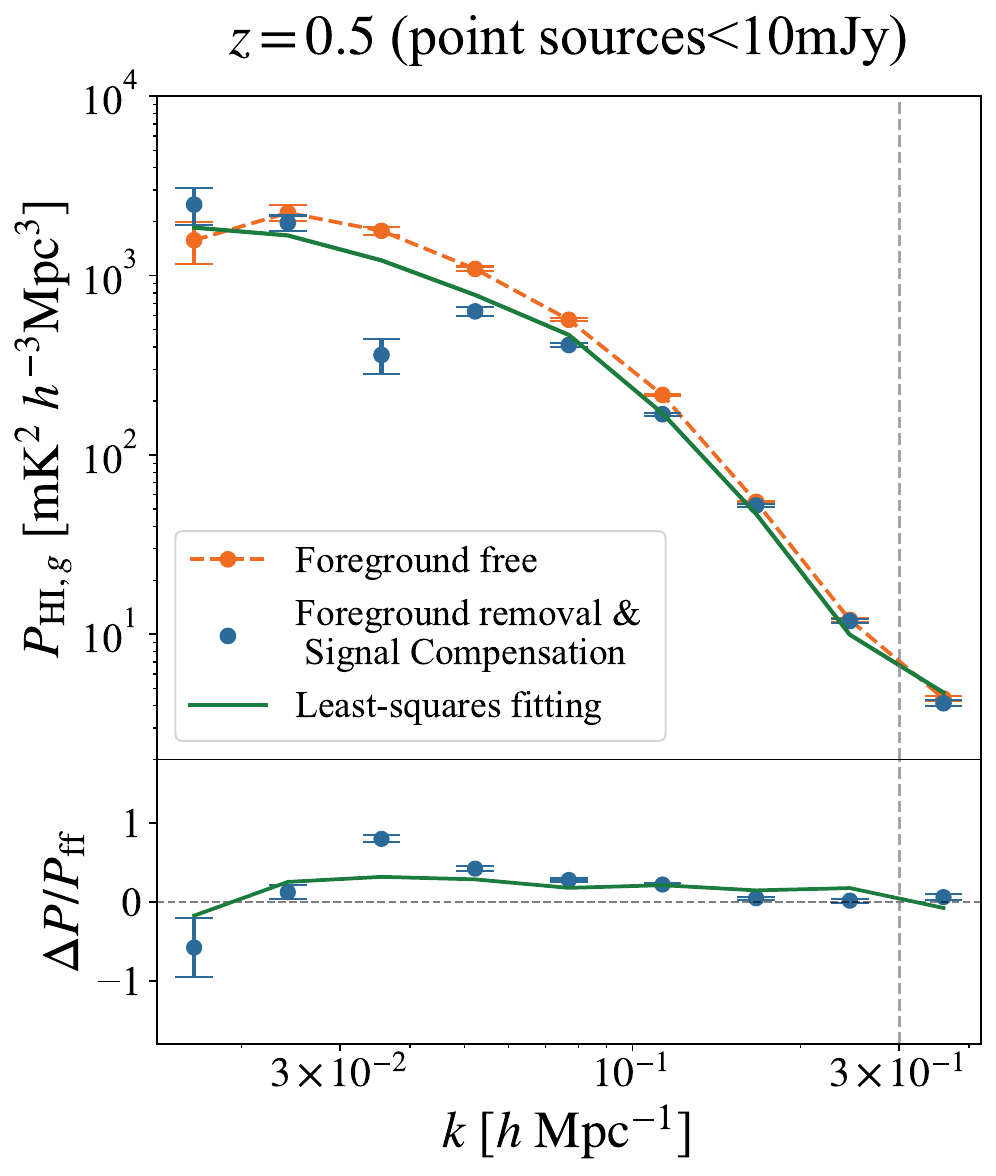}
    \label{fig:a}
  \end{minipage}
  \hfill
  \begin{minipage}{\figwidth}
    \includegraphics[width=\linewidth, keepaspectratio]{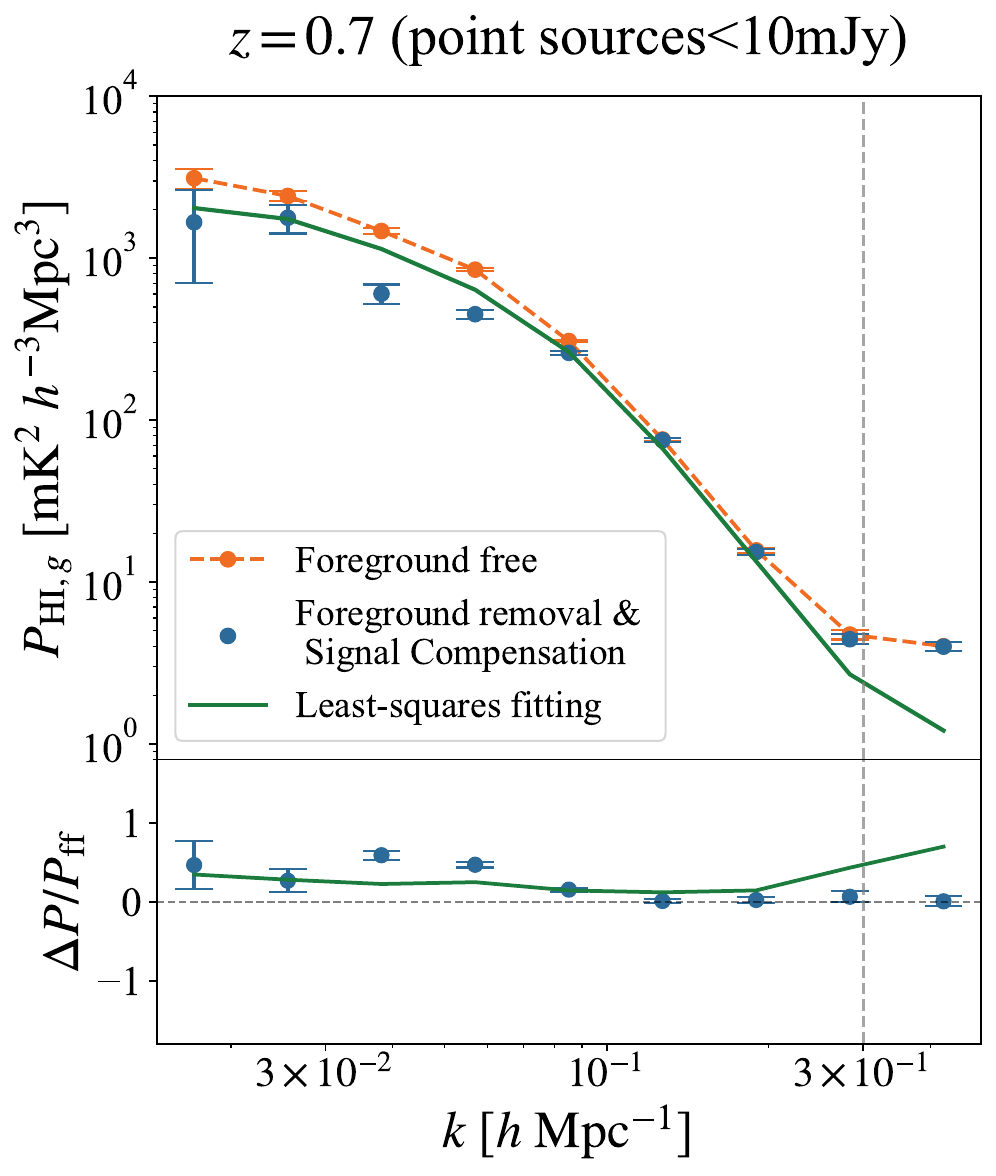}
    \label{fig:b}
  \end{minipage}
  \hfill
  \begin{minipage}{\figwidth}
    \includegraphics[width=\linewidth, keepaspectratio]{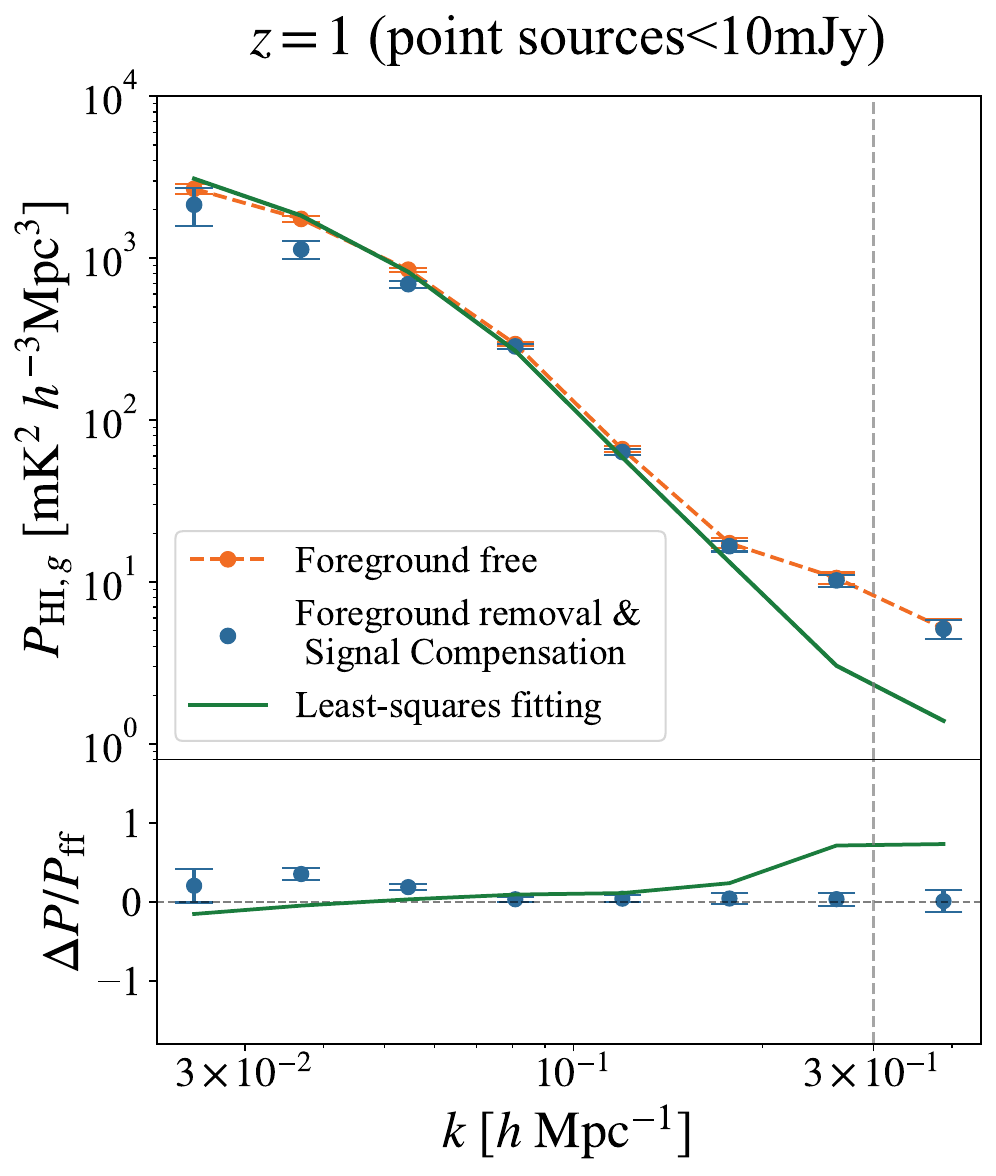}
    \label{fig:c}
  \end{minipage}
  
  \caption{The CSST-MeerKAT cross-power spectra at $z=0.5$, 0.7, and 1. The upper and lower rows show the results of the non-masking and masking H\textsc{i} intensity maps, respectively. The orange dashed curves are the power spectra of foreground-free H\textsc{i} maps $P_{\mathrm{ff}}$, the blue points are the results after foreground removal and signal compensation $P_{\mathrm{obs}}$, and the green curves are the least-squares fitting results $P_{\mathrm{fit}}$. Note that we only consider the scales at $k<0.3\, \mathrm{Mpc^{-1}}h$ in the fitting process. The lower panels of the figures show the the level of signal loss, the blue points denote $\Delta P = P_{\mathrm{ff}}-P_{\mathrm{obs}}$ and the green lines denote $\Delta P = P_{\mathrm{ff}}-P_{\mathrm{fit}}$.}
  \label{fig:fitting}
\end{figure*}

\subsection{Signal Compensation} \label{sec:TF}
In practice, after the foreground removal process, part of H\textsc{i} signal will be inevitably removed along with the foregrounds. As we shown in Figure~\ref{fig:PCA}, severe signal loss is shown in the cross-power spectra at all three redshifts, especially in the scale range we are interested in ($k<0.3\, \mathrm{Mpc^{-1}}h$). Therefore, the over-eliminated signal must be compensated.

\begin{figure*}[ht]
  \centering
  \setlength{\figwidth}{0.32\textwidth} 
  \setlength{\figheight}{0.2\textheight} 

  \begin{minipage}{\figwidth}
    \includegraphics[width=\linewidth, height=\figheight, keepaspectratio]{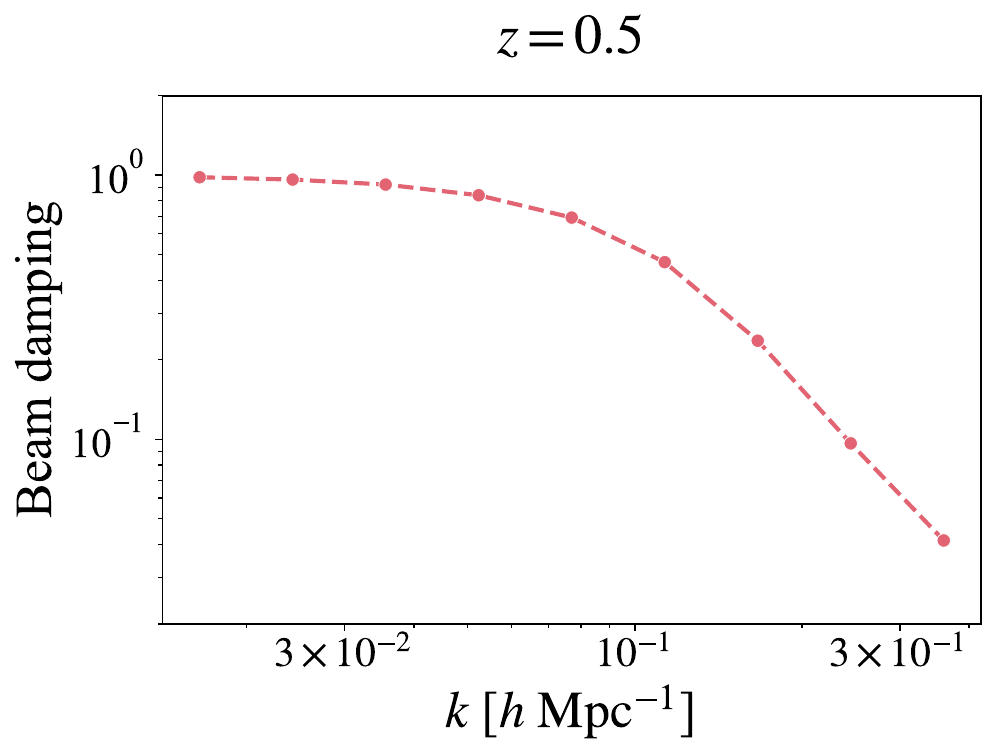}
    \label{fig:c-a}
  \end{minipage}
  \hfill
  \begin{minipage}{\figwidth}
    \includegraphics[width=\linewidth, height=\figheight, keepaspectratio]{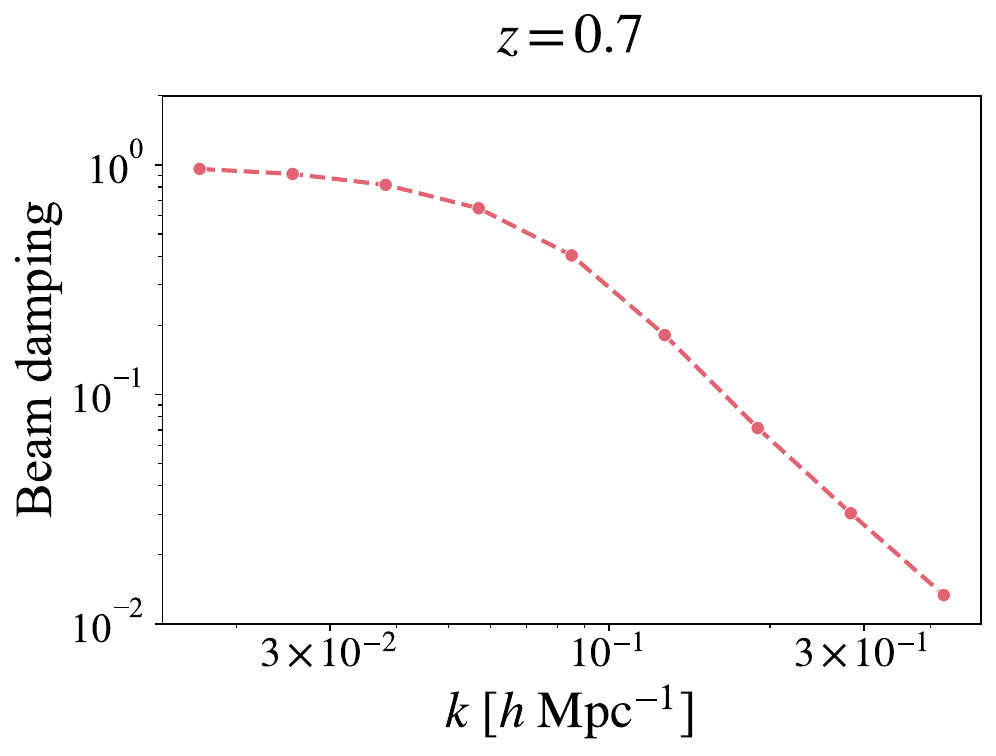}
    \label{fig:c-b}
  \end{minipage}
  \hfill
  \begin{minipage}{\figwidth}
    \includegraphics[width=\linewidth, height=\figheight, keepaspectratio]{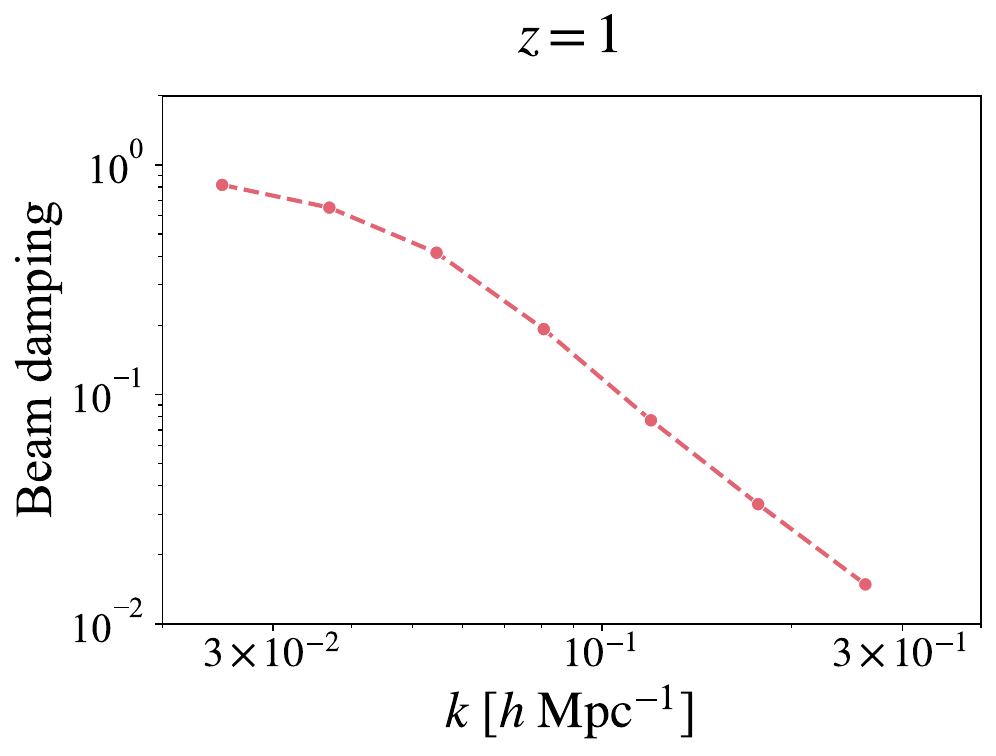}
    \label{fig:c-c}
  \end{minipage}
  \caption{The beam damping effect we model for the cross-power spectra at $z=0.5$, 0.7, and 1.}
  \label{fig:beam damping}
\end{figure*}

We follow the the method given in \citep{transfer_function} to construct the transfer functions for signal compensation of the cross-power spectra. The procedure are described as follows:
\begin{enumerate}
    \item First, we generate the mock data of H\textsc{i} intensity mapping. We use the \textsc{Python} toolkit \textsc{powerbox} to generate high-resolution Gaussian random fields of H\textsc{i} brightness temperature $T_{\rm H\textsc{i}}(\bm{x})$ with the same size as the survey volumes from the neutral hydrogen auto-correlation power spectrum of Jiutian-1G $P_{\rm H\textsc{i}}(k)$. Then we convolve the high resolution H\textsc{i} maps with the beam pattern at each frequency and downgrade the resolution to be same as the corresponding observation maps. The H\textsc{i} mock data will further be transformed into mock data matrix $Y$ with the same dimensions of data matrix $X_{\mathrm{obs}}$.
    \item Then the mock data matrix $Y$ is injected into the data matrix $X_{\mathrm{obs}}$, and PCA is applied with the same projection matrix $W'$ as in the previous foreground removal. Then the foreground removed mock data can be obtained by
    \begin{align}
        Y_c=\left[Y+X\right]-S_{\rm H\textsc{i}}.
    \end{align}
    \item Finally, the signal loss can be determined by transfer function \citep{transfer_function} 
    \begin{align}
        \mathcal{T}(k) = \bigg\langle \frac{\mathcal{P}(Y_c, Y)}{\mathcal{P}(Y, Y)} \bigg\rangle,
    \end{align}
    where $\mathcal{P}()$ denotes the cross-correlation power spectrum and the angled brackets represent the average value over a large number of mocks (in this work, we use 100 mocks for each map).
 \end{enumerate} 

 In Figure~\ref{fig:transfer function} and Figure~\ref{fig:fitting}, we show the transfer function $\mathcal{T}(k)$ and the signal compensation results at all three redshifts, respectively. In Figure~\ref{fig:transfer function}, the differences among the transfer functions are mainly caused by the number of removed principle components $N_{\mathrm{fg}}$, which reflects the mixture of H\textsc{i} signal in the foreground dominated components. In Figure~\ref{fig:fitting}, we can find that the range and level of H\textsc{i} signal loss are becoming wider and higher as we are removing more principle components. We also notice that the behavior of signal compensation is good at all three redshifts, and the level of signal loss is effectively reduced, especially in the $k$ range we are focusing on. Besides, comparing the power spectra in the upper and lower rows in Figure~\ref{fig:fitting}, we can also find that the masking of bright point sources is not sensitive to the compensation result, since they are showing similar results at all three redshifts.

\section{ Cosmological Constraint} \label{sec:Obbr}

\subsection{Theoretical model}

In this section, we explore the cosmological constraint from the MeerKAT and CSST joint surveys. At a given redshift, the theoretical models of the galaxy and 21cm auto-power spectra and galaxy-21cm cross-power spectrum are given by 
\begin{align}
    P_{g}(k) &= P_{g}^{\mathrm{clus}}(k) + P_{g}^{\mathrm{SN}} \nonumber \\
    & = b^2_{g}\,P_{\rm m}(k) + P_{g}^{\mathrm{SN}},
\end{align}
\begin{align}
    P_{\rm H\textsc{i}} (k) &= \overline{T}^2_{\rm H\textsc{i}}b^2_{\rm H\textsc{i}}P_{\rm m}(k) + P_{\mathrm{H\textsc{i}}}^{\mathrm{SN}}\nonumber \\
    & = \overline{T}^{2}_{0}\Omega^2_{\rm H\textsc{i}}b^2_{\rm H\textsc{i}}P_{\rm m}(k) + P_{\mathrm{H\textsc{i}}}^{\mathrm{SN}},
\end{align}
and
\begin{align}
    P_{\mathrm{H\textsc{i}},g}({k}) = &\overline{T}_{0}\Omega_{\rm H\textsc{i}}b_{\rm H\textsc{i}}b_{g}r_{\mathrm{H\textsc{i}},g}P_{\rm m}(k)\times \nonumber \\
    &\hspace{2em}\mathrm{exp}\left[\frac{-(1-\mu^2)k^2R^2_{\rm beam}}{2}\right] + P_{\mathrm{H\textsc{i}},g}^{\mathrm{SN}},
\end{align}
where $\overline{T}_{\rm H\textsc{i}}$ is the mean brightness temperature of H\textsc{i}, $b_g$ and $b_{\rm H\textsc{i}}$ are the galaxy bias and H\textsc{i} bias, respectively, and $r_{\mathrm{H\textsc{i}},g}$ is the cross-correlation coefficient. $P_{\rm m}(k)$ is the linear matter power spectrum which is generated by $\tt CAMB$ with the same cosmology as Jiutian-1G \citep{CAMB}. The exponential factor describes the beam damping effect of the MeerKAT beam on the perpendicular modes. 
$\mu$ is the cosine of the angle from line-of-sight and $R_{\rm beam}$ is the standard deviation of beam profile. $P_{g}^{\mathrm{SN}}$, $P_{\mathrm{H\textsc{i}}}^{\mathrm{SN}}$ and $P_{\mathrm{H\textsc{i}},g}^{\mathrm{SN}}$ are the shot noise terms of the galaxy and 21cm auto-power spectra, and galaxy-21cm cross-power spectrum, respectively.

Modeling the beam damping effect is a challenging part of parameter constraint, since the beam pattern in our simulation has frequency dependence and non-central symmetry. However, since we aim to approach the real data processing, conservatively, we model the beam damping as if we have only a basic understanding of beam patterns. We use Gaussian profile to model the beam profile, and its standard deviation $R_{\rm beam}$ at redshift $z$ can be written as \citep{MeerKAT_WiggleZ}
\begin{align}
    R_{\rm beam}(z) =\frac{1}{2\sqrt{2\mathrm{ln}2}}\frac{\lambda_{0}(1+z)}{D}, 
\end{align}
where $\lambda_{0}=0.21\, \rm m$ is the rest frequency of 21cm emission line and $D$ is the diameter of MeerKAT dish. Here we set a cutoff at $k=\max(k_{\rm perp})$, since the beam only smooths the map on perpendicular modes. The beam damping effect we model are shown in the Figure~\ref{fig:beam damping}.

\subsection{Constraint result}
We fit the mock CSST galaxy auto-power spectrum and 21cm-galaxy cross-power spectrum of MeerKAT and CSST in the  linear scale range ($k<0.3\, \mathrm{Mpc^{-1}}h$) using the least-squares method. We assume that there is no effective detection of the MeerKAT 21cm auto power spectrum, which suffers huge foreground contamination for signal extraction. The best-fit curves of the cross-power spectra can be find in Figure~\ref{fig:fitting}. The best-fit values of $P_{\mathrm{H\textsc{i}},g}^{\mathrm{SN}}$ at $z=0.5, 0.7$ and $1$ are $3.00\pm1.66~\mathrm{mK^2}h^{-3}\mathrm{Mpc^3}$, $2.11\pm0.97~ \mathrm{mK^2}h^{-3}\mathrm{Mpc^3}$ and $6.66\pm1.99~\mathrm{mK^2}h^{-3}\mathrm{Mpc^3}$ for the non-masking H\textsc{i} maps, and $3.05 \pm1.71~\mathrm{mK^2}h^{-3}\mathrm{Mpc^3}$, $2.21\pm1.14~ \mathrm{mK^2}h^{-3}\mathrm{Mpc^3}$ and $6.69 \pm2.07~\mathrm{mK^2}h^{-3}\mathrm{Mpc^3}$ for the bright sources masked H\textsc{i} maps.

\begin{figure}[ht]
  \centering
  \includegraphics[width=\columnwidth]{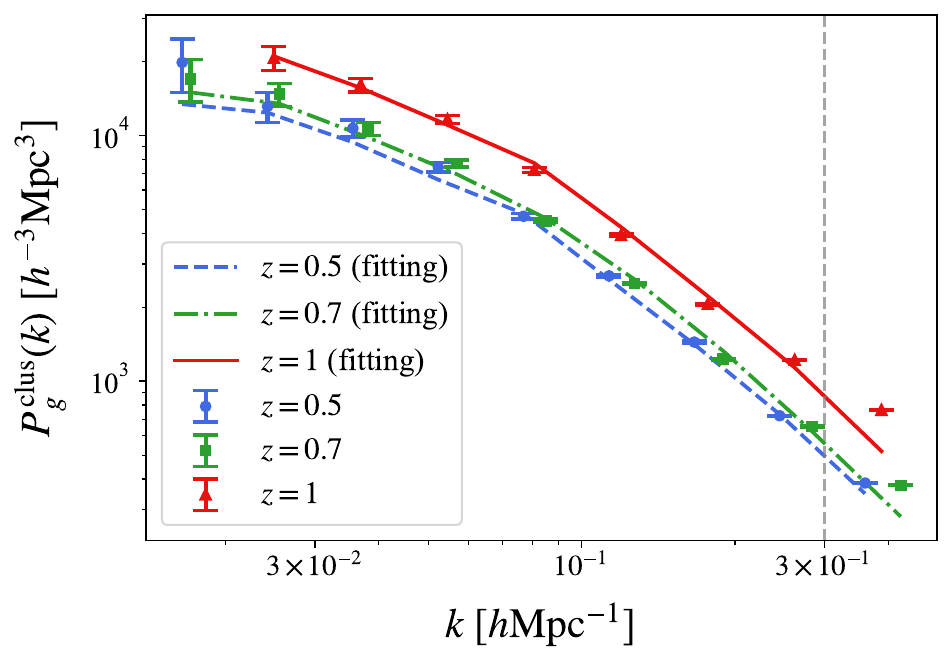} 
  \caption{The mock data of the auto-power spectra of CSST spectroscopic galaxy survey at $z=0.5$ (blue dots), 0.7 (green squares) and 1 (red triangles). The blue dashed, green dash-dotted, and red solid curves are the fitting results at the three redshfits with $k<0.3\, \mathrm{Mpc^{-1}}h$.}
  \label{fig:galaxy_power_spectrum}
\end{figure}

In practice, we can first constrain the parameter product $\Omega_{\rm H\textsc{i}}b_{\rm H\textsc{i}}b_{g}r_{\mathrm{H\textsc{i}},g}$. Besides, since the galaxy bias $b_{g}$ can be estimated individually using the mock data of $P_g^{\rm clus}(k)$ and calculating $P_{\rm m}(k)$ with $\tt CAMB$  (as shown in Figure~\ref{fig:galaxy_power_spectrum}), the parameter product $\Omega_{\rm H\textsc{i}}b_{\rm H\textsc{i}}r_{\mathrm{H\textsc{i}},g}$ can also be constrained through propagation of uncertainty.
In Table.\ref{tab:parameter constraint}, we listed the parameter constraint results of $\Omega_{\rm H\textsc{i}}b_{\rm H\textsc{i}}b_{g}r_{\mathrm{H\textsc{i}},g}$, $\Omega_{\rm H\textsc{i}}b_{\rm H\textsc{i}}r_{\mathrm{H\textsc{i}},g}$ and $b_{g}$.
We can find that the galaxy bias $b_g$ increases with redshift, which matches our knowledge of galaxy distribution,  
and the relative accuracies of $\Omega_{\rm H\textsc{i}}b_{\rm H\textsc{i}}b_{g}r_{\mathrm{H\textsc{i}},g}$ and $\Omega_{\rm H\textsc{i}}b_{\rm H\textsc{i}}r_{\mathrm{H\textsc{i}},g}$ are about $6\%\sim8\%$ for all three redshifts,
which is 3$\sim$4 times smaller than the existing results of MeerKAT \citep{MeerKAT_WiggleZ, MeerKAT_GAMA}. This constraint accuracy level will be helpful to distinguish different models of H\textsc{i} assignment in halos and constrain the co-evolution of galaxies and H\textsc{i} \citep{10.1093/mnras/stx979, Villaescusa-Navarro_2018, 2019MNRAS.484.1007W}.

We also notice that, although masking the bright point sources will lessen the difficulty of foreground removal to some extent, it doesn't have too much affect on the constraint of cosmological parameters. Besides, the constraint accuracy decreases as the increasing of redshift, which is due to the decreasing of both survey area and number density of observable galaxies in our simulation.

Compared to the result obtained using a simplified instrumental and foreground model in \citep{Jiang_2023}, the uncertainty of $\Omega_{\mathrm{HI}} b_{\mathrm{HI}} r_{\mathrm{HI},g}$ in this work is larger by a factor of $\sim6$. 
Although the Statistical error should decrease as the simulated survey area expands, this increase in uncertainty is expected.
Because in this work, we improved the simulation by incorporating more realistic foreground components and more precise instrument effects, such as non-Gaussian beam patterns and polarization leakages. In our simulations at $z=0.5$, the foreground is sufficiently subtracted by PCA with $N_{\mathrm{fg}}=10$, which is close to the latest result of real observations of MeerKAT \citep{MeerKAT_GAMA}, suggesting a good reliability of our forecast on cosmological parameter constraints. The frequency-dependent structures of the beam patterns and polarization leakages work as the key terms of getting close to real observations. Therefore, this work provides a more realistic reference for future studies.

\begin{table*}
\setlength{\tabcolsep}{2mm}{
    \caption{ The best-fit values, errors, and relative accuracies of $b_g$, $\Omega_{\rm H\textsc{i}}b_{\rm H\textsc{i}}b_{g}r_{\mathrm{H\textsc{i}},g}$ and $\Omega_{\rm H\textsc{i}}b_{\rm H\textsc{i}}r_{\mathrm{H\textsc{i}},g}$.} \label{tab:parameter constraint}
    \renewcommand\arraystretch{1.15}
    \begin{tabular}{clccc}
    \hline\hline
    Redshift & \multicolumn{1}{c}{Parameter} & Best-fit value & Error & Relative accuracy \\ \hline
    \multirow{5}{*}{$z=0.5$} & 
        \begin{tabular}{l}$b_g$\end{tabular} 
        & 0.9478 & $\pm 0.0041$ & 0.44\% \\
        & \begin{tabular}{l}
             $\Omega_{\rm H\textsc{i}}b_{\rm H\textsc{i}}b_{g}r_{\mathrm{H\textsc{i}},g}$
        \end{tabular}
        & 0.5455 & $\pm0.0363$ & 6.65\% \\
        & \begin{tabular}{l}
            $\Omega_{\rm H\textsc{i}}b_{\rm H\textsc{i}}r_{\mathrm{H\textsc{i}},g}$
        \end{tabular}
        & 0.5756 & $\pm0.0384$ & 6.66\%\\
        &
        \begin{tabular}{l}
             $\Omega_{\rm H\textsc{i}}b_{\rm H\textsc{i}}b_{g}r_{\mathrm{H\textsc{i}},g}$ \\(masking point sources $>10\mathrm{mJy}$)
        \end{tabular}
        & 0.5489 & $\pm0.0370$ & 6.75\%\\
        & \begin{tabular}{l}
            $\Omega_{\rm H\textsc{i}}b_{\rm H\textsc{i}}r_{\mathrm{H\textsc{i}},g}$ \\(masking point sources $>10\mathrm{mJy}$)
        \end{tabular}
        & 0.5791 & $\pm0.0392$ & 6.76\% \\ \hline
    \multirow{5}{*}{$z=0.7$} & \begin{tabular}{l}$b_g$\end{tabular}  & 1.1058 & $\pm0.0118$ & 1.07\%\\
        & \begin{tabular}{l}
             $\Omega_{\rm H\textsc{i}}b_{\rm H\textsc{i}}b_{g}r_{\mathrm{H\textsc{i}},g}$
        \end{tabular} & 0.6775 & $\pm0.0470$ & 6.94\%\\
        & \begin{tabular}{l}
            $\Omega_{\rm H\textsc{i}}b_{\rm H\textsc{i}}r_{\mathrm{H\textsc{i}},g}$
        \end{tabular} & 0.6127 & $\pm0.0430$ & 7.02\%\\
        & \begin{tabular}{l}
             $\Omega_{\rm H\textsc{i}}b_{\rm H\textsc{i}}b_{g}r_{\mathrm{H\textsc{i}},g}$ \\(masking point sources $>10\mathrm{mJy}$)
        \end{tabular}
        & 0.6701 & $\pm0.0492$ & 7.34\%\\
        & \begin{tabular}{l}
            $\Omega_{\rm H\textsc{i}}b_{\rm H\textsc{i}}r_{\mathrm{H\textsc{i}},g}$ \\(masking point sources $>10\mathrm{mJy}$)
        \end{tabular}
        & 0.6060 & $\pm0.0450$ & 7.42\%\\ \hline
    \multirow{5}{*}{$z=1$}   & \begin{tabular}{l}$b_g$\end{tabular}  & 1.5754 & $\pm0.0231$ & 1.46\%\\
        & \begin{tabular}{l}
             $\Omega_{\rm H\textsc{i}}b_{\rm H\textsc{i}}b_{g}r_{\mathrm{H\textsc{i}},g}$
        \end{tabular} & 1.4595 & $\pm0.0965$ & 6.61\%\\
        & \begin{tabular}{l}
            $\Omega_{\rm H\textsc{i}}b_{\rm H\textsc{i}}r_{\mathrm{H\textsc{i}},g}$
        \end{tabular} & 0.9264 & $\pm0.0627$ & 6.77\%\\
        & \begin{tabular}{l}
             $\Omega_{\rm H\textsc{i}}b_{\rm H\textsc{i}}b_{g}r_{\mathrm{H\textsc{i}},g}$ \\(masking point sources $>10\mathrm{mJy}$)
        \end{tabular}
        & 1.4595 & $\pm0.1005$ & 6.88\%\\
        & \begin{tabular}{l}
            $\Omega_{\rm H\textsc{i}}b_{\rm H\textsc{i}}r_{\mathrm{H\textsc{i}},g}$ \\(masking point sources $>10\mathrm{mJy}$)
        \end{tabular} 
        & 0.9265 & $\pm0.0652$ & 7.04\%\\ \hline
    \end{tabular}
}
\end{table*}

\section{Summary} \label{sec:conclusion}

In this study, we present a comprehensive analysis of the cross-correlation between MeerKAT H\textsc{i} intensity mapping and CSST spectroscopic galaxy surveys at $z = $ 0.5, 0.7, and 1. 
We use Jiutian-1G cosmological simulation to generate the mock MeerKAT H\textsc{i} intensity maps and CSST galaxy catalogs in survey areas from $\sim1600$ to $600$ deg$^2$ at the three redshifts. The H\textsc{i} distribution is obtained by SAM model, and the voxel of the simulation is divided by the angular and frequency resolution based on the MeerKAT receivers. Then the H\textsc{i} brightness temperature of each voxel is estimated to get the H\textsc{i} intensity maps. The CSST galaxy catalog is constructed by applying the $\mathrm{SNR}>5$ threshold to four emission lines (H$\alpha$, H$\beta$, [O\textsc{ii}], and [O\textsc{iii}]) of galaxies, incorporating the instrumental design. To simulate observational conditions, we generate TOD using the on-the-fly scanning strategy, including instrumental noise, beam effects, and polarization leakage. Foreground contamination comprising Galactic synchrotron emission and extragalactic point sources is modeled and added to the mock data.

The foreground removal of H\textsc{i} intensity mapping is performed using the PCA/SVD method. The case of masking bright point sources ($S>10\, \mathrm{mJy}$) to reduce contamination is also explored. After the foreground removal, we construct  transfer functions from 100 mock realizations to compensate for signal loss in the cross-power spectra. Finally, we derive the mock data of the galaxy auto- and 21cm-galaxy cross-power spectra of CSST and MeerKAT, which are used in the cosmological constraints.

While cross-correlation detections between 21\,cm intensity mapping and galaxy surveys have already been achieved with smaller survey areas and lower galaxy number densities and redshifts, our work provides a realistic and quantitative forecast for the forthcoming MeerKAT-CSST joint observations. 
We employ the least-squares fitting method to constrain the cosmological parameter products, i.e. $\Omega_{\rm H\textsc{i}}b_{\rm H\textsc{i}}b_{g}r_{\mathrm{H\textsc{i}},g}$ and $\Omega_{\rm H\textsc{i}}b_{\rm H\textsc{i}}r_{\mathrm{H\textsc{i}},g}$. The relative accuracy is about $6\%\sim 8\%$ at $z=0.4\sim1.2$, which is $3\sim4$ times smaller than the existing cross-correlaiton results of MeerKAT at low redshifts.

By incorporating frequency-dependent beam effects, polarization leakage, realistic foregrounds, our simulation indicates that the full MeerKAT-CSST joint observation with a several thousand square degrees overlapping survey area, the cross-correlation of MeerKAT H\textsc{i} intensity mapping and CSST spectroscopic galaxy survey will provide promising detection of the evolution of neutral hydrogen and its connection to galaxy formation in a wide redshift range, beyond a mere detection of the cross-correlation signal. 
And our pipeline provides a robust framework for analyzing future MeerKAT-CSST joint observations.

\begin{acknowledgments}
Y.E.J. and Y.G. acknowledge the support from the CAS Project for Young Scientists in Basic Research (No. YSBR-092), the National Key R\&D Program of China grant Nos. 2020SKA0110402 and 2022YFF0503404. Y.E.J. and Z.Y.Y are supported by the Program of China Scholarship Council Grant No. 202404910398 and No. 202404910329. X.L.C. acknowledges the support of the National Natural Science Foundation of China through grant Nos. 11473044 and 11973047 and the Chinese Academy of Science grants ZDKYYQ20200008, QYZDJ- SSW-SLH017, XDB 23040100, and XDA15020200. 
Y.-Z. Ma acknowledges the support from South Africa’s National Research Foundation under grant Nos. 150580 and CHN22111069370. Q.G. acknowledges the support from the National Natural Science Foundation of China (NSFC No. 12033008). B.Y. and X.L.C. also acknowledge the support by National Natural Science Foundation of China (NSFC) grants 12361141814. The Jiutian simulations were conducted under the support of the science research grants from the China Manned Space Project with grant No. CMS- CSST-2021-A03. This work is also supported by science research grants from the China Manned Space Project with grant Nos. CMS-CSST-2025-A02, CMS-CSST-2021-B01, and CMS-CSST-2021-A01.
\end{acknowledgments}

\bibliography{sample7}{}
\bibliographystyle{aasjournal}



\end{document}